\documentclass[12pt]{article}
\usepackage{amsmath, amsfonts, amscd}

\newcommand{\be}[1]{\begin{equation}\label{#1}}
\newcommand{\bra}[1]{{\langle #1 |}}
\newcommand{\braket}[2]{{\langle #1 |\, #2\rangle}}

\newcommand{\ee}{\end{equation}}

\newcommand{\ket}[1]{{|\, #1\rangle}}
\newcommand{\ketbra}[2]{{|\, #1\rangle\langle #2|}}

\newcommand{\omg}{\Omega}

\newcommand{\spn}{{\rm span}}

\newcommand{\amg}{\vec{\Omega}}

\newcommand{\N}{\mathbb{N}}
\newcommand{\R}{\mathbb{R}}
\newcommand{\com}{\mathbb{C}}
\newcommand{\as}{\mathfrak{P}}
\newcommand{\rz}{\mathfrak{R}}
\newcommand{\quan}{\mathfrak{Q}}
\newcommand{\loc}{\mathfrak{L}}
\newcommand{\topl}{\mathfrak{T}}

\newcommand{\cem}{{\cal{M}}}

\title{\bf \Large\bf Non-Commutative Topology\\ for Curved Quantum Causality}

\author{Ioannis Raptis\footnote{Department of Mathematics, University of Pretoria, Pretoria 0002, Republic of South Africa; e-mail: iraptis@math.up.ac.za}}

\date{}

\begin{document}

\maketitle

\begin{abstract}

\noindent{\footnotesize A quantum causal topology is presented. This
is modeled after a non-commutative scheme type of theory for the
curved finitary spacetime sheaves of the non-abelian incidence
Rota algebras that represent `gravitational quantum causal sets'. The
finitary spacetime primitive algebra scheme structures for quantum
causal sets proposed here  
are interpreted as the kinematics of a curved and reticular local quantum
causality. Dynamics for quantum causal sets is then
represented by appropriate scheme morphisms, thus it has a purely categorical description that is manifestly
`gauge-independent'. Hence, a schematic version of the Principle of General
Covariance of General Relativity is formulated for the dynamically variable quantum causal sets. We compare our non-commutative scheme-theoretic curved quantum causal
topology with some recent  
$C^{*}$-quantale models for non-abelian generalizations of classical commutative
topological spaces or locales, as well as with some relevant recent results
obtained from applying sheaf and topos-theoretic ideas
to quantum logic proper. Motivated by the latter, we organize our finitary spacetime primitive
algebra schemes of curved quantum causal sets into a topos-like
structure, coined `quantum topos', and argue that it is a sound model
of a structure that Selesnick has anticipated to underlie Finkelstein's reticular and curved quantum causal net. At the end 
we conjecture 
that the fundamental quantum time-asymmetry that Penrose has expected
to be the main characteristic of the elusive `true
quantum gravity' is possibly of a kinematical or
structural rather than of a dynamical character, and we also discuss the
possibility of a unified description of quantum logic and quantum
gravity in quantum topos-theoretic terms.}

\end{abstract}

\newpage

\section*{\normalsize\bf 1. INTRODUCTION CUM PHYSICAL MOTIVATION}

\indent It has been suggested for quite some time now that not only the
geometry of spacetime should be subject to quantum dynamical
fluctuations, but also its topology (Wheeler, 1964). Such
a proposal may be understood as saying that not only the spacetime metric, which physically
represents the gravitational potential in General Relativity (GR), should
be subjected to some sort of `quantization' thus become a `quantum
observable' at Planck scales where quantum gravitational effects are
expected to be significant, but also that the spacetime topology should be
regarded as such a quantum dynamical variable that can in principle be
observed (Isham, 1989, Finkelstein and Hallidy,
1991, Zapatrin, 1998, Breslav {\it et al.}, 1999). Thus, the need for a quantum theory of dynamical spacetime topology has become an
integral and indispensable 
part of our apparently never ending quest for a cogent quantum theory of gravity.

>From a mathematical perspective, while a non-commutative geometry has
already been proposed and significantly developed with main aim to model a quantal
version of the classical spacetime geometry-a project that is
supposed to be the preliminary, albeit important, step towards arriving at a
mathematically well founded quantum gravity (Connes, 1994), an analogous
non-commutative topology has been rather slow in coming. This may be partly
attributed to our apparent inability or inertia so far at posing the `proper' physical
questions about `quantum spacetime topology'. For instance, a
seemingly reasonable question one would be tempted to ask is whether the sought after  
quantum spacetime 
topology is a somehow quantized version of the classical locally Euclidean
manifold topology of the spacetime of General Relativity in the same
way that, say, Canonical Quantum GR purports to quantize the
metric field of the classical theory of gravity ({\it ie}, GR). And
if so, in what sense the spacetime topology may be regarded as a
measurable quantum 
dynamical entity ({\it ie}, a quantum observable proper) ?
Confusion about what constitutes the `right' way of approaching the
problem of quantum spacetime topology may be precisely due to the
unphysicality, hence `inappropriateness', of the questions posed in the first place.

For example, the inappropriateness of the two questions mentioned
above, which are motivated by
analogy to how the metrical features of classical spacetime are
usually quantized, may be {\it prima facie} due to the following reasons: first, the fixed continuous manifold topology of GR is the main
culprit for the
unphysical infinities in the form of singularities that plague it long
before the need to quantize gravity becomes an issue. Second, while GR
provides a classical dynamics for the spacetime metric, but it fixes
the spacetime topology, it does not give us even hints, let alone a
classical theoretical paradigm,  
of how to dynamically vary the spacetime topology\footnote{Equivalently posited, the
dynamical spacetime metric $g_{\mu\nu}$ varies against a `frozen' smooth
background spacetime manifold, so that the geometry and the topology
of spacetime are two fundamentally different and independent of each
other structures: the first is 
variable, relativistic and, in principle, `measurable', while the
second is
constant, absolute and effectively unobservable (Einstein, 1924). Furthermore, $g_{\mu\nu}$, as a
smooth field, depends not only on the fixed continuous topological
({\it ie}, $C^{0}$) structure of spacetime, but also on its
$C^{\infty}$-smooth one which is also 
postulated or fixed up-front in GR for the sake of differential
locality (Einstein, 1924).}. Third, it has been seriously proposed
that the `true quantum gravity' or at least a sound theory for the dynamics of spacetime at quantum scales must account for the fundamental
quantum time-asymmetry (Penrose, 1987, Finkelstein, 1988, Haag,
1990)\footnote{Penrose (1987), for example, maintains that ``{\it the true quantum
gravity is a time-asymmetric theory}''. For more on this, see section 5.}, so that the non-dynamical
character of the classical manifold topology aside, its undirected or 
spatial traits also point to its unphysical and rather non-fundamental nature. Fourth, it is altogether conceptually doubtful
whether the epithet  
`quantum'-the `processual' conception of Nature as being innately dynamical and
ever-fluxing, and the noun `topology'-the mathematical theory for an
inert, static space(time) 
`out there'-can go hand in hand after all\footnote{Thus it is perhaps
better to use the term
`quantum topology' tongue-in-cheek from now on.}. The latter doubt
is pronounced especially in quantum gravity research where the assumption of a
fixed background spacetime manifold has manifested its pathological
nature both classically, due to the singularities that infest GR, and
quantum mechanically, for the weaker but still troublesome infinities
that assail Quantum Field Theory (QFT)\footnote{Furthermore, when GR is treated
as another QFT, like in the QGR approach to quantum gravity, the
non-renormalizable infinities that plague the latter appear to be
insurmountable obstacles on the way to a cogent and finite quantum
gravity. This has prompted many researchers in quantum gravity to
regard the problem of the quantum structure of spacetime to be a 
first step of utmost relevance to the problem of its quantum dynamics
({\it ie}, quantum gravity proper). Thus, a thorough understanding of
the kinematical structure of quantum spacetime will also give us
invaluable clues for its dynamics (Sorkin, 1995, Mallios and Raptis,
2000) (see below and section 5).}. And fifth, from the very general classical conception of
`topology' as `the study of the global features of space', there is
already a tension in terms between the `quantum', which is supposed to
be an
effective way of looking at things at the fine, small-scale, `micro-local' level, and `topology', which is more
likely to be of pragmatic value at the coarse, large-scale,  `macro-global' level of
description of the world-the
aforementioned 
temporal
and spatial semantic differences of the two terms aside.

The causal set (causet) approach to the small-scale structure and dynamics of
spacetime was initially conceived more-or-less with an eye towards
evading the five physical  
impediments to theory contsruction presented above (Bombelli
{\it et al.}, 1987, Sorkin, 1990{\it a,b}, 1995, Rideout and Sorkin, 2000). For the quest of
developing a cogent theory for quantum spacetime topology in particular, perhaps the
most significant feature of Causet Theory (CT) is about the third
point made above, namely, its insistence on a
directed,
causal, hence time-like or temporal conception of topology, rather
than rely on the undirected, static, space-like or spatial undertones
that the usual 
mathematical term `topology' carries with it (Sorkin,
1995)\footnote{See also (Finkelstein, 1988, Raptis, 2000{\it a},
Mallios and Raptis, 2000). Perhaps one should replace the
contradictory nomer `quantum (spacetime) topology' by `quantum causal
topology', or `quantum causality' for short, thus evoke straight-away ideas about
time, chronological structure, processes of change and, ultimately, dynamics.}. Interestingly enough, the causet idea was
born out of considerations of finitary discretizations of the
classical ({\it ie}, $C^{0}$) continuous manifold topology of spacetime (Sorkin, 1991,
1995) in that the basic mathematical structures-the partially ordered sets (posets)-involved in the
latter remained the same, while their physical interpretation changed
fundamentally 
from topo- or choro-logical (spatial) to chrono-logical or causal (temporal) (Sorkin, 1995, Raptis,
2000{\it a}, Mallios and Raptis, 2000). The partial order, when interpreted
causally as the `after' relation between events rather than topologically as
set-theoretic inclusion between `elementary spatial objects' (point-sets), is able to
account for many macroscopic attributes of Lorentzian spacetime such as
its topological ($C^{0}$) and differential ($C^{\infty}$) structure,
its dimensionality ($4$), its
signature ($\pm2$), as well as for its othochronous
spin-Lorentzian ({\it ie}, $SL(2,\com)$) local relativity (Bombelli
{\it et al.}, 1987, Mallios and Raptis, 2000).

On the other hand these `emergent classical properties' of spacetime
from a microscopic realm consisting of fundamental causet substrata may be characterized as
`kinematical' ({\it ie}, of a static structural, non-dynamical kind)
(Mallios and Raptis, 2000). It is understood that if CT, or its quantum descendant
QCT\footnote{That is, Quantum Causal Set (or Qauset) Theory.}
(Raptis, 2000{\it a}, Mallios and Raptis 2000), is supposed to be a
promising candidate for a finite quantum theory of gravity, then one must
be able to describe a quantum dynamics for causets and their
qauset relatives. Albeit, a quantum dynamical scenario for causets, and {\it in extenso}
for qausets, has been quite slow in coming. This may be partly
attributed to our difficulty in conceiving of a way of varying a poset or
equivalently the
incidence Rota algebra corresponding to its associated qauset (Raptis, 2000{\it
a})\footnote{Ray Sorkin in private communication (2000). Also,
in (Rideout and Sorkin, 2000) the authors attribute this persistent lack of a 
dynamics for causets to ``{\it the sparseness of the fundamental mathematical
structure}'' in that ``{\it when all one has to work with is a
discrete set and a partial order, even the notion of what we should
mean by a dynamics is not obvious}''.}. Similarly, the finitary spacetime sheaves (finsheaves) of
qausets model for a locally finite, causal and quantal version of
Lorentzian gravity presented in (Mallios and Raptis, 2000) has been
criticized as being `too kinematical' to qualify as some kind of quantum gravity
proper\footnote{Chris Isham and Lee Smolin in private communication (2000).}. The essentially kinematical
character of the model was already recognized by the authors
in (Mallios and Raptis, 2000) who, on the other hand, also stressed 
the importance of first understanding and developing the kinematics of
a physical theory before tackling the problem of how to formulate the dynamics as Sorkin
had previously suggested and argued for (Sorkin, 1995). 

Having commented on the kinematical character of the finsheaves of qausets
model for quantum gravity, at the end of (Mallios and Raptis, 2000) the following rather
suggestive analogy for finding a dynamics for qausets was mentioned: as the classical topos ${\bf Sh}(X)$
of sheaves of sets over a topological spacetime manifold $X$ can be
thought as a universe of continuously variable sets (Selesnick, 1991,
Lambek and Scott, 1986, Bell, 1988, Mac Lane and Moerdijk, 1992), so a (quantum ?) topos of
finsheaves of qausets may be thought of as a universe of dynamically
variable qausets perhaps varying due to a locally finite, causal and quantal
version of Lorentzian gravity still to be discovered. For conceptual reasons and interpretations coming from modern
algebraic geometry and topos theory {\it per se}, such a topos may be
regarded as a (mathematical) universe of the 
variable non-abelian Rota incidence algebras that model
qausets-physical interpretations ({\it ie},
quantum 
dynamical interpretations of this variability) aside. In any case, the
richer algebraic structure of qausets relative to the
sparse\footnote{See quotation from (Rideout and Sorkin, 2000) in
footnote 7.}
locally finite poset one of
`classical' causets\footnote{The characterization of the causets of Sorkin
{\it et al.} as being `classical' structures is based on (Raptis, 2000{\it a}).} may be used not only to `enhance' the quantum and
operationally sound intepretation
of these structures and their finsheaves (Raptis, 2000{\it a,b}, Mallios and Raptis, 2000), 
but it also gives us hope, mainly inspired by deep ideas and results
in purely mathematical disciplines such as
algebraic geometry and category theory, that topoi of finsheaves of
qausets are the `proper' realms for formulating and studying the
quantum dynamics of spacetime\footnote{See also (Butterfield and
Isham, 2000) for some possible roles that topos theory can play in
quantum gravity.}. For instance, in (Mallios and Raptis,
2000) a finitary and quantal version of the Principle of General
Covariance (PGC) of GR was formulated in terms of finsheaf morphisms such as
the connection ${\cal{D}}$. This, it was emphasized there, implied
independence or gauge invariance of the causal-topological qauset dynamics
from the background parameter base spacetime $X$ on which the
corresponding finsheaves are supposed to be soldered. This sheaf-theoretic
formulation of the PGC is in complete analogy to the discrete version
of the same principle for the dynamics of
classical causets given in (Rideout
and Sorkin, 2000)\footnote{The classical sequential growth-dynamics suggested in this paper
is seen to be independent of the natural number ($\N$) labeling of the
causets' vertices, which labeling is physically interpreted as a `gauge of external
time'. In other words, two causets with the same $\N$-order of labeling of
their vertices are `physically indistinguishable' ({\it ie}, there is no external
clock/time 
to parametrize 
the causets' dynamics for it has been `gauged away'). Similarly, in (Mallios and Raptis, 2000) the
generator of dynamics in a curved finsheaf of qausets was taken to be the
sheaf morphism $\cal D$ corresponding to a finitary spin-Lorentzian
connection. Being a sheaf morphism, $\cal D$ was seen to be 
causal-topologically independent of the `coarse gauges'
${\cal{U}}_{n}$ that one could use to localize approximately or
`coarsely measure' the quantum causal
topology of the dynamically variable qausets.}. 

In this paper we will not go as far as to give an explicit dynamics for qausets in
the aforementioned topos of their finsheaves. We postpone this for 
another couple of papers that are currently in preparation (Raptis, 2000{\it
c,d}). Instead, we will give a scheme-theoretic description of the
kinematics of the curved quantum causality presented in (Mallios and
Raptis, 2000) based on some deep ideas about ring or algebra
localizations in algebraic geometry. The resulting scenario may be
characterized as a first approach to a non-commutative topology for
spacetime at small scales that can also be regarded as a rigorous
mathematical formulation of the physical conception of a `quantum
spacetime topology'. 

So, the present paper is organized as follows: in section 2 we present, at
the physical level of rigour, some rudiments of sheaf and
scheme theory for ring and algebra localizations. We define the
central notion of `primitive algebra schemes'.  In section 3 we recast the finsheaves of qausets
presented in (Mallios and Raptis, 2000) in those basic non-commutative
scheme-theoretic terms. Basically, we will hold that these finsheaves are
in fact primitive algebra schemes. In section 4 we discuss the physical semantics
of this scheme-theoretic model of quantum causal topology with special
emphasis placed on its kinematical character which is seen to be
fundamentally non-commutative and directed. We also compare our scheme-theoretic
non-commutative topology representing the kinematics of the curved
quantum causality presented in (Mallios and Raptis, 2000) against a
recent definition of `quantum points' and a similar non-commutative
$C^{*}$-quantale topology between them by Mulvey and Pelletier
(2000)\footnote{Albeit, without a directly causal
interpretation for this topology like our schematic qausets have (Raptis,
2000{\it a}, Mallios and Raptis, 2000). In fact, the
$C^{*}$-quantales are in a very strong sense `spatial' structures
(Mulvey and Pelletier, 2000).}. We also discuss in some detail some close similarities
between our primitive finitary spacetime schemes of quantum causal
sets models for non-commutative curved quantum causal topology with certain
results about a `warped quantum logic' obtained from applying sheaf and topos-theoretic ideas to
quantum logic proper (Butterfield and Isham, 1998, 1999, Butterfield
{\it et al.}, 2000). Motivated by these similarities we organize our
schemes into a topos-like structure and we argue that it should be
called `quantum topos' after a structure that Selesnick (1991) had
anticipated to underlie Finkelstein's reticular and curved quantum causal net. In section 5 we entertain the idea that the
fundamental quantum time-asymmetry expected of the `true quantum
gravity' (Penrose, 1987) may be of a purely kinematical character if
we assume that the kinematics of a dynamical quantum causal topology
is soundly represented by the curved primitive
finschemes\footnote{`Finitary spacetime schemes'.} of qausets of section 3 and their aforementioned organization into a
quantum topos in section 4. We also argue that this quantum topos
structure may prove to be a unifying platform for quantum logic and
quantum gravity, thus vindicate Lawvere's (1975) deep mathematical insight that
algebraic geometry is in fact geometric logic and Finkelstein's (1969,
1979, 
1996) similarly fundamental physical insight that the world's quantum
logic has its origin in the 
dynamics of quantum spacetime\footnote{See also Selesnick (1991) for a nice discussion of
the close similarities between Lawvere's Topos Theory vision of
unifying logic and geometry, and Finkelstein's Quantum Relativity
vision of unifying the basic principles
of the quantum and relativity theories of the world.}.

\section*{\normalsize\bf 2. SHEAVES, SCHEMES AND ALGEBRA LOCALIZATIONS}

\indent Below we present at a level of rigour suitable and sufficient for
our physical elaborations in the next three sections some basic
elements of sheaf and scheme theory, as well as the main idea  of ring
and algebra localizations for which the latter
theories\footnote{Especially scheme theory.} were primarily  
developed.

First, we recall from (Raptis, 2000{\it b}) that a sheaf $\cal S$ of some
mathematical objects  $O$ over a topological space $X$, written as
${\cal{S}}(X)$\footnote{Or as ${\cal{O}}(X)$.}, may be defined as a local homeomorphism $s$ from the base space
$X$ to the sheaf or `sheafified' fiber or stalk space ${\cal{S}}:=\{ S\}$\footnote{The
term `sheafification' pertaining to the assignment of a suitable
topology to the fiber space $\cal S$ consisting of stalks or fibers of the objects
$O$ (see below). The reader should note the use of calligraphic
letters for a sheaf as a collection or bundle of stalks ({\it ie}, ${\cal{S}}(X)$ or ${\cal{O}}(X)$) and of
non-calligraphic for (the objects dwelling in) these stalks ({\it ie}, $S$
or $O$, respectively). This distinction will be used subsequently
when we define algebra schemes as sheaf-theoretic localizations of
algebraic objects.}: $s:~X\rightarrow
{\cal{S}}(X)$\footnote{Since $s$ is a local homeomorphism ({\it ie}, a
bicontinuous bijection), its inverse $s^{-1}=\pi$ is also a local
homeomorphism (Raptis, 2000{\it b}). The map $\pi:~{\cal{S}}(X)\rightarrow X$ is usually
called `the projection of the sheaf on its base space' or `the
localization of the sheaf in $X$', or even `the soldering of the sheaf
on $X$' (see below).}.  When the objects $O$ residing in the stalks $S$ of ${\cal{S}}(X)$ have `extra'
algebraic
structure, as for example when
one considers sheaves of groups, rings, modules or algebras, it is
understood that this `vertical' structure raised stalk-wise in the
sheaf over the points of the base space $X$ respects or is compatible with the
latter's 
`horizontal continuity' ({\it ie}, its topology). Due to the
aforementioned 
definition of a sheaf as a local homeomorphism $s$, the latter may be
taken as saying that the algebraic operations in a, say, algebra sheaf
${\cal{A}}(X)$\footnote{That is, when the objects $O$ in the stalks $S$ of
${\cal{S}}(X)$ are algebras $A$.}, are `locally continuous'\footnote{For example, the algebraic
product $A_{x}\otimes A_{x}\rightarrow A_{x}$ vertically along the
stalk $A_{x}$ of ${\cal{A}}(X)$ ($x\in X$) is a continuous operation relative to how $x$'s neighbouring points in $X$ are connected to it
({\it ie}, with respect to $x$'s local topology or
connectivity). One
may equivalently say that the $\otimes_{x}$ structure of the stalk
$A_{x}$ of ${\cal{A}}(X)$ is continuous relative to the latter's 
`$\pi$ point-localization index' $x$ in $X$. This is secured by the $\pi$ localization map, since $\pi=s^{-1}$.}. Alternatively, due to the fact that
as an `unsheafified' or non-topologized set the algebra 
sheaf is the disjoint union or direct sum of its stalks, 
${\cal{A}}(X)=\bigcup_{x}A_{x}$, one could also say that the algebra sheaf
space is so topologized, or that the stalks of the sheaf are so `glued
together', that the local algebraic operations in each stalk $A_{x}$
respect $X$'s local connectivity or topology. We may summarize this to
the following motto: ``locally, the sheaf space ${\cal{S}}$ and the underlying base space
$X$ are topologically equivalent or indistinguishable regardless of the
extra algebraic structure that the objects $O$ in the stalks
$S$ of the former may carry''.

In sheaf theory an important notion is that of a section of a
sheaf $S(X)$. Let ${\cal{U}}({\cal{T}})=\{ U\}$ be an open cover for $X$'s topology ${\cal{T}}$,
that is to say, a collection of open sets $U$ the countable unions of
finite intersections of which `generate' $X$ as a topological
space\footnote{Recall that a topology ${\cal{T}}$ for a space $X$, regarded as a
non-topologized point-set, is a
collection of
subsets $V$ of $X$, the so-called `open sets', such that: i) the empty set $\emptyset$ and
the space $X$ itself belong to
${\cal{T}}$, ii) ${\cal{T}}$ is closed under countable unions and finite
intersections. A space $X$ equipped with a topology ${\cal{T}}=\{ V\}$ is called a
topological space, symbolized as ${\cal{T}}(X)$. An open cover
${\cal{U}}$ of a topological space ${\cal{T}}(X)$ is a
collection of open subsets $U$ of $X$ such that every $V$ in the latter's topology
${\cal{T}}$ can be written as (or generated by) countable unions of
finite intersections of the sets in ${\cal{U}}$. One says that
arbitrary unions of finite intersections of the sets $U$ in $\cal U$
generate a subtopology ${\cal{T}}({\cal{U}})$ of ${\cal{T}}(X)$ (Sorkin, 1991, Raptis, 2000{\it b}).}. Assign to each $U$ in ${\cal{U}}$, that is to say,
`locally' in $X$'s subtopology ${\cal{T}}({\cal{U}})$ generated by
$\cal U$, a class $\Gamma$ of continuous
maps $s_{U}$ from $U$ to the sheaf space ${\cal{S}}$, usually written as
$\Gamma(U,{\cal{S}})=\{ s_{U}\}$. The $s_{U}$s in
$\Gamma(U,{\cal{S}})$ are the local, with respect to $X$'s subtopology ${\cal{T}}({\cal{U}})$, sections of ${\cal{S}}(X)$. It is a basic result in sheaf theory that a sheaf is
completely determined by its (local) sections, so that the following slogan
pervades the theory: ``{\it a sheaf is its sections}'' (Mallios,
1998, Raptis, 2000{\it b}). Intuitively speaking, a sheaf ${\cal{S}}(X)$, as a local bicontinuous
bijection 
between the topological base space $X$ and the sheaf space
${\cal{S}}$, is determined solely by the local `basic'\footnote{In the
sense that the $U$s in ${\cal{U}}$ constitute a basis or generating
set for $X$'s subtopology ${\cal{T}}({\cal{U}})$ as explained above.} continuous
maps $s_{U}$ in
$\Gamma(U,{\cal{S}})$ for all basic open sets $U$ in ${\cal{U}}$
covering $X$ as a topological space in the sense described above.

Our decision to present above the sections $\Gamma(U,{\cal{S}})$ of
${\cal{S}}(X)$ relative to an open cover ${\cal{U}}$ of $X$'s topology ${\cal{T}}$, was intended with an eye
towards briefly discussing the notion of `localization of the sheaf's objects
$O$  
with respect to the underlying space $X$' as presented in (Raptis,
2000{\it b}). Central role in this
discussion is played by the notion of germ $[s]_{x}$ of a (continuous)
section $s_{U}$ at the point $x$ of $X$ ($x\in U$). Again, we recall
from (Raptis, 2000{\it b}) that the finest basis for the topology of
${\cal{S}}(X)$ consists of `irreducible' basic open sets of the following sort $(x,[s]_{x})$,
where the second entry of the pair corresponds to the germ of a
continuous section of ${\cal{S}}(X)$. Germs are obtained by a direct limit process of `infinite
localization' or refinement of a net of sets of sections of continuous functions
$\Gamma(U\in{\cal{U}},{\cal{S}})$ defined on a corresponding inverse system of
open covers for the Euclidean manifold ({\it ie}, $C^{0}$) topology of $X$. Briefly\footnote{See
(Raptis, 2000{\it b}) for a more detailed and more technical  account of this `direct limit
process' defining germs of continuous sections of sheaves and how it 
compares to its dual `inverse limit process' of localizing the points
of $X$ from a net  or inverse system of the latter's finitary open
covers as originally presented in (Sorkin, 1991). In the next two
sections we will see that this `categorical' duality between the
inverse and direct limit processes of localization (Raptis, 2000{\it b}) will prove to be a
very fruitful fact indeed.}, by the latter
we mean 
that as $x$ is the product of maximum localization or refinement of
(nested by inclusion) 
open subsets of $X$\footnote{Fancy way of saying that the points of
$X$ are its irreducible `{\it ur}-subsets' obtained at the limit of infinite
refinement of the `fatter' or `coarser' open neighborhoods $U$ about
them. In this sense, ``points are the elementary carriers of
$X$'s topology, and the $C^{0}$-manifold topology for $X$ is its finest
one attained at the inverse limit of an inverse system (poset) of its subtopologies partially ordered by inclusion''
(Sorkin, 1991, Raptis, 2000{\it b}).}, so a germ of a continuous section of ${\cal{S}}(X)$ is
simply the maximum restriction of an element $s_{U}$ of $\Gamma(U,{\cal{S}})$
to (take its values in) the stalk $S_{x}$ of ${\cal{S}}(X)$ over $x$ ($x\in U\subset
X$)\footnote{Fancy way of saying that the germ of a continuous section
of a sheaf is the section evaluated at a point of the base space and
taking values in the stalk of the sheaf over it. In this sense,
``{\it the stalks of the sheaf are the carriers of its topology}''-its
irreducible or finest point-like elements 
(Raptis, 2000{\it b}).}. Ultimately, the germs of continuous sections of the sheaf ${\cal{S}}(X)$
of continuous functions on $X$,
that take values in the irreducible, finest, `ultra local' point-like elements of the sheaf,
namely, its stalks ${\cal{S}}_{x}$, together with the finest elements of the
base space $X$,
namely, its points $x$, generate ({\it ie}, they constitute a
basis for) the topology of ${\cal{S}}(X)$. It follows that if the
underlying topological base space $X$ is replaced by a topologically
equivalent relational structure\footnote{Like when the
finitary subtopologies ${\cal{T}}({\cal{U}})$ of the (bounded region
of the) Euclidean
continuum $X$, themselves generated by locally finite open covers
$\cal U$ of $X$ as described above, are effectively substituted by locally
finite $T_{0}$ poset topologies in (Sorkin, 1991, Raptis and Zapatrin, 2000, Raptis, 2000{\it b}).} the germs of
the continuous sections of the sheaf ${\cal{S}}$ over the
latter\footnote{The so-called `finitary spacetime sheaves of
continuous functions over the locally finite poset substitutes of $X$'
presented in (Raptis,
2000{\it b}).} preserve the local generating relations for the topology
of this relational topological base space\footnote{In the case of
Sorkin's finitary poset substitutes of $X$, these generating relations
for the poset topology,
the so-called `local germs of the poset topology', correspond to the
transitive reduction of the partial order about each of the poset's vertices
and they are precisely the immediate arrows between the vertices of
the poset in its Hasse diagram (Breslav {\it et al.}, 1999, Raptis and Zapatrin, 2000, 
Raptis, 2000{\it a}, Mallios and Raptis, 2000). In (Rideout and
Sorkin, 2000) these germs of the poset topology are called `links' and
in the mathematical literature they are also known as `covering
relations of the poset' in the sense that the poset topology is
generated as the transitive closure of the latter (Breslav {\it et
al.}, 1999, Raptis and Zapatrin, 2000, 
Raptis, 2000{\it a}).}. This last remark will prove to be important for the physical
applications in the next section of the abstract scheme theory to be presented below.   

It is clear from the discussion above that the concept of a sheaf ${\cal{S}}$ of
some (algebraic) objects $O$ over a topological space $X$ and the
associated process of localization of these objects over the points of $X$ are notions
intimately related to each other. However, we saw briefly how the
algebraic structure of the objects $O$ of a sheaf ${\cal{S}}(X)$
does not play any essential role in defining the latter's topology in
that the topology of the background base space $X$ is `externally
prescribed' ({\it ie}, a fixed given\footnote{In the case described
above, the locally Euclidean $C^{0}$-manifold topology {\it a priori} fixed for $X$.}) and all that is required of the
(algebraic) structure of the objects of the sheaf that are soldered on
it is that it respects or preserves
this `fixed background local connectivity about $X$'s points'\footnote{For example, in (Raptis,
2000{\it b}), where only the topological features of the sheaf ${\cal{S}}(X)$
of continuous functions over a bounded region $X$ of a spacetime $C^{0}$-manifold
were of particular interest, the algebraic structure of the stalks of the sheaf
played no role whatsoever in the characterization of its
topology. Thus, 
the topology of the ${\cal{S}}(X)$ considered in (Raptis, 2000{\it b})
merely `immitates' the locally Euclidean topology of the given base $C^{0}$-manifold $X$ in that the sheaf space's topology is also locally
Euclidean since, by definition, ${\cal{S}}(X)$ is a local homeomorphism from $X$
to ${\cal{S}}$.}. Indeed, it would be nice to have some kind of a 
sheaf of algebraic objects whose topology and local properties
derive from the algebraic structure of the objects themselves without
any 
essential 
commitment to or dependence on a given fixed external space `out there'. In other
words, space and its (local) properties ({\it ie}, `local topology') should be
ideally derived from the algebraic structure of the objects of
the sheaf itself and not be given up-front, {\it a priori} fixed once
and for all by assuming {\it ab initio} a background topological base space on whose
points 
these algebras are soldered (localized). From a
physical point of view, and in accord with the general algebraic-operational
philosophy of quantum theory, such an idea seems very attractive, since
`spacetime' as a given, fixed, inert, geometric state space `out
there', assumed only to serve as an inert background parameter
space-an external stage that indexes the dynamical propagation and interaction of
physical fields, has
revealed to us its `metaphysical', `chimerical'
nature\footnote{Especially in numerous attempts to unite quantum
mechanics with relativistic spacetime physics, as in `quantum gravity' for instance, the assumption of a fixed
background geometric spacetime manifold is regarded as the main culprit for the
non-renormalizable infinities that afflict the theory thus render it
physically unacceptable, conceptually unsatisfactory, thus deeply
problematic and fundamentally incomplete.} and has prompted many thinkers to value as physically significant only (the algebraic mechanism) of our own
dynamical actions of observing `it'\footnote{That is, `spacetime'.}
(Finkelstein, 1996), which actions
can, in turn, be conveniently organized into algebra sheaves (Mallios,
1998, Raptis, 2000{\it b}, Mallios and Raptis, 2000).

The discussion in the last paragraph motivates the definition of a
general 
scheme (of algebras) as follows\footnote{The following material is taken mainly from
(Hartshorne, 1983, Shafarevich, 1994), but with occasional alterations
of terminology and symbolism, as well as relevant additions and discussion to suit our
exposition here. For instance, here we are interested in schemes of  
(not necessarily commutative)   
algebras and not just of abelian rings as presented in both of these books
(see below).}: let
$A$ be an associative, but not necessarily commutative, algebra. The prime spectrum of $A$, denoted as
$Spec\, A$, is the set consisting of all prime ideals of $A$\footnote{Recall that an ideal ${\cal{P}}$ in an algebra $A$ is prime whenever $ab\in
{\cal{P}}\Rightarrow (a\in {\cal{P}})\vee(b\in {\cal{P}});~(a,b\in A)$.}. The basic idea in
general scheme theory is first to `appropriately' topologize $Spec\, A$, and then localize
the algebra $A$, as a sheaf ${\cal{A}}$ of objects $O$ isomorphic to
$A$, over the `points' of its prime 
spectrum, that is to say, over its own prime ideals. In this way, the
ideas in the last paragraph about an algebra sheaf over a
topological space that derives from the algbera itself and which is not
`externally prescribed', are realized. Indeed, the underlying base
space of a scheme of algebras $A$ is taken to be its own prime
spectrum $Spec\, A$ suitably topologized. Then, over every prime ideal
${\cal P}$ in $Spec\, A$ an isomorphic copy of $A$ is
raised, written as $A_{{\cal{P}}}$. Subsequently, an $A$-algebra sheaf
${\cal{A}}$\footnote{An $A$-algebra sheaf will also be called an
`$A$-sheaf' for short.} is defined, as a non-topologized or non-sheafified set of stalks,
to be the disjoint union of stalks of the form  $A_{{\cal{P}}}$ over
each of 
$Spec(A)$'s points $\cal P$; ${\cal{A}}:=\bigcup_{{\cal{P}}\in
Spec\, A}A_{\cal{P}}$-in complete analogy to the sheaf ${\cal{S}}(X)$ described
before. Then, by giving to $Spec\, A$ a fairly `natural'
topology\footnote{The epithet `natural' pertaining to a topology
defined by using solely the algebraic structure of $A$ as explained
above. For $R$-ring sheaves ${\cal{R}}$ ($R$-sheaves) a natural topology ${\cal{T}}$ for
$Spec\, R$ is the
so-called Zariski topology (Hartshorne, 1983). Here we need not digress and analyze further this most interesting of `closed
subset topologies' on a ring's prime spectrum. In section 5 we will
encounter another closed subset topology, albeit a non-commutative one, associated with a non-abelian
$C^{*}$-algebra called a `quantale' (Mulvey and Pelletier, 2000).}, one defines the sheaf
${\cal{A}}(Spec\, A)$ as a local homeomorphism
$\alpha:~Spec\, A\rightarrow {\cal{A}}(Spec\, A)$ again in complete
analogy to the ${\cal{S}}(X)$ case above. Thus, like in the case of the sheaf
${\cal{S}}(X)$ 
of continuous functions on the topological manifold spacetime $X$, the
stalks $A_{\cal{P}}$ of ${\cal{A}}(Spec\, A)$ are called
`$A$-algebra localizations at the points $\cal P$ of $Spec\, A$'. It
follows that the germs $[\alpha]_{\cal{P}}$ of continuous (in $Spec\, A$'s Zariski topology for example) 
sections of ${\cal{A}}(Spec\, A)$ take values in the latter's stalks,
so that, like in ${\cal{S}}(X)$, the basic open sets generating
${\cal{A}}(Spec\, A)$'s topology are of
the form $({\cal{P}},[\alpha]_{\cal{P}})$ and ${\cal{A}}(Spec\, A)$ is
generated by its germs of continuous sections at the points $\cal
P$ of
its topological base space $Spec\, A$. Thus, like ${\cal{S}}(X)$,
``${\cal{A}}(Spec\, A)$ is its sections'' (Mallios, 1998,
Raptis, 2000{\it b}). 

Now we are in a position to give three formal definitions,
more-or-less taken from (Hartshorne, 1983, Shafarevich, 1994), in
order to arrive at the abstract, but important for our study here and
the physical applications to come, notion
of a `primitive $A$-scheme' which will be amply used in the next section:

\vskip 0.1in

\noindent{\bf Definition.} An $A$-algebraized space is a pair
$(X,{\cal{A}})$ consisting of a topological space $X$ and 
a sheaf ${\cal{A}}$ of algebras $A$ over it. The sheaf, denoted by
${\cal{A}}(X)$, is called `the structure sheaf of
$X$'\footnote{Subsequently, after we have defined primitive algebra
schemes, we will abuse the standard terminology and instead of calling
${\cal{A}}(X)$ `the structure sheaf of the base space $X$ of the  $A$-scheme',
we will simply call it `the structure sheaf of the $A$-scheme'.}.

\vskip 0.1in

A `primitive $A$-spectrum' defined next is a particular instance
of an $A$-algebraized space when the topological base space $X$ of the
structure sheaf ${\cal{A}}(X)$ is identified with $A$'s own primitive
spectrum ${\bf Spec}\, A$ provided the latter is suitably topologized\footnote{Recall that the primitive
spectrum ${\bf Spec}\, A$ of an algebra $A$ is defined to be the collection of
primitive ideals ${\cal{I}}$ in $A$. Also recall that an ideal $\cal
I$ of an algebra $A$ is called primitive if the factor algebra
$A/{\cal{I}}$ is simple which, in turn, means that it has no subideals
other than $0$ and itself. The primitive ideals of an algebra $A$ are in $1$-$1$
correspondence 
with the kernels of (equivalence classes of) its irreducible representations (irreps)
(Zapatrin, 1998, Breslav {\it et al.}, 1999, Raptis
and Zapatrin, 2000).}:

\vskip 0.1in

\noindent{\bf Definition.} Let $A$ be an algebra\footnote{Associative,
but not necessarily commutative. In fact, in the following two sections the
non-abelian case will interest us.}. A primitive $A$-spectrum is the
pair $({\bf Spec}\, A,{\cal{A}})$ consisting of the primitive spectrum
${\bf Spec}\, A$
of an algebra $A$, which is suitably topologized, and an 
$A$-sheaf ${\cal{A}}$ over it.
   
\vskip 0.1in

With an eye towards applying the abstract definitions of this section
to our particular physical model in the next, we give the definition of a primitive
$A$-scheme as follows:

\vskip 0.1in

\noindent{\bf Definition.} A primitive $A$-scheme is an $A$-algebraized
space which locally looks like a primitive $A$-spectrum\footnote{The
three definitions above may be found almost {\it verbatim} in either
Hartshorne (1983) or Shafarevich (1994), but with rings ($R$) and
$R$-sheaves ${\cal{R}}$ instead of algebras ($A$) and $A$-sheaves
${\cal{A}}$, hence also with `$R$-ringed' spaces instead of `$A$-algebraized' ones, as
well as with prime spectra $Spec\, R$ of $R$s instead of primitive
ones ${\bf Spec}\, A$ of $A$s. With these substitutions, the
third definition of primitive schemes above corresponds to the
standard definition of a (general) scheme in (commutative) algebraic geometry.}.

\vskip 0.1in

The adverb `locally' in the definition of a primitive scheme above
requires some further explanation. Again, we follow the general lines
of (Hartshorne, 1983, Shafarevich, 1994), keeping the differences in nomenclature 
mentioned in the last footnote in mind, and call an $A$-algebraized
space a `locally $A$-algebraized space' if for each point $x$ of the
base topological space $X$ the
stalk ${\cal{A}}_{x}$ is an isomorphic copy of $A$\footnote{The
definition of a `locally primitive $A$-spectrum', as a particular kind 
of a locally $A$-algebraized space, follows directly from
this.}. Thus, the definition of a primitive $A$-scheme may be
re-expressed as follows:

\vskip 0.1in

\noindent{\bf Definition.} A primitive $A$-scheme is a locally
primitive $A$-spectrum.  

\vskip 0.1in

The last definition essentially means that for every point ({\it ie}, primitive
ideal) ${\cal{I}}$ in ${\bf Spec}\, A$ there is an open neighborhood
$U({\cal{I}})\subset {\bf Spec}\, A$ about it such that the
`restriction subsheaf'\footnote{For the notion of a subsheaf of a
sheaf see (Hartshorne, 1983, Mallios, 1998).}
${\cal{A}}({\bf Spec}\, A)|_{U({\cal{I}})}$ of ${\cal{A}}({\bf Spec}\,
A)$ is isomorphic to a primitive $A$-spectrum\footnote{The analogue in
(Hartshorne, 1983, Shafarevich, 1994) of our primitive $A$-spectrum is
a so-called `affine $R$-scheme', while that of our primitive
$A$-scheme, a `general $R$-scheme' or simply `$R$-scheme'. Thus a general scheme
is a locally affine scheme.}.

To complete the definition of primitive $A$-schemes above we need to
explain a bit more the word `isomorphic' in the last sentence. For
this we first have to define the notion of $A$-scheme morphisms again along
the general 
lines of Hartshorne (1983) and Shafarevich (1994):

\vskip 0.1in

\noindent{\bf Definition.} A morphism of $A$-algebraized spaces
$(X,{\cal{A}}^{1}(X))$ and $(Y, {\cal{A}}^{2}(Y))$ is a pair $(f,f^{\#})$ of a
continuous map $f:~X\rightarrow Y$ between the underlying 
topological base spaces, and a map
$f^{\#}:~{\cal{A}}^{1}(X)\rightarrow{\cal{A}}^{2}(Y)$ which is a sheaf
morphism\footnote{For the definition of sheaf morphisms see
(Hartshorne, 1983, Mallios, 1998).}.

\vskip 0.1in

\noindent{\bf Definition.} A morphism of locally $A$-algebraized
spaces is a morphism $(f,f^{\#})$ of $A$-algebraized spaces such that
for each point $x\in X$, the induced map of local algebras (stalks)
$f_{x}^{\#}:~{\cal{A}}^{2}(Y)|_{y=f(x)}\rightarrow{\cal{A}}^{1}(X)|_{x}$
in the respective structure sheaves is a local homomorphism of local algebras\footnote{See (Hartshorne, 1983) for
more details on this. Note in particular that if ${\cal{A}}^{1}_{x}$ and
${\cal{A}}^{2}_{y=f(x)}$ are the localizations of an algebra $A^{1}$ over
$x\in X$
and of another algebra $A^{2}$ over $y\in Y$ in the respective locally $A$-algebraized spaces ({\it
ie}, the stalks of the corresponding structure sheaves ${\cal{A}}^{1}(X)$
and ${\cal{A}}^{2}(Y)$ over $x$ and $y$, respectively), the algebra homomorphism
$\phi:~{\cal{A}}^{1}_{x}\rightarrow{\cal{A}}^{2}_{y}$ between them is called
`local' if it preserves their maximal ideals ${\cal{M}}$ as follows:
$\phi^{-1}({\cal{M}}^{2}_{y})={\cal{M}}^{1}_{x}$
(${\cal{M}}^{1}_{x}\in{\rm Max}\,{\cal{A}}^{1}_{x}$, ${\cal{M}}^{2}_{y}\in{\rm Max}\,{\cal{A}}^{2}_{y}$).}. 

\vskip 0.1in

A morphism $(f,f^{\#})$ between two locally $A$-algebraized spaces is
called an `isomorphism' if $f$ is a homeomorphism of the underlying
topological base spaces and $f^{\#}$ an isomorphism of
sheaves\footnote{That is, $f^{\#}$ is a bijective sheaf morphism
(Mallios, 1998).}. 

The definitions above, when applied to the case of primitive
$A$-schemes, define primitive $A$-scheme morphisms and isomorphisms,
respectively\footnote{Note in particular that a primitive $A$-scheme
morphism is simply required to carry, stalk-wise, primitive ideals from
the $A^{2}_{{\cal{I}}_{2}}$ stalk to primitive ideals in
the $A^{1}_{{\cal{I}}_{1}}$ stalk in such a way that the local
topology of the respective primitive spectra base spaces is 
preserved.}. Finally, a pair $(f,f^{\#})$ of
endomorphic 
maps $f:~{\bf Spec}\, A\rightarrow{\bf Spec}\, A$ and
$f^{\#}:~{\cal{A}}({\bf Spec}\, A)\rightarrow {\cal{A}}({\bf Spec}\,
A)$, so that $f$ is a homeomorphism and $f^{\#}$ a sheaf isomorphism, is called
a `primitive $A$-scheme automorphism'. From the discussion about 
germs of sheaves before it
follows that a primitive $A$-scheme automorphism preserves the germs
of the local topology of the topological base space ${\bf Spec}\, A$,
and in the structure $A$-sheaf ${\cal{A}}$ over the latter,  the
germs of its continuous (in ${\bf Spec}\, A$'s topology) local sections. In particular, when a relational
topology ${\cal{T}}$, such as a poset topology, is given to ${\bf
Spec}\, A$\footnote{As will be given to ${\bf Spec}\, A$ in the
following section.}, a primitive $A$-scheme automorphism preserves the latter's
germ-relations or links\footnote{See footnote 28.}, as well as the corresponding
germs of ${\cal{A}}$'s continuous local sections which, by the definition of
${\cal{A}}$ as a local homeomorphism, map the germ-relations of the
topological base space to similar generating relations in the
structure sheaf space ${\cal{A}}$.

In closing this section we follow (Hartshorne, 1983) and alternatively
refer to the topological base space $X$ without its overlying
scheme structure as the `space of $X$', and write $sp(X)$. It is the
essence of scheme theory that for schematic $A$-algebra localizations, 
$sp(X)$ derives from the algebra itself and it is not given `from
outside' as explained above. In the next section we will see how
point-events and a 
quantum causal topology on them may be extracted by a so-called
`Gel'fand spatialization procedure' from the incidence algebras representing
qausets which, in turn, will be seen to be localized, as primitive
$A$-schemes, over ${\bf Spec}\, A$. Hence in the latter schemes,
$sp(X)\equiv{\bf Spec}\, A$, and the space of the scheme is extracted
by spatialization from the very $A$-stalks of its structure sheaf
$\cal{A}$.  
  
As we said earlier, in the following section we present the curved finsheaves of qausets in (Mallios and
Raptis, 2000) as primitive $A$-schemes as defined above.

\section*{\normalsize\bf 3. NON-COMMUTATIVE PRIMITIVE $\amg$-SCHEMES OF QAUSETS}

\indent In this section we plan to give a brief history of QCT
picking from each stage of its development the elements of the theory
that are of immediate interest to our work here, also giving  the
corresponding references to the literature. This selective review of
QCT will culminate in presenting the curved finsheaves of qausets in (Mallios and
Raptis, 2000) as primitive $A$-schemes and it will highlight the way in
which very general physical considerations are able to determine the
precise mathematical structure of a model for (at least the kinematics
of\footnote{See (Mallios and Raptis, 2000) and section 5.}) a physical theory-in our case `Quantum Causal Dynamics'
(QCD)\footnote{QCD is the locally finite, causal and quantal version of
Lorentzian gravity (Mallios and Raptis, 2000) whose acronym should not be confused with the standard one for
Quantum Chromodynamics.}.

As it was mentioned in the introduction, CT (Bombelli {\it et al.},
1987, Sorkin, 1990{\it a,b}) was originally motivated by discrete approximations of the
continuous ({\it ie}, $C^{0}$) topology of a spacetime manifold (Sorkin, 1995). These
so-called `finitary substitutes' of the $C^{0}$-topology of spacetime
were seen to be posets having the structure of $T_{0}$ topological
spaces (Sorkin, 1991)\footnote{These posets may be called `topological
posets' (Raptis, 2000{\it a}) and the topological spaces that they
stand for `relational topologies', the relation being a partial
order. See discussion earlier around footnote 27.}. The physical interpretation of these structures were as
locally finite approximations of the locally Euclidean manifold
topology of (a bounded region $X$ of) classical spacetime, whereby, a
spacetime point-event in the latter is effectively 
substituted by a `coarse' open neighborhood about it and the region $X$ is
covered by a locally finite number of the latter. From such locally finite
open covers $\cal U$ of $X$\footnote{An open cover $\cal U$ of $X$ is
said to be locally finite if for each point $x$ of $X$ there is an
open neighborhood about it that meets a finite number of the covering open sets in
$\cal U$.}, Sorkin (1991) extracted by a suitable `algorithm' the aforementioned
`topological posets'. The soundness of the interpretation of
topological posets as finitary replacements of the continuum $X$ rests
on the fact that an inverse system of the latter possesses an inverse
limit topological space which is homeomorphic to $X$. This limit may
alternatively be stated as follows: the $C^{0}$-topology of $X$ is
recovered at maximum localization or refinement of $X$'s
point-events (Raptis, 2000{\it b}).

Indeed, with every finitary poset substitute of $X$ a sheaf of
appropriately defined continuous functions (on $X$) was defined in
such a way that, as a space on its own, was seen to be locally
homeomorphic to the poset, thus, technically speaking, a sheaf over it
(Raptis, 2000{\it b}). These structures were coined `finitary spacetime
sheaves' (finsheaves) and their physical interpretation was as locally finite or
`coarse' 
approximations of the $C^{0}$-spacetime observables. The soundness of
this interpretation rests on the fact that an inverse system of such
finsheaves was seen to yield, again at the limit of maximum
localization or refinement of $X$ into its point-events, a space
homeomorphic to ${\cal{S}}(X)$-the sheaf of continuous functions on
$X$. Thus, there is a significant change of emphasis in the physical
interpretation of finsheaves in comparison to that of finitary substitutes: from rough approximations of spacetime
point-events in (Sorkin, 1991), to coarse approximations of (algebras
of) operations of localization of spacetime point-events and of their
locally Euclidean manifold topology, as these (algebraic) operations  
reside in the (algebra) stalks of the corresponding finsheaves
(Raptis, 2000{\it b})\footnote{The words `algebra' and `algebraic' are
put in parentheses above because, as it was mentioned in the previous section, in (Raptis, 2000{\it b}) no allusion to
the particular algebraic structure of the stalks of the sheaf ${\cal{S}}(X)$ was
made. We just note with an eye towards the next section that ${\cal{S}}(X)$
is usually taken to be the `commutative sheaf' ${\cal{C}}^{0}(X)$ of abelian $C^{*}$-algebras $C^{0}(X,\com)$
of continuous complex-valued functions on $X$, so that the locally
Euclidean manifold topology of $X$ is usually identified with that of the sheaf
${\cal{C}}^{0}$. Due to this identification we used `locally Euclidean
manifold topology' and `$C^{0}$-topology' interchangeably above. Also due to
this identification, the locally Euclidean manifold topology of $X$
may be characterized as `commutative' or `classical'-effectively a
`locale' (Mac Lane and Moerdijk, 1992, Mulvey and Pelletier, 2000, see
next 
section)
which is regarded as a generalization of classical topological spaces. It
follows that a sheaf of non-commutative algebras, playing the role of
the structure sheaf of a non-commutative scheme, may be associated
with a `non-commutative' or `quantal' topological base space-a
`quantale' (Mulvey and Pelletier, 2000; see next section).}. From the discussion in the previous section about sheaves whose base
spaces are relational topologies, it follows that the germs of the
finsheaves in (Raptis, 2000{\it b}) preserve the links or immediate
arrows of their underlying $T_{0}$ topological poset base spaces.

In (Raptis and Zapatrin, 2000) an algebraic quantization
procedure for the finitary relational poset topological spaces of (Sorkin,
1991) was presented. This essentially involved the association with
every finitary poset substitute $P$ of the continuous manifold $X$ of
an algebra $\omg(P)$, the so-called `incidence algebra of the poset'
(Rota, 1968), and the `dual' assignment\footnote{The epithet `dual' is
given in the categorical sense of the word, that is to say, it means
that the two 
maps corresponding to these dual to each other assignments will be seen subsequently to be contravariant
functors between the respective categories of finitary posets/poset
morphisms and of 
incidence algebras/algebra homomorphisms associated with them.} of a poset $P(\omg)$ to an
arbitrary finite dimensional algebra $\omg$. The first association we
may 
formally represent by the arrow  
$P{\stackrel{\omega}{\rightarrow}}\omg(P)$, while its dual or
`opposite'\footnote{The epithet `inverse' could also be used instead
of `opposite'. We will see later that in a subtle sheaf-theoretic sense the maps
$\omega$ and $p$ are inverse of each other and are the finitary
correspondents of the `sheafifying map' $s:~X\rightarrow{\cal{S}}(X)$
and its inverse `soldering map' $\pi:~{\cal{S}}(X)\rightarrow X$, respectively, that we
saw in the previous section in connection with continuous spacetime sheaves.} by the arrow $\omg{\stackrel{p}{\rightarrow}}P(\omg)$. We
first present the $p$-correspondence, then the $\omega$ one.

The $p$-association $\omg{\stackrel{p}{\rightarrow}}P(\omg)$ is of
special interest to us here and corresponds to a construction originally presented in detail in (Zapatrin,
1998) called `Gel'fand spatialization procedure'\footnote{See also
(Breslav {\it et al.}, 1999, Raptis 2000{\it a}).}. We briefly review
it below: let $\omg$ be a finite dimensional, associative, non-abelian algebra. Let $p$ and $q$ be two (equivalence classes of) irreps of
$\omg$\footnote{Explicit matrix irreps of incidence algebras can be found in (Zapatrin, 1998).} whose kernels $p^{-1}(0)$ and $q^{-1}(0)$
are primitive ideals ${\cal{I}}_{p}$ and ${\cal{I}}_{q}$ in it ({\it
ie}, ${\cal{I}}_{p},{\cal{I}}_{q}\in{\bf Spec}\,\omg$). One regards as
vertices of $P(\omg)$ the points of ${\bf Spec}\,\omg$ and builds
the immediate arrows or links between them according to the following rule

\begin{equation}
p{\stackrel{*}{\rightarrow}} q\Leftrightarrow {\cal{I}}_{p}\rho{\cal{I}}_{q}:= 
{\cal{I}}_{p}{\cal{I}}_{q}(\not={\cal{I}}_{q}{\cal{I}}_{p}){\stackrel{\not=}{\subset}}{\cal{I}}_{p}\cap{\cal{I}}_{q}
\end{equation}

\noindent where ${\cal{I}}_{p}{\cal{I}}_{q}$ is understood as the
product of subsets of $\omg$\footnote{That is, the `product ideal' of ${\cal{I}}_{p}$ and
${\cal{I}}_{q}$ in $\omg$. Note in the parenthesis in $(1)$ that
primitive ideals (to be subsequently identified with `quantum points') do not commute:
${\cal{I}}_{p}{\cal{I}}_{q}\not={\cal{I}}_{q}{\cal{I}}_{p}$. In the
next section this observation will prove to be invaluable both for the
quantum and the local quantum time-directed interpretation of the structure
of the $\amg^{s}$s, as well as for their comparison with the
non-commutative 
$C^{*}$-quantales of Mulvey and Pelletier (2000).} and ${\cal{I}}_{p}\cap{\cal{I}}_{q}$ is
their `intersection ideal' in $\omg$. ${\stackrel{*}{\rightarrow}}$ represents
links in $P$, while $\rho$ is the generator or germ of $\omg$'s Rota
topology ${\cal{T}}_{\rho}({\bf Spec}\,\omg)$\footnote{`$\rho$', as an
index of ${\cal{T}}$, indicates that it is the generating or `germ' 
relation of this topology.}
(Breslav {\it et al.}, 1999, Raptis and Zapatrin, 2000, Raptis,
2000{\it a}, Mallios and Raptis, 2000). 

In fact, the identification in $(1)$ of $P$'s immediate
arrows or links ${\stackrel{*}{\rightarrow}}$ with the germ $\rho$
of ${\cal{T}}({\bf Spec}\,\omg)$ is a theorem\footnote{See (Breslav {\it et al.}, 1999).} which can be stated thus: the Sorkin
$T_{0}$ 
poset topology of $P(\omg)$ is obtained as the transitive closure of
${\stackrel{*}{\rightarrow}}$ when the latter is identified with the
germ $\rho$ of $\omg$'s Rota topology\footnote{Equivalently stated,
$\rho$ is the transitive reduction of the partial order relation 
`$\rightarrow$' ({\it ie}, formally
$\rho\equiv{\stackrel{*}{\rightarrow}}$) that defines the Sorkin poset
topology of $\omg(P)$ (Raptis, 2000{\it a}).}.  Thus, the Gel'fand
spatialization procedure $\omg{\stackrel{p}{\rightarrow}}P(\omg)$ first 
consists of
suitably topologizing ${\bf Spec}\,\omg$\footnote{That is, give to its
points ${\cal{I}}_{p}$ the Rota topology ${\cal{T}}_{\rho}({\bf Spec}\,\omg)$
generated by $\rho$.}, then builds a topological poset on it
whose links between its points\footnote{That is, the (kernels of equivalence
classes of) irreps of
$\omg$.} are drawn precisely when the generating relation $\rho$ holds
between the corresponding ideals in ${\bf Spec}\,\omg$. It is
important to remark that when the algebra $\omg$ is commutative, the
Rota topology is trivial in the sense that it is discrete ({\it ie},
no linked pairs $p{\stackrel{*}{\rightarrow}}q$), so that interesting
topologies arise only when $\omg$ is non-commutative as we assumed above (Zapatrin, 1998,
Breslav {\it et al.}, 1999, Raptis and Zapatrin, 2000)\footnote{For
instance, in the commutative case the Gel'fand topology is always
discrete. Also, for finite
dimensional algebras $\omg$ that we consider as being `physically pragmatic'
(Raptis and Zapatrin, 2000), another topology, the so-called Jacobson
one, that can
be imposed on ${\bf Spec}\,\omg$ is always discrete (Breslav {\it et
al.}, 1999); hence our choice
of the Rota topology for our finite dimensional, non-abelian Rota
algebras $\omg$.}. The reader should also note in $(1)$
that homomorphic copies of $\omg$, namely $p$ and $q$, index the points of
its primitive spectrum. By `inverting' the symbols to
$\omg({\cal{I}}_{p})\equiv\omg_{p^{-1}(0)}$, one gets a first
impression of an $\omg$-localization procedure over ${\bf Spec}(\omg)$
implicit in the above $p$-construction\footnote{This remark will prove to be crucial in
our subsequent identification of the curved finsheaves of qausets in 
(Mallios and Raptis, 2000) with primitive $A$ or $\omg$-schemes.  The
reader should also note that the name `spatialization' given to the
$p$-construction above (Zapatrin, 1998) is an appropriate one, because
effectively one extracts from $\omg$ a topological space
${\cal{T}}_{\rho}({\bf Spec}\,\omg)$, which will later serve as the
base `space' $sp({\bf Spec}\,\omg)$ over which a primitive
$\omg$-scheme will be erected (see previous section).}.

Which brings us to the dual construction
$P{\stackrel{\omega}{\rightarrow}}\omg(P)$ whereby to a finitary poset
substitute $P$ {\it \`a la} Sorkin (1991) one associates the incidence
algebra $\omega(P)$ according to the following steps:

\noindent a. Represent $p\rightarrow q$ arrows in $P$, by using the Dirac
`ket-bra' operator notation, as $\ketbra{p}{q}$ in $\omg$ (Breslav {\it et al.},
1999, Raptis and Zapatrin, 2000,
Raptis, 2000{\it a}).

\noindent b. Define $\omg(P)$ as a vector space over a field
$F$\footnote{We may take $F$ to be the field $\com$ of complex
numbers. This seems to enhance the quantum interpretation of the
incidence algebras constructed by $\omega$ (Raptis and Zapatrin, 2000,
Raptis, 2000{\it a}).} as follows 

\begin{equation}
\omg(P)=\spn_{F}\{\ketbra{p}{q}:~p\rightarrow q\in P\}
\end{equation}

\noindent c. Define an associative product structure `$\circ$' on
$\omg(P)$ as follows

\begin{equation}
\ketbra{p}{q} \circ \ketbra{r}{s} =
\ket{p} \braket{q}{r} \bra{s}
=
\braket{q}{r} \cdot \ketbra{p}{s} =
\left\lbrace
\begin{array}{rcl}
\ketbra{p}{s} &,& \mbox{if } q=r \cr
0 &&
\mbox{otherwise.}
\end{array} \right.
\end{equation}

\noindent where `$.$' is the usual $F$-scalar multiplication of vectors.

\noindent d. Topologize ${\bf Spec}\,\omg(P)$ by first identifying the
primitive ideals or points in it with subsets of the form

\begin{equation}{\cal{I}}_{p}=\spn_{F}\{\ketbra{q}{r}:~\ketbra{q}{r}\not=\ketbra{p}{p}\}
\end{equation}

\noindent and then $\rho$-relate them as in $(1)$, thus give a
topological space structure to $\omg(P)$'s primitive spectrum 
${\bf Spec}\,\omg(P):=\{{\cal{I}}_{p}\}$ ($\forall p\in P$) as follows

\begin{equation}
{\cal{I}}_{p}\rho{\cal{I}}_{q}:= 
{\cal{I}}_{p}{\cal{I}}_{q}(\not={\cal{I}}_{q}{\cal{I}}_{p}){\stackrel{\not=}{\subset}}{\cal{I}}_{p}\cap{\cal{I}}_{q}
\end{equation}

\noindent One can interpret this $\omega$-process as a lifting
over $\omg(P)$'s points\footnote{That is, its primitive ideals which
are in $1$-$1$ correspondence with the point-vertices of the
`underlying' $P$.} of local isomorphs of $\omg(P)$\footnote{Actually,
`homomorphic' copies of $\omg(P)$ since the underlying points of $P$
correspond to kernels (of equivalence classes) of irreducible homomorphic images ({\it ie}, irreps) of
$\omg$.}. Thus, in a sense, the $\omega$-lifting of $\omg$ over $P$ is
a process `inverse'\footnote{Recall from footnotes 56 and 57 that the words `opposite' or `dual' could also
be used instead of `inverse' (see below).} to the $p$-soldering or localization of an
$\omg$ on the poset $P$ extracted from it {\it \`a la} Gel'fand as it was briefly described in
the passage after $(1)$. 

Now we will comment briefly on the $(\omega ,p)$-pair of `dual constructions' $P(\Omega)
\genfrac{}{}{0pt}{}{\underrightarrow{\rule[-2pt]{0pt}{2pt}\omega}}{\overleftarrow{\rule[-2pt]{0pt}{2pt}p}}
\Omega(P)$. $(\omega ,p)$ may be regarded as a pair of contravariant
functorial correspondences 
between the Sorkin poset category $\as$ of finitary topological posets
 and continuous (in the Sorkin topology) injections between
them\footnote{These are simply the partial order preserving maps $f_{ij}$
between the finitary topological posets $P_{i}$ in (Sorkin, 1991) that
are partially ordered by refinement `$\preceq$'. See
also (Raptis and Zapatrin, 2000, Raptis,
2000{\it b}).}, and
the Rota poset category $\rz$ of incidence algebras and surjective algebra homomorphisms,
provided the posets in the former category are simplicial complexes or
`nerves' in
the sense of (Alexandrov, 1956)\footnote{This is the case for `nice'
topological spaces that one discretizes {\it \`a la} Sorkin
(1991). Thus, the Sorkin poset category $\as$
should be properly called `the Alexandrov-Sorkin category' and it
consists 
of finitary posets or simplicial complexes, and injective poset
morphisms or injective simplicial maps between them.}. The latter condition was shown to
hold in (Raptis and Zapatrin, 2000)\footnote{See also (Zapatrin,
2000).}, thus the Alexandrov-Sorkin poset category is
indeed dual or `opposite' to the Rota
one\footnote{In the sense that the functorial
correspondences between them are in fact contravariant functors
(Raptis and Zapatrin, 2000, Raptis, 2000{\it b}, Zapatrin, 2000). We will comment
further on this duality later in this section when sheaf
and scheme-theoretic localizations of Rota algebras will be involved,
as well in the next section when the primitive finschemes of qausets
will be organized into a quantum topos structure.}. It should be mentioned that the discovery of the dual functoriality of the
$(\omega ,p)$-pair in (Raptis and Zapatrin, 2000), which enables
transitions between $\as$ and $\rz$, was essentially discovered first when Zapatrin
(1998) showed that the composition of the $(\omega ,p)$ constructions
$p\circ\omega:~P{\stackrel{\omega}{\rightarrow}}\omg(P){\stackrel{p}{\rightarrow}}P(\omg(P))$
is an isomorphism between the poset $P$ one starts with and the poset 
$P(\omg(P))$ resulting from the composite construction. One would
normally 
expect that a $P$-morphism should correspond to an
$\omg$-homomorphism, but this is not always the case. As we just said, provided the
original finitary topological poset $P$ {\it \`a la} Sorkin (1991) is
a simplicial complex {\it \`a la} Alexandrov (1956), this would indeed
be the
case\footnote{Thus, it would
be perhaps more accurate to call the Rota poset category $\rz$ `the Rota-Zapatrin category'.}.

One can see more easily that the $P$s in $\as$ are simplicial
complexes if one realizes that the incidence Rota algebras $\omg$
associated with them by the $\omega$-map above are in fact graded
 algebras. The latter is to say that every 
$\omg(P)$ splits into linear subspaces as follows

\begin{equation}
\omg=\omg^{0}\oplus\omg^{1}\oplus\cdots
\end{equation}

\noindent where $\omg^{0}=\spn\{\ketbra{p}{p}:~{\rm deg}(\ketbra{p}{p})=0\}$,
$\omg^{1}=\spn\{\ketbra{p}{q}:~{\rm deg}(\ketbra{p}{q})=1\}$, and so
on (Raptis and Zapatrin, 2000)\footnote{Where `${\rm deg}(p)$' is the
degree or grade or cardinality of the simplex $p$, so that  
${\rm deg}(\ketbra{p}{q})$ counts the difference in cardinality of $p$ and $q$ ({\it ie}, the number of vertices in $P$
mediating between $p$ and $q$).}. In simplicial parlance, the relation
$[p\rightarrow q\in
P]\genfrac{}{}{0pt}{}{\underrightarrow{\rule[-2pt]{0pt}{2pt}\omega}}{\overleftarrow{\rule[-2pt]{0pt}{2pt}p}}[\ketbra{p}{q}\in\omg]$
may be read as `$p$ is a face of $q$'\footnote{See (Raptis and Zapatrin, 2000) for more
details about the simplicial character of the $\omg(P)$s.}. 

This
graded algebra character of the $\omg(P)$s also facilitates their
interpretation as discrete differential manifolds (Dimakis and
M\"uller-Hoissen, 1999, Raptis and
Zapatrin, 2000, Breslav and Zapatrin, 2000). The $\omg^{0}$ subspace of $\omg(P)$\footnote{From
now we will occasionally omit the poset $P$ from which $\omg$ derives by the
$\omega$-map above and simply write $\omg$.} is an abelian subalgebra
of $\omg$ consisting of the algebra's self-incidences\footnote{That is, the
identity arrows (reflexive relations) $p\rightarrow p$ in the underlying $P$ regarded as a
`poset category' ({\it ie}, the arrow product in the latter is simply
the 
$\circ$-concatenation of the partial order arrows in $P$ as in $(3)$).} and the linear combinations thereof, and was coined `the space
of stationaries' in (Raptis and Zapatrin, 2000). The linear subspace
$\omg^{1}$ of $\omg$ consists of linear combinations of the immediate arrows
or links in $P$ and was called `the space of transients' in (Raptis
and Zapatrin, 2000). Finally, the linear subspaces $\omg^{i}$ ($i\geq
2$) of $\omg$ were called `the spaces of paths of length (or duration)
$i$' in (Raptis and Zapatrin, 2000).  

It is the categorical $(\omega ,p)$-duality above between Sorkin's finitary
posets and their associated incidence Rota algebras that enhances the physical
interpretation of the latter as quantum spaces of discrete
differential forms (Raptis and Zapatrin, 2000)\footnote{Briefly,
stationaries are the discrete quantum analogues of the $C^{\infty}$
coordinate functions of events in the smooth classical spacetime
manifold, transients the discrete quantum correspondents of vectors
cotangent to those classical events, and paths the discrete quantum analogues of higher
degree differential forms on the smooth continuum. All in all, $\omg$
is a discrete and quantum analogue of the usual module of differential
forms on the smooth continuum. The epithet
`quantum' refers to the interpretation of the linear structure ({\it
ie}, `$+$') of the $\omg^{i}$s as `coherent quantum superposition'.}. Furthermore, 
the inverse limit that the 
Alexandrov-Sorkin poset category $\as$ possesses when it is equivalently regarded as an inverse system
of finitary topological posets
(Sorkin, 1991, Raptis and Zapatrin, 2000, Raptis, 2000{\it b}), which limit yields the classical
$C^{0}$-manifold topology for spacetime as mentioned above, was interpreted for the
contravariant Rota-Zapatrin poset category $\rz$, now it also regarded as an
inverse system or net, as Bohr's Correspondence Principle by
Zapatrin and this author (2000). Thus, a net of quantum spaces of
discrete differential forms, at the physically non-pragmatic limit of
infinite localization of spacetime events, `decoheres' to a classical
event living in a smooth continuum, the commutative algebra of its smooth coordinates, and the space of differential forms
or covariant tensors tangent to it\footnote{In particular, $\omg^{0}$
decoheres to the abelian algebra of the classical coordinates of
spacetime events (usually taken to be the commutative $C^{*}$-algebra
$C^{\infty}(X,\com)$ of holomorphic functions on $X$), 
$\omg^{1}$ to the Lie algebra of covariant derivations at every 
point-event of the continuum, and the $\omg^{i}$s ($i\geq2$) to forms
or covariant tensors 
of higher grade or degree or rank.}. 

It is also worth mentioning that it is exactly due to the richer algebraic
structure of incidence algebras relative to that of posets that,
first, the former 
have a sound quantum interpretation (Raptis and Zapatrin, 2000, Raptis,
2000{\it a}), second, that they can be interpreted as discrete
differential manifolds that reproduce at the inverse limit not only
the continuous ({\it ie}, $C^{0}$) topology of classical
spacetime, but also its differential ({\it ie},
$C^{\infty}$-smooth) structure (Raptis and Zapatrin, 2000, Raptis, 2000{\it b}, Mallios and
Raptis, 2000), and third, that their
localizations can be studied using powerful concepts, constructions
and results from scheme theory\footnote{This paper.}. This should be
compared against the remarks in (Rideout and Sorkin, 2000) about the
apparent 
sparseness of the locally finite posets' mathematical structure in the
particular case that the 
latter represent causets (Bombelli {\it et al.},
1987)\footnote{See quotation in footnote 7.}.

These last remarks bring us straight to the definition of
qausets. Qausets have been defined as the causally and quantally
interpreted incidence Rota algebras
$\omg$ associated with Sorkin's finitary topological posets $P$ (Raptis,
2000{\it a}) as described above. This definition was mainly inspired
by Sorkin's (1995) insistence on a fundamental change of
physical interpretation for the partial orders involved in his
finitary substitutes for continuous topological spacetimes (Sorkin,
1991) `from topological to causal', while their basic
mathematical structure ({\it ie}, the locally finite poset) was
essentially retained. This
change in the physical semantics of locally finite posets had already resulted
in the definition of causets (Bombelli, 1987, Sorkin, 1990{\it a,b}),
only that qausets, being algebraic and not merely
relational structures, can also afford a sound quantum interpretation
as mentioned above (Raptis and Zapatrin, 2000, Raptis, 2000{\it
a}). Thus, by emulating the sound semantics given to finitary topological
posets in (Sorkin, 1991, Raptis, 2000{\it b}) as `locally finite
approximations of the continuous, locally Euclidean topological
relations between events in a bounded region $X$ of a $C^{0}$-manifold
spacetime $M$', qausets were subsequently interpreted as `locally finite
and quantal replacements of the causal relations between events in
a bounded region $X$ of a (possibly curved) and smooth Lorentzian spacetime
manifold $M$' (Mallios and Raptis, 2000)\footnote{Again, we should
emphasize the important
change in the meaning of the incidence algebras involved: from
quantum topological discretizations of `spatial' nature (Zapatrin, 1998, Breslav {\it et
al.}, 1999, Raptis and Zapatrin, 2000), to reticular quantum causal
topologies of `temporal' character (Finkelstein, 1988, Raptis,
2000{\it a}, Mallios and Raptis, 2000).}. At the end of (Raptis,
2000{\it b}) it was explicitly anticipated that finsheaves of qausets
could play an important role in formulating rigorously and entirely in
algebraic terms a reticular, causal and quantal version of gravity.

Indeed, this possible application of curved finsheaves of qausets to represent a locally
finite, causal and quantal version of Lorentzian gravity\footnote{A
`finitary and causal quantum gravity' so to speak.}, was the main theme
in (Mallios and Raptis, 2000). In that paper it was explicitly shown
how a straightforward translation of the Classical Equivalence
Principle (CEP) of GR\footnote{GR is formulated on a classical smooth
spacetime continuum.} in the reticular, causal and quantal algebraic realm of
qausets,
mandates that the latter be localized or gauged, hence curved. Thus we arrive swiftly at the most
important `structural' result about a sound mathematical model of
the kinematics of a curved (thus dynamical) local quantum causal topology that was the main
goal of this section:

\vskip 0.1in

\noindent {\bf Fact.} The localized or gauged, thus curved, (principal) finsheaves
of qausets presented in (Mallios and Raptis, 2000) are
(principal)\footnote{The epithet `principal' will be discussed shortly
in connection with the local finitary spin-Lorentzian structure
(gauge) 
symmetries of our primitive $\amg$-finschemes of qausets.} 
$A$-finschemes\footnote{Again, the term `finschemes' meaning `finitary spacetime
schemes', and the algebras $A$ being localized in it are the incidence
Rota ones $\omg$ 
modeling qausets. Thus in our context, the $A$-schemes of the previous
section are $\omg$-schemes, issues of finitarity aside.} in the sense defined in the previous section.

\vskip 0.1in

Below we explain analytically this fact by basing our arguments on
material taken directly from (Mallios and Raptis, 2000), the
definitions of the previous section and the brief presentation of posets and
their incidence algebras above:

\vskip 0.1in

\noindent i) In (Mallios and Raptis, 2000) the base spaces for the
finsheaves of qausets were initially taken to be the finitary topological poset
substitutes $P$ of a bounded region $X$ of the curved smooth classical spacetime
continuum $M$.

\vskip 0.1in

\noindent ii) Finsheaves of the incidence Rota algebras $\omg$ modeling
quantum discretized manifolds as in (Raptis and Zapatrin, 2000) over
the finitary topological posets $P$ of i) were defined in the
manner of (Raptis, 2000{\it b}). This was achieved essentially by recognizing that
point-wise in the topological poset base space $P$\footnote{That is,
locally 
over each of the point-vertices of the finitary poset $P$.} the $\omega$-map
$\omega:~P\rightarrow\omg(P)$ can be viewed as a local homeomorphic
`lifting' of a homomorphic copy of  
$\omg$ over $P$\footnote{See discussion following $(5)$.}. In turn, the latter can be thought of as  
defining ${\bf\omg}(P)$ as an $\omg$-sheaf over $P$, since $\omega$, by
construction\footnote{See equations $(1)$ and $(5)$.},  
maps the germ $\stackrel{*}{\rightarrow}$ of the Sorkin
topology of $P$ to the germ $\rho$ of the Rota topology of $\omg(P)$,
thus it may be regarded as a local homeomorphism from the `base space'
$P$ to the `sheaf space' $\omg^{\omega}(P)$\footnote{Now the latter consists
of independent stalks isomorphic (better, `homomorphic' since reps
of $\omg$ were used) to $\omg$ over $P$'s points. Note that the
superscript `$\omega$' of $\omg$ indicates `sheaf' since it lifts individual
vertices of the base poset to `germs' in the stalks of $\omg^{\omega}$
over them 
by the local homeomorphism $\omega$.}. 

\vskip 0.1in

\noindent iii) Now, by topologically identifying $P$ (and its Sorkin
topology) with ${\bf Spec}\,\omg(P)$ (and its Rota
topology)\footnote{Since they are topologically indistinguishable
({\it ie}, homeomorphic) by the $p$-construction which is
categorically dual to the $\omega$ one.}, we take the latter to be the 
topological base space of the $\omg$-sheaf $\omg^{\omega}(P)$ in ii). But
the latter is nothing else than $\omg^{\omega}({\bf Spec}\, \omg)$, thus
it corresponds to the structure sheaf of the primitive $\omg$-scheme
$\omega^{s}=({\bf Spec}\,\omg(P),\omg^{\omega}({\bf Spec}\,\omg(P)))$
whose space $sp(\omega^{s})$ is clearly ${\bf Spec}\,\omg(P)$ equipped
with the Rota topology\footnote{The superscript `$s$' of $\omg$
denoting `scheme'.}. It is crucial to note that the $p$-map that
solders or localizes stalks in the structure sheaf space
$\omg^{\omega}$ to points in the topological base space ${\bf
Spec}\,\omg(P)$ is precisely 
the inverse of the $\omega$-map that `lifted' the structure sheaf
$\omg^{\omega}({\bf Spec}\,\omg(P))$ of the primitive $\omg$-scheme $\omg^{s}$ and defined it as a local homeomorphism in ii). Formally
this can be written as $\omega= 
p^{-1}$ and it is in complete analogy to the definition of the sheaf
map $s$ and its inverse projection or localization or `soldering' map $\pi$ for the
continuous sheaf ${\cal{S}}(X)$ in the previous section\footnote{See
footnotes 21 and 57.}. That
$\omega=p^{-1}$ is, of course, a (local in $\omg^{s}$) consequence of Zapatrin's
theorem (1998) that $P(\omg(P))\simeq P$\footnote{Or in terms of $p$
and $\omega$: $p\circ\omega(P)\simeq
P\Leftrightarrow p\circ\omega={\rm Id}\Leftrightarrow\omega=p^{-1}$,
`${\rm Id}$' standing for `isomorphism' or `algebraic identity
map' (see above).}. Thus, our earlier calling the maps $(\omega ,p)$ inverse
(or dual) of each
other is literal in the primitive $\omg$-scheme $\omg^{s}$.  

\vskip 0.1in

\noindent iv) Subsequently in (Mallios and Raptis, 2000) the authors
re-interpreted the underlying topological base posets $P$ as causets
$\vec{P}$ and their associated topological incidence algebras $\omg$
as qausets $\amg$\footnote{Note the arrow over their symbols that
reminds one of the causal meaning of these structures.} as in (Raptis,
2000{\it a}). Thus we arrive at the notion of a primitive
scheme $\amg^{s}$ of qausets over a (possibly curved) finitary causal
base space\footnote{In fact, if like in (Mallios and Raptis, 2000) we
assume that the causal base space is curved, we cannot identify it
with a flat classical causet $\vec{P}$. Instead, it will be a
flat transitive causet only locally (Mallios and Raptis, 2000).}.  
We must emphasize here that this mathematical structure is a sound
model of a finitary, causal and quantal version of the CEP of GR\footnote{Called FEP in
(Mallios and Raptis, 2000).}, as it was argued in
(Mallios and Raptis, 2000). This model, the authors explained in that
paper, is a sound model of the kinematical structure of QCD. Below we
recall from (Mallios and Raptis, 2000) the basic arguments that led to
the formulation of the FEP and its `corollary', the FPGC\footnote{The `Finitary
Principle of General Covariance'.}, for the curved finsheaves of
qausets by presenting them in the new light of scheme theory.

If the underlying finitary causal base space is to be regarded as
being `curved', at least in the geometrical sense of this word, then
these $\amg$-scheme-theoretic localizations can be physically interpreted as some
kind of `gauging of qausets' (Mallios and Raptis, 2000). This gauging is
foreshadowed by the CEP of GR which, we recall, may be taken as
saying that independent flat isomorphs of Minkowski space $\cem$ are
raised over every point-event of the curved classical spacetime continuum
$M$. Similarly, in our primitive finschemes\footnote{That is,
`finitary spacetime schemes' as `finsheaves' stands for `finitary
spacetime sheaves' (Mallios and Raptis, 2000).} $\amg^{s}$, independent flat qauset
stalks are raised over every point-vertex of a curved finitary causal
space which is then thought of as serving as a base space for the
localization or soldering of these $\amg$-stalks of the structure sheaf
of $\amg^{s}$. 

As a result of this localization or gauging of qausets,
the subsheaf morphisms $d :~\amg^{i}\rightarrow\amg^{i+1}$ effected
by the flat\footnote{Because nilpotent (Dimakis and M\"uller-Hoissen,
1999, Mallios and Raptis, 2000).} K\"ahler-Cartan differential $d$
(Mallios, 1998, Raptis, 2000{\it b}, Mallios and Raptis, 2000), are `gauged' into a
`curved' ({\it ie}, non-flat) connection operator
${\cal{D}}=d+{\cal{A}}$ (Mallios and Raptis, 2000). ${\cal{D}}$ is
now a curved finscheme morphism in the gauged $\amg^{s}$, and its
non-flat part ${\cal{A}}$\footnote{This symbol for the `gauge potential'
of quantum causality (Mallios and Raptis, 2000)  
should not to be confused with same one used for symbolizing algebra
sheaves in the previous section.} is a section of the `subscheme'\footnote{A subscheme $B$ of a
given scheme $C$ may be defined as a scheme whose structure sheaf is a
subsheaf of that of $C$. It follows that the stalks of $B$ are
substructures of the stalks of $C$.} $\amg^{1}$ ({\it ie},
${\cal{A}}\in\Gamma(U,\amg^{1})$)\footnote{All this follows from $(6)$
which shows that the reticular and quantal analogue of the space
$\omg^{1}$ of differential forms over the smooth spacetime continuum
is a linear subspace of the qauset stalks $\omg$ of the structure sheaf of $\amg^{s}$.}, albeit, not
a global section (Mallios, 1998, Mallios and Raptis, 2000)\footnote{In
the next section we are going to present and discuss in some detail a deep similarity between
this non-existence of a global section for the gauge potential of
quantum causality $\cal{A}$ in the curved $\amg^{s}$ and the
non-existence of a global section in a certain topos of presheaves
employed to model truth-valuations in quantum logic proper (Butterfield and
Isham, 1998, 1999, Butterfield {\it et al.}, 2000).}. In (Mallios and
Raptis, 2000) this non-existence of a global section for the curved
finsheaves (here finschemes) of qausets was attributed to the
non-transitivity of causality in the underlying curved causal base
space due to gravity. Thus, the quantum causal topology encoded in the
primitive finscheme of qausets $\amg^{s}$ is not fixed once and forever as
if the latter consists of a single qauset algebra $\omg$ over a single
flat
causet $\vec{P}$. Rather, over the points of the latter (now it
regarded as being curved), local
independent $\amg$ isomorphs, `twisted' or `warped' relative to each other, are erected and the pattern of
quantum causal connections between qauset elements of these stalks of
$\amg^{s}$ is a local dynamical quantum variable\footnote{In our case this is the local quantum
causality represented by $\vec{\rho}$. Note the arrow over the germ of
the Rota topology $\rho$ of $\amg$ that reminds one of the latter's
causal interpretation.} (Mallios and Raptis, 2000).

This
localization or gauging of the qausets in $\amg^{s}$ and the
resulting definition of the covariant derivative $\cal D$ in the latter may be physically
interpreted as follows: the dynamical variation of qausets from stalk
to stalk in $\amg^{s}$ may be attributed to the local
({\it ie}, point-wise) gauge
freedom of selecting a qauset\footnote{By a germ of the
${\cal{T}}_{\vec{\rho}}\,$-continuous sections of the structure sheaf of
$\amg^{s}$.} from the independent stalks of the
structure sheaf of 
$\amg^{s}$. If this `point-wise selection' dynamical process for qausets is to
respect the sheaf structure of $\amg^{s}$\footnote{Physically
speaking: to be in some sense `covariant' with respect to the local
quantum causal topology ${\cal{T}}_{\vec{\rho}}$ encoded
in the structure sheaf of $\amg^{s}$ (Mallios and Raptis, 2000).},
then it must be formulated categorically in terms of sheaf or, in our
case, scheme 
morphisms\footnote{For the definition of these, see previous section.}. The connection $\cal{D}$ is conveniently such a scheme
morphism that is readily seen to stitch the stalks of the structure
sheaf of $\amg^{s}$ in a way that respects the aforementioned
`stalk-wise gauge
freedom' (Mallios, 1998, Mallios and Raptis, 2000)\footnote{In this
sense $\cal D$, which respects the local quantum causal topology ({\it
ie}, the sheaf of qausets) generates a dynamics for qausets that is
`locally causal'. This is the finitary analogue of the classical
differential locality ({\it ie}, infinitesimal local causality) of the
gravitational spacetime continuum (Einstein, 1924, Mallios and Raptis,
2000).}. Thus, the 
scheme-theoretic version of the sheaf-theoretic FPGC formulated in (Mallios and Raptis, 2000)
reads: dynamical laws for the qausets in $\amg^{s}$ must be equations
between finscheme morphisms such as $\cal D$. 

It follows that if one
regards $\amg^{s}$ as a principal primitive finscheme of qausets
having for 
structure group ${\bf G}$ a reticular version of the orthochronous Lorentz
group $L^{+}=SO(1,3)^{\uparrow}$ (or its local isomorph $SL(2,\com)$)
as in (Mallios and Raptis, 2000), one is able to interpret the FPGC
above as some kind of local gauge invariance of the dynamics of
qausets under ${\bf G}$ ({\it ie}, as the local orthochronous Lorentzian
relativity of the dynamics of the gauged qausets in the scheme). The
primitive ${\bf G}$-scheme ${\cal{G}}^{s}=({\bf Spec}\,
\amg,{\cal{G}}({\bf Spec}\, \amg))$ associated with $\amg^{s}$\footnote{In (Mallios
and Raptis, 2000) the ${\bf G}$-sheaf was also called `adjoint to the
$\amg$-sheaf'.} is the latter's local gauge symmetry structure and has as
structure sheaf ${\cal{G}}({\bf Spec}\, \amg))$ the localized or
gauged reticular orthochronous Lorentz invariances of SR in accordance
with the schematic version of the FEP above. Then $\cal A$, the non-flat part of $\cal D$, was
seen in (Mallios and Raptis, 2000) to 
take values in the Lie algebra stalks of the structure sheaf of the
finscheme 
${\cal{G}}^{s}$ associated with $\amg^{s}$\footnote{That is, it was
seen to be a $so(1,3)^{\uparrow}\simeq
sl(2,\com)$-valued reticular $1$-form on the principal primitive
$\omg$-scheme $(\amg^{s},{\cal{G}}^{s})$.}. This was seen to be in complete
analogy to the classical smooth Lorentzian manifold case $M$ on which (the
kinematics of) GR may be effectively represented by a ${\bf G}$-bundle
and a ${\bf g}$-valued gravitational connection $1$-form ${\cal{A}}$
on it, which, in turn, is a section of its associated bundle of modules of smooth
Cartan forms on $M$ (Mallios and Raptis, 2000). Finally, we mention
again\footnote{See footnote 12 in the introduction.} that precisely by the way that curved finsheaves (here
finschemes) of qausets were constructed, a dynamics for them expressed
in terms of the sheaf (here scheme) morphism $\cal D$ will be, by
definition of the latter, `gauge independent' if one physically interprets the open sets in the locally finite open covers of
$X$ with `coarse local gauges' ({\it ie}, `local coordinate patches or frames or
laboratories of approximate measurements or localizations of the
dynamical quantum 
causal relations between spacetime
point-events') in $X$ (Mallios, 1998, Mallios and Raptis, 2000). In this way the dynamics of qausets is
independent of the background inert `parameter base space' $X$ whose sole
purpose is to 
serve as a scaffolding for discretizing and subsequently soldering
the fundamentally `a-local'\footnote{The characterization of qausets as
being alocal structures follows from the same interpretation given to
their quantum discretized topological space relatives in (Raptis and
Zapatrin, 2000).} qausets, but in itself is of no physical significance (Mallios and Raptis, 2000). Such an
independence of the qauset dynamics from an inert geometrical
background spacetime is welcome for reasons discussed earlier in this
paper and in
more detail in (Mallios and Raptis, 2000)\footnote{See also concluding
physical remarks on possible curved finsheaves of qausets in (Raptis,
2000{\it b}.)}.

At this point it must be noted that in view of the fact that the
qausets in $\amg^{s}$ coherently superpose with each other
locally\footnote{In (Mallios and Raptis, 2000) this was the content of
the Finitary Local Superposition Principle (FLSP) for the curved finsheaves
of qausets.}, their local spin-Lorentzian structure symmetries in the adjoint
principal primitive finscheme ${\cal{G}}^{s}$ must also be
quantum\footnote{In the next section we will see how these local
coherent quantum superpositions of qausets determine a quantum
kind of subobject classifier, thus also a `local quantum logic', for a
topos organization of the $\amg^{s}$s. Essentially due to this, the resulting
topos-like structure will be called a `quantum topos'.}. In
turn, the latter entails that the finitary ${\bf g}$-valued non-flat
spin-Lorentzian connections ${\cal{A}}$ representing the qausets'
dynamics will also be quantum variables; hence, as it was also
anticipated in (Mallios and Raptis, 2000), perhaps a finitary and
causal version of the usual `covariant path-integral over connection space' quantization of
Lorentzian gravity (Baez and Muniain, 1994) will be a good candidate for modeling
QCD\footnote{See also remarks in section 5.}. 

Now one must realize that having a non-trivial $\cal D$ on $\amg^{s}$ may 
enable us to write a finitary, causal and quantal analogue of the left
hand side of the classical Einstein equations of GR\footnote{Recall:
$G_{\mu\nu}:=R_{\mu\nu}-{\frac{1}{2}}g_{\mu\nu}R=\kappa T_{\mu\nu}$.} by
expressing the Einstein curvature tensor $G_{\mu\nu}$ in terms of the reticular Lorentzian
connection $\cal D$\footnote{In turn, this expressiom may derive from
varying with respect to $\cal A$ a Lagrangian that is appropriately defined in terms of $\cal D$.} as it is done
in the usual connection-based gauge-theoretic approaches to GR\footnote{The so-called
Palatini formulation of GR or its recent spinorial formulation in
terms of new variables (Ashtekar, 1986, Baez and Muniain, 1994). See
also (Mallios and Raptis, 2000) for some comments on this in
connection with the curved finsheaves of qausets, and also look at the
discussion in the last section of the present paper.} (Raptis,
2000{\it c}). This can be used then to represent the so-called
vacuum Einstein 
equations\footnote{That is, in the absence of the matter tensor
$T_{\mu\nu}$.} by
equating with zero the left 
hand side of Einstein's equations ({\it ie}, $G_{\mu\nu}=0$; $T_{\mu\nu}=0$). However, in
a non-vacuum situation ({\it ie}, in the presence of matter) it was
until recently quite  
doubtful whether the
right hand side of Einstein's equations\footnote{The so-called energy-momentum
tensor of matter $T_{\mu\nu}$.} could be derived strictly from
causal considerations and arguments. Rideout and Sorkin (2000)
actually derive matter-like `fields' 
 from causets, thus it seems reasonable, following Einstein's
fundamental insight to equate the tensorial aspect of the spacetime
geometry ($G_{\mu\nu}$) with a similar tensor expression for 
the dynamical actions of matter ($T_{\mu\nu}$), to represent matter
actions in $\amg^{s}$ again by appropriate finscheme morphisms between
our gauged qausets (Raptis,
2000{\it c})\footnote{Thus in this way we will possess a reticular, causal and
quantal analogue of Einstein's famous `action-reaction' interpretation
of the equations of GR which holds that geometry acts on matter in the form
of gravity, and matter reacts back by curving geometry.}. 

In the next section we highlight the fundamentally non-commutative
character of our incidence algebra localizations in $\amg^{s}$ that
model (the kinematics of) a curved (thus dynamical) local quantum causality and prepare the
ground for a comparison between the non-commutative topology that
$\amg^{s}$ stands for and the $C^{*}$-quantale models of
non-commutative topological spaces presented in (Mulvey and Pelletier,
2000). We will also compare our curved primitive finschemes of qausets
against 
the `warped' topos of presheaves of sets over the Stone space of a
quantum lattice $\cal L$ as presented recently in a series of papers on
quantum logic proper (Butterfield and Isham, 1998,
1999, Butterfield {\it et al.}, 2000). 

\section*{\normalsize\bf 4. NON-COMMUTATIVE ASPECTS OF PRIMITIVE $\amg$-FINSCHEMES}

First we describe briefly the Gel'fand spatialization method for
extracting topological spaces from  
commutative and involutive ({\it ie}, $*$) algebras $A$ alluded to in the previous
section\footnote{See also (Zapatrin, 1998, Breslav {\it et al.}, 1999).},
then we pass to the non-commutative case of our particular interest here. 

Let $A$ be a
commutative $*$-algebra. A natural or canonical way $D$ to represent
$A$ 
is as a functional algebra on a topological space
$X$\footnote{In particular, Gel'fand worked with the $*$-representations $D$ of commutative
algebras in the abelian $C^{*}$-algebra $C^{0}(X,\com)$ of
continuous complex valued functions on $X$. The latter
is regarded as the `standard representation' of commutative
involutive algebras.}. One considers the following ideals in the 
representation algebra $D(A)$ associated with every point $x$ of $X$

\begin{equation}
M_{x}:=\{ f\in D(A):~f(x)=0\}
\end{equation}

\noindent The factor algebra
$D(A)/M_{x}$ is seen to be $1$-dimensional, hence it is 
simple. Thus, the kernels $D^{-1}(M_{x})$ of $A$'s $1$-dimensional
irreps $D$\footnote{Such $D$s are usually called `characters'.} are maximal ideals in $A$ and, provided the underlying topological space
$X$ is reasonably `nice'\footnote{We will not go into the
technicalities of what kind of topological space is regarded as being `nice'. For example, a locally Euclidean topological ({\it ie}, $C^{0}$) manifold $X$ is considered to be
`nice'  
and it is usually taken as the `standard domain' for the functional
representation of abelian algebras. The intimate relation between the
locally Euclidean 
classical topological space $X$ and the commutative structure of
algebras that can be functionally represented on it will be discussed
shortly in connection with the work of (Mulvey and Pelletier, 2000).}, they are in $1$-$1$
correspondence with the points of $X$. This correspondence is
effectively used in the Gel'fand spatialization
procedure whereby the points of $X$ are first substituted by the
maximal ideals of $A$, and then one imposes a
suitable topology ${\cal{T}}$ on the `maximal spectrum' ${\rm Max}\,
A$ of $A$. In this way, one `extracts' in
some sense a `classical' topological space from a
commutative algebra $A$\footnote{Thus, the Gel'fand spatialization
method may also be called `algebra geometrization'.}. It is also
interesting to note in connection with the commutative case that this 
considering in $(7)$ of the `ideals of zeros' of the representation algebra $D(A)$
point-wise in the representation domain $X$ in order to define the
point-set ${\rm
Max}\, A$ to be subsequently topologized, is in complete analogy with how affine
schemes are defined in the algebraic geometry of commutative
polynomial rings over affine (Euclidean) space (Hartshorne,
1983, Shafarevich, 1994)\footnote{As we noted in section 2, the prime
spectra of these commutative polynomial rings are first topologized
according to the so-called Zariski topology ${\cal{T}}_{Z}$, and then sheaves of such
rings $R$ are erected over the Zariski topological space
${\cal{T}}_{Z}({\bf Spec}\, R)$ (see footnote 37).}.  

Now Gel'fand's
method translates straightforwardly to the non-commutative case by
passing from characters to equivalence classes of irreps, and accordingly, 
from maximal ideals to primitive ones. Then, there are many ways of
imposing a topology on the points of the primitive spectrum ${\bf
Spec}\, A$ of a non-abelian algebra $A$. As we mentioned in the
previous section, for our non-commutative and finite dimensional
incidence Rota algebras $\amg$ representing qausets (Raptis, 2000{\it a}), we chose the so-called Rota
topology\footnote{Which is 
generated by the relation $\vec{\rho}$ on the point-events of 
of ${\bf Spec}\,\amg$ as $(1)$ and $(5)$ show. Note again here the arrow
that is put over the generator $\rho$ to remind one of its directly (local) causal meaning.} instead of the more `standard' ones of Jacobson
of Gel'fand, because the latter reduce to the trivial, totally
disconnected case\footnote{That is, the discrete topology on 
${\bf Spec}\,\omg$.} (Zapatrin, 1998, Breslav {\it et al.}, 1999,
Raptis and Zapatrin, 2000). Then, primitive $\amg$-finschemes are
defined by suitably localizing the $\amg$s as sheaves $\vec{\omega}$ over
${\cal{T}}_{\vec\rho}({\bf Spec}\,\amg)$\footnote{Which by the inverse
`soldering' local homeomorphism or 
sheaf $\vec{p}=\vec{\omega}^{-1}$ corresponds to the curved causal base
space $\vec{P}(\amg)$ whose constant transitive partial order causality $\rightarrow$ is
`cut-off' by gravity to an intransitive dynamically variable germ of
the causal topology 
$\stackrel{*}{\rightarrow}$ (Raptis, 2000{\it a}, Mallios and Raptis,
2000).}.   

The essentially non-commutative, thus quantal, character of the
structure of $\amg^{s}$ which models the kinematics of a gauged thus
curved and 
dynamically variable local  
quantum causality $\vec{\rho}$ (Mallios and Raptis, 2000), is
explicitly manifested in the very definition of the latter. Apart from
the fact that the stalks $\amg$ of the structure sheaf
$\amg^{\vec{\omega}}({\bf Spec}\,\amg)$ of $\amg^{s}$ are non-abelian
Rota 
algebras, the very generating relation $\vec{\rho}$ of
quantum causality\footnote{Occasionally we will call $\vec{\rho}$ `local quantum
causality'. Local quantum causality is the local dynamical variable
defining by its dynamics the `observable quantum causal topology'
${\cal{T}}_{\vec{\rho}}$ of $\amg^{s}$  (Mallios and Raptis, 2000).} is defined as a directed line
segment ({\it ie}, an arrow) $\stackrel{*}{\rightarrow}$ joining the
point-events in ${\bf Spec}\,\amg$ exactly `because' a non-commutative
relation\footnote{That is,
${\cal{I}}_{p}{\cal{I}}_{q}\not={\cal{I}}_{q}{\cal{I}}_{p}$.} holds
between them as $(1)$ and $(5)$ show. 

In order to see more clearly the quantum physical interpretation of this
non-commutativity\footnote{That is, interpret it as some kind of
Heisenberg uncertainty
relation built-in fundamentally at the germ-level of the (kinematical)
structure for (dynamical) quantum
causal topology that $\amg^{s}$ stands for.}, we recall from (Mallios and Raptis, 2000) the quantum
semantics given to the ideals in $\amg(\vec{P})$ related by 
$\vec{\rho}$\footnote{See footnote 128 in (Mallios and Raptis, 2000).}. 

First observe that when events $p$ and $q$ are
immediately related by causality in $\vec{P}$ ({\it ie},
$p\stackrel{*}{\rightarrow}q$), the corresponding ideals
$\vec{{\cal{I}}}_{p}$ and $\vec{{\cal{I}}}_{q}$ in $\amg(\vec{P})$ are
$\vec{\rho}$-related which, in turn, by the definition of the latter
as in $(1)$ and $(5)$,  
is equivalent to 

$$\vec{{\cal{I}}}_{p}\vec{{\cal{I}}}_{q}\not=\vec{{\cal{I}}}_{q}\vec{{\cal{I}}}_{p}\stackrel{\not=}{\subset}\vec{{\cal{I}}}_{p}\cap\vec{{\cal{I}}}_{q}$$

\noindent Now the physical interpretation that can be given to
$\vec{\rho}$-related primitive ideals in $\amg(\vec{P})$ is as
`transients'\footnote{The elements of $\amg^{1}$ as we saw in $(6)$ of
the previous section.}, that
is to say, elementary quantum dynamical processes of propagation of a quantum
of causality\footnote{Which may be called `causon'. We do not wish to use the
name `chronon' for the `elementary particle' of quantum causality, because it
has been used in Finkelstein's Quantum Relativity theory of quantum
spacetime and its dynamics, and there it stands for 
the elementary quantum of time (Finkelstein, 1988, 1996). Of course,
one expects that causons and chronons are intimately related to each
other in view of the close structural similarities between our QCD and
Finkelstein's curved quantum causal nets (Finkelstein, 1988, Raptis,
2000{\it a}, Mallios and Raptis, 2000).} thus defining
immediate reticular and quantal processes of `energy-momentum transfer'
between quantum events (Raptis and Zapatrin, 2000). On the other hand, the `position
determinations'\footnote{Or quantum acts of localization of the
aforementioned causon.} of these $\vec{\rho}$-related events, namely, the
`stationaries' $\ketbra{p}{p}$ and $\ketbra{q}{q}$ (Raptis and
Zapatrin, 2000), are excluded by
the very definition of the `quantum point-events'
$\vec{{\cal{I}}}_{p}$ and $\vec{{\cal{I}}}_{q}$ \footnote{See
subsequent comparison with the `spatial' 
quantum points defined by Mulvey and Pelletier (2000).}  in
$\amg(\vec{P})$, as $(4)$ shows. Thus, the non-commutativity relation
holding between quantum events in $\amg(\vec{P})$ can be physically
interpreted as an indeterminacy or uncertainty relation between quantum actions of localization
({\it ie}, determination of `position') of a causon and their dual or complementary quantum
actions of its momentous propagation defining the directed 
$p\stackrel{*}{\rightarrow}q\equiv\vec{{\cal{I}}}_{p}\vec{\rho}\vec{{\cal{I}}}_{q}$
immediate quantum causal
connections\footnote{The energy-momentum-like `transients' in $\amg^{1}$ of $(6)$.}. This may be viewed as a finitary and causal analogue of
the usual kinematical 
Heisenberg 
uncertainty relations between the complementary time/position-energy/momentum $(t,x)$-$(E,p)$
observables of relativistic matter quanta propagating on a classical
Minkowski continuum\footnote{The reader may also like to refer to
(Zapatrin, 1998, Breslav {\it et al.}, 1999) to see some possible
connections between topological incidence Rota algebras (albeit,
without a directly causal physical interpretation like our $\amg$s) and
Noncommutative Geometry (Connes, 1994), especially when the latter is approached via
$C^{*}$-algebraic non-commutative lattices as in (Balachandran {\it et
al.}, 1996, Landi, 1997, Landi
and Lizzi, 1997). See also remarks below in connection with the
$C^{*}$-quantales of Mulvey and Pelletier (2000).}.   

It must be also emphasized that the non-commutativity relations between
the quantum points in the primitive spectrum base space of $\amg^{s}$
above reflect not only the `quantumness' but also the `temporal directedness' of the local quantum causality
relations 
$\vec{\rho}$ that bind them\footnote{See anticipating remarks in
footnote 129 of (Mallios and Raptis, 2000).}, as follows

\begin{equation}
\begin{array}{ll}
&\vec{{\cal{I}}}_{p}\vec{\rho}\vec{{\cal{I}}}_{q}\equiv
\vec{{\cal{I}}}_{p}\vec{{\cal{I}}}_{q}(\not=\vec{{\cal{I}}}_{q}\vec{{\cal{I}}}_{p})\stackrel{\not=}{\subset}\vec{{\cal{I}}}_{p}\cap\vec{{\cal{I}}}_{q}(=\vec{{\cal{I}}}_{q}\cap\vec{{\cal{I}}}_{p})\Leftrightarrow
p\stackrel{*}{\rightarrow}q\cr
&\vec{{\cal{I}}}_{q}\vec{\rho}\vec{{\cal{I}}}_{p}\equiv
\vec{{\cal{I}}}_{q}\vec{{\cal{I}}}_{p}(\not=\vec{{\cal{I}}}_{p}\vec{{\cal{I}}}_{q})\stackrel{\not=}{\subset}\vec{{\cal{I}}}_{q}\cap\vec{{\cal{I}}}_{p}(=\vec{{\cal{I}}}_{p}\cap\vec{{\cal{I}}}_{q})\Leftrightarrow
q\stackrel{*}{\rightarrow}p\cr
\end{array}
\end{equation}

\noindent To explain $(8)$ in more detail, define $\vec{P}^{\rm op}$ to be the
causet obtained from $\vec{P}$ by reversing all of its
arrows\footnote{What is also commonly known as the poset
category opposite to $\vec{P}$, hence the superscript `${\rm op}$'.}. This
`causality reversal' unary map `${\rm op}$' induces an algebra map
$\dagger:~\amg(\vec{P})\rightarrow\amg(\vec{P})^{\dagger}=\amg(\vec{P}^{\rm
op})$
whose action on an arbitrary element $\omega$ of $\amg(\vec{P})$ is
given by

$$\omega=z_{1}\cdot\omega_{1}\circ\omega_{2}+z_{2}\cdot\omega_{3}\circ\omega_{4}+\cdots\stackrel{\dagger}{\longrightarrow}\omega^{\dagger}=z_{1}^{*}\cdot\omega_{2}^{\dagger}\circ\omega_{1}^{\dagger}+z_{2}^{*}\cdot\omega_{4}^{\dagger}\circ\omega_{3}^{\dagger}+\cdots$$

\noindent where $z_{1}^{*}$ and $z_{2}^{*}$ are the complex conjugates
of the complex coefficients  $z_{1}$ and $z_{2}$ over which the
quantum causal arrows $\omega$
in $\amg(\vec{P})$ coherently superpose (Raptis, 2000{\it
a})\footnote{Thus we assume indeed that the $F$ in $(2)$ is
$\com$, as footnote 68 fixed.}, so that for $\omega_{1}=\ketbra{p}{q}$ in
$\amg(\vec{P})$ for example:
$\omega_{1}^{\dagger}=\ketbra{q}{p}$. $\omega_{1}^{\dagger}$ is called `the
conjugate of $\omega_{1}$' and it is an element of `the Rota incidence
algebra $\amg(\vec{P})^{\dagger}$ conjugate to
$\amg(\vec{P})$'\footnote{`$\omega_{1}^{\dagger}$' is simply the Dirac
ket-bra corresponding to the arrow $q\rightarrow p$ in $\vec{P}^{\rm op}$
of opposite direction to the arrow $p\rightarrow q$ in
$\vec{P}$.}. Note that the stationaries of $\amg(\vec{P})$ in $\amg^{0}$ are
self-conjugate elements since $\forall
p\in\vec{P}:~(\ketbra{p}{p})^{\dagger}=\ketbra{p}{p}$\footnote{This
points to the direction of a possible `explanation' of the reality
($\R$) of the spacetime coordinates in the continuum Bohr 
correspondence limit of an inverse system of qausets (Raptis and
Zapatrin, 2000, Mallios and Raptis, 2000). However we are not going to
discuss further this emergence of the $\R$-valuedness of the classical spacetime coordinates in the present paper.}.

Having defined $\amg(\vec{P})^{\dagger}$, we regard the opposite
order of multiplication ${\cal{I}}_{q}{\cal{I}}_{p}$ of quantum
point-events in $\amg(\vec{P})$ in the second line of
$(8)$\footnote{An order which is also held to be equivalent to the
order or direction of the arrow  
$q\stackrel{*}{\rightarrow}p$ in the base causet $\vec{P}$.},  when
`canonically' mapped by $\dagger$, as actually defining `immediate quantum causal 
arrows\footnote{Again, the arrows in the base causet $\vec{P}$ may be regarded as being
classical and their $\vec{\omega}$-correspondent qausets in $\amg(\vec{P})$
as being quantum (Raptis, 2000{\it a}).} 
in $\amg(\vec{P})^{\dagger}$ over $\vec{P}^{\rm op}$'. This is because $\vec{P}$ is a poset,
thus if $p\stackrel{*}{\rightarrow}q\in\vec{P}$ then
$q\stackrel{*}{\rightarrow}p\not\in\vec{P}$, since the transitive
reduction 
$\stackrel{*}{\rightarrow}$ of the partial order $\rightarrow$ of
$\vec{P}$ is by definition antisymmetric\footnote{That is,
$p\stackrel{*}{\rightarrow}q\Rightarrow
q\not\!\!{\stackrel{*}{\rightarrow}}p$ (antisymmetry property of $\stackrel{*}{\rightarrow}$).}.  

Thus the idea is to regard the product of two quantum points in the scheme
$\amg^{s}$ 
as being `locally directed' in the sense that perhaps we should consider as being
`physically significant' the product of two
primitive ideals in the scheme's base space only in the same order that
these points appear to be quantum causally $\vec{\rho}$-related in the
latter. At the same time, we should regard the opposite order of
multiplication of these quantum point-events as defining immediate quantum causal arrows of opposite direction in the conjugate scheme
$\amg^{s\dagger}$\footnote{This is just the primitive scheme of
qausets obtained by totally reversing the direction of the  quantum
arrows in the stalks of the structure sheaf of $\amg^{s}$ and complex
conjugating their $\com$-amplitudes (c-coefficients).}. In this way the non-commutative
product between the quantum point-events in $\amg^{s}$'s base space
represents a distinction between the direction of quantum causality in
the scheme. Thus the non-commutativity of the algebraic product in our
primitive scheme of quantum causal sets reflects not only its quantal
nature, but also its `temporal directedness' which, we will
hypothesize in the next section, lies at the heart of the
fundamental\footnote{Albeit, `kinematical' in our case.}  
quantum time-asymmetry that is expected of the `true quantum gravity'
(Penrose, 1987). 

Then we may call
$\amg^{s}$ `the curved quantum time-forward' or `gauged future quantum
causal topology', and its
conjugate $\amg^{s\dagger}$ `the curved quantum time-backward' or
`gauged past quantum causal 
topology'\footnote{Represented by the local quantum causal topological
(dynamical) variables $\vec{\rho}$ and $\vec{\rho}\,^{\dagger}$, respectively.}. Of course, it is a matter of convention or `external choice' which scheme one
takes to be future and which past, but once this freedom of choice is fixed
by a decision to adopt $\amg^{s}$ or $\amg^{s\dagger}$ for the
kinematics of a dynamical local quantum causal topology,
the dynamics of qausets in the respective schemes, which are
represented by scheme morphisms as discussed earlier, will respect or
preserve 
this `structural microlocal directedness of quantum causality' or
`kinematical local quantum arrow of time'. As we said in the last
paragraph of the next section, we will return to
discuss in some detail how this `local quantum arrow of time' may be thought of as
the characteristic feature of the kinematics of a time-asymmetric
quantum gravity which then the dynamics of qausets ({\it ie}, our
finitary and causal version of Lorentzian quantum
gravity {\it per se}) that is formulated categorically in terms of
scheme morphisms should conserve. Thus we will `justify' Penrose's (1987) 
claim that the true quantum gravity must be a time-asymmetric theory on the
grounds that the QCD is
time-asymmetric `because' its kinematical structure, as encoded in
$\amg^{s}$, is also locally quantum time-directed.

Now that we have discussed the fundamentally non-commutative character
of the localized or gauged thus  
curved quantum qausality modeled after the primitive
finschemes $\amg^{s}$, we will briefly compare the latter with
the recently proposed by Mulvey and
Pelletier (2000) non-commutative topological
spaces modeled after so-called $C^{*}$-quantales. We will first
present abstract quantales in their dual role as non-commutative
topological spaces and quantal logics, then we will discuss relevant
elements from the (Mulvey and Pelletier, 2000) paper which presents a
particular paradigm of quantales deriving from non-commutative
$C^{*}$-algebras (especially by focusing on how these $C^{*}$-quantales may be viewed
as non-commutative topological spaces similar to our $\amg$-finschemes
of qausets), and finally we will comment briefly on the entries of table 1 
attached at the end which
summarizes the comparison between $C^{*}$-quantales and primitive
$\amg$-finschemes.

An abstract quantale $\quan$ may be thought of as a lattice
${\cal{L}}(\vee ,\wedge)$ together with an `extra structure' 
$\&$ which is an associative but non-commutative multiplication
between its elements\footnote{For a more thorough treatment of
quantales with various applications the reader is referred to
(Rosenthal, 1990).}. We write the quantale as the following triplet of
structures $\quan(\vee ,\wedge ,\&)$. 
 For reasons to become
transparent shortly, let us think of a quantale $\quan(\vee ,\wedge,\&
)$ as splitting into two lattice-like substructures

$$\quan(\vee ,\wedge,\& )=\left\lbrace
\begin{array}{rcl}\topl(\vee ,\&)&,& \mbox{topological lattice}\cr
{\cal{L}}(\vee ,\wedge) &,&
\mbox{logical lattice}\end{array} \right.$$ 

\noindent called `topological lattice' $\topl(\vee ,\&)$ and 
`logical lattice' ${\cal{L}}(\vee ,\wedge)$, respectively. 

Let us call $\wedge$ `the logical meet of the quantale' and $\&$
`the topological meet of the quantale'. The topological meet is
non-commutative and distributes over arbitrary joins $\vee$ of
$\quan$'s elements, while the logical meet is commutative, but it is
usually taken to be non-distributive over $\vee$. In this sense a
quantale plays the following dual role as mentioned above: first, as
an abstract topological space 
$\topl(\vee,\&)$, $\quan$ may be thought of as a non-commutative generalization of
`classical', because commutative, topological
spaces or locales $\loc(\vee ,\wedge)$ (Mac Lane and Moerdijk, 1992)\footnote{The usual example of a
$\loc(\vee ,\wedge)$ being the $\wedge$-commutative lattice of open subsets
of a given topological space with $\vee$ and $\wedge$ standing for the
commutative set-theoretic union $\cup$ and intersection $\cap$
operations, respectively.}, and second, as a logical lattice
${\cal{L}}(\vee ,\wedge)$, $\quan$ may be thought of as an abstract
quantum logic in the sense of Birkhoff and Von Neumann
(1936)\footnote{That is, $\quan$ is an abstract non-distributive
quantum lattice.}. The particular $C^{*}$-quantales presented  in
(Mulvey and Pelletier, 2000) may be regarded as being both non-commutative
topological spaces and quantum logics as we shall see below. 

In (Mulvey and Pelletier, 2000) the quantales considered derive from
non-commuta- tive $C^{*}$-algebras and are interpreted as particular
realizations of 
non-commutative generalizations of classical topological
spaces ({\it ie}, realizations of abstract quantales). The central motivation for deriving quantales from non-abelian
$C^{*}$-algebras is the possibility of an extension to the non-commutative case of the Gel'fand method of extracting classical topological spaces from
commutative $C^{*}$-algebras that was briefly discussed in the
beginning of this section. The
motivating analogy with the commutative case is the following: as
general 
`classical' topological spaces, namely locales $\loc(\vee, \wedge)$\footnote{Any complete
distributive and $\wedge$-commutative lattice $\loc(\vee ,\wedge)$, otherwise known as a complete Heyting algebra, is called a
`locale'. As mentioned before, locales are viewed as generalizations
of the lattices of open subsets of a given space $X$ that classical
topological spaces ${\cal{T}}(X)$ are modeled after 
(Mac Lane and Moerdijk, 1992).}, are the results of
applying the Gel'fand procedure to commutative $C^{*}$-algebras, due to which
locales, in turn, may be called `commutative topological
spaces'\footnote{As noted above in discussing abstract quantales, at
the set-theoretic or topological level commutativity pertains to
$\cap$'s abelianess. In this sense, the set theory underlying locales ({\it ie}, classical
topologies) is `classical'.}, so
quantales\footnote{The non-commutative extension or generalization of locales.} may
be obtained from applying an analogous Gel'fand procedure to non-commutative
$C^{*}$-algebras\footnote{Hence at the set-theoretic or topological level, the `quantal' character of the
resulting non-commutative topological spaces (from which they derive
their name `quantales'), consists precisely in the
definition of a non-commutative intersection-like operation $\&$
between the elements of a quantale. Thus the set theory
underlying quantales may be called `quantal set
theory'. It would certainly be an interesting project to
compare this
non-commutative extension of classical set theory and its 
resulting `quantalic' topology with Finkelstein's
anticommutative ({\it ie}, Grassmannian) hence nilpotent extension of not only
 the classical, but also of the usual non-distributive Birkhoff-Von Neumann quantum logic, to a
so-called `quantum set theory' and its concomitant `quantum spacetime
topology' (Finkelstein and Hallidy, 1991, Finkelstein, 1996). In the
same line of thought, it would certainly be worthwhile also to
investigate whether 
our $\amg$-schematic non-commutative topologies are particular
instances of the abstract 
non-commutative topologies modeled after Grothendieck-type of schemes
of associative non-abelian Polynomial Identity (PI) rings and
algebras in what is called `Non-Commutative Algebraic Geometry' (Van
Oystaeyen and Verschoren, 1981, Van Oystaeyen,
2000{\it a,b}), since our
incidence Rota algebras may be viewed as such non-commutative PI-rings
(Fred Van Oystaeyen in private communication).}.  

The central issue in the quest for a sound quantal
extension of locales {\it \`a la} Gel'fand is the question of what constitutes a point
in the non-commutative $C^{*}$-algebraic
context\footnote{Equivalently, the questions `what is a quantum
point ?' or `what represents a point in a $C^{*}$-quantale ?'}. The Gel'fand procedure by which one can define `quantum
points' within non-commutative $C^{*}$-algebras is the main theme
in (Mulvey and Pelletier, 2000). As we saw earlier, the Gel'fand
method for `spatializing' or constructing a topological space ${\cal{T}}(X)$ from a
commutative $C^{*}$-algebra $A$ consists first in identifying the points of
the space $X$ to be extracted from $A$ with the
maximal ideals of $A$. Thus one identifies $X$ with the spectrum ${\rm
Max}\, A$. Then one topologizes ${\rm Max}\, A$ by defining a
suitable topology between its points. As Mulvey and Pelletier point
out, ${\rm Max}\, A$ may be straightforwardly constructed as a locale
by considering the `propositional geometric theory' of closed prime
ideals of the commutative $A$ and the logic that this theory
represents. In this case ${\rm Max}\, A$ is the distributive
lattice of propositions of the logic of the theory commonly known as the
Lindenbaum algebra of the theory. 

Now, it seems natural to assume that for a non-abelian $C^{*}$-algebra
$A$, ${\rm Max}\, A$ in the previous paragraph should be substituted
by $A$'s primitive spectrum ${\bf Spec}\, A$ and a non-commutative quantalic topology should be defined on its points. The latter as we
saw previously in the context of our $\amg$-schemes are no other than
the kernels of (equivalence classes) of $A$'s irreps.

Indeed, as it is shown in (Mulvey and Pelletier, 2000), a `good'
spectrum ${\rm Max}\, A$ for defining a quantale in the non-abelian
case is the set of closed two-sided ideals of $A$ or, what it effectively amounts
to the same, the space $\{ M_{i}\}$ of closed linear subspaces of $A$. The latter
may be regarded as the `quantum points' of the quantale. The
topological lattice $\topl(\vee ,\&)$ structure defined on them is
taken to be distributive and its 
joins ($\vee$) and meets ($\&$) are defined algebraically as follows

\begin{equation}
\begin{array}{ll}
&\bigvee_{i}M_{i}=\overline{\sum_{i}M_{i}}\cr
&M_{1}\& M_{2}=\overline{M_{1}\cdot M_{2}}\cr
&M\& (\bigvee_{i}N_{i})=\bigvee_{i} (M\& N_{i})\cr 
&(\bigvee_{i} M_{i})\&
N=\bigvee_{i}(M_{i}\& N)\cr
\end{array}
\end{equation}

\noindent where the join corresponds to the `closure of the linear
span', the meet to the `closure of the product' and the meet
left-right distributes over arbitrary joins. 

One notices immediately that the topology defined on this quantale
${\rm Max}\, A$ is non-commutative exactly because the \& operation is
not so. In return, the latter is not commutative, because it derives from the
non-commutative product of the algebra $A$, as $(9)$ shows. This
non-commutativity should be compared directly with the non-commutative Rota topology
defined between the `quantum points'\footnote{In our case these are
physically interpreted as quantum spacetime events as we saw earlier.} in the base primitive spectra of our primitive $\amg$-schemes
which is generated by the relation $\vec{\rho}$ that also depends crucially on the non-abelianess of the product of the incidence Rota
algebras dwelling at the stalks of the structure sheaves of these
schemes, as $(1)$, $(5)$ and $(8)$ depict.

The reader can refer now to table 1 at the back which summarizes the
comparison between our non-commutative $\amg$-schemes with Mulvey and
Pelletier's $C^{*}$-quantales. It must be said that we have not
commented yet on the last two rows of table 1. That is to say, we
have not commented on how
`the underlying logic and geometry' and `the physical interpretation'
of the two mathematical models compare against each other. 

The underlying or `internal' logic ${\cal{L}}(\vee ,\wedge)$ of $C^{*}$-quantales is seen to be `quantal', analogously to
how the internal logic of a locale is classical intuitionistic (Mac
Lane and Moerdijk, 1992), in a rather strict
technical sense that the reader can find in (Mulvey and Pelletier,
2000)\footnote{Briefly, as the points of a locale constructed by the
Gel'fand procedure from an abelian $C^{*}$-algebra $A$ ({\it ie},
$A$'s maximal ideals) correspond to the classical models of its
underlying classical constructivistic logic ({\it ie},
`intuitionistic' Heyting-Lindenbaum algebra), so the points of a quantale obtained by a similar
Gel'fand procedure from a non-abelian $C^{*}$-algebra $A$ represent
the classical models of its propositional geometric theory
within its `intrinsically quantal' logic. This `intrinsically quantal'
characterization of the logic of the $C^{*}$-quantales considered in
(Mulvey and Pelletier, 2000) may be `justified' on the grounds that
quantum ({\it ie}, 
 Hilbert space) representations of the $C^{*}$-algebras were 
considered there, so that the underlying logical lattice
${\cal{L}}(\vee ,\wedge)$ of these Gel'fand-Hilbert $C^{*}$-quantales
is a non-distributive quantum logic proper in the sense of Birkhoff and Von Neumann (1936). For more
technical details, refer to
(Mulvey and Pelletier, 2000).}. Similarly, when we compare
our $\amg$-schemes to Isham {\it et al.}'s topos-theoretic treatment of quantum
logic proper shortly, we will argue that the internal logic of a topos-like
organization of our $\amg^{s}$s is locally `quantum'\footnote{In the sense of
Birkhoff and Von Neumann (1936), that is to say, 
`non-distributive'.}. This is due to the local coherent quantum
superpositions of qausets in the stalks of the structure sheaves of
our $\amg$-schemes (Raptis, 2000{\it a}, Mallios and Raptis,
2000)\footnote{See footnote 118 in the previous section.}. Thus we
will infer that the aforementioned topos of qausets is `locally
quantalic' indeed.

About the not localized or ungauged hence flat geometrical character of the $C^{*}$-quantales of
Mulvey and Pelletier, as opposed to the localized or gauged thus curved geometry that
our $\amg$-finschemes support, we bring the reader's attention to the
fact that the non-commutative $C^{*}$-algebras $A$ considered in (Mulvey and
Pelletier, 2000) are not localized (as sheaves) over their spectra like our
$\amg$s are in $\amg^{s}$. By our remarks earlier in this
section it follows that the $C^{*}$-quantales as presented in (Mulvey
and Pelletier, 2000) are not suitable to model a variable ({\it ie}, dynamical)
non-commutative topology, while our primitive finschemes of qausets, 
and the non-trivial spin-Lorentzian connections that they host, 
are\footnote{For more on the curved character of a topos organization
of our primitive finschemes of qausets the reader must wait until we
compare our scenario with Isham {\it et al.}'s warped topos model of
quantum logic proper.}. 

Finally, about the physical interpretation of the two mathematical
models for non-commutative topology: $C^{*}$-quantales are constant and  
characteristically `spatial' topologies defined between quantum
`space' points (Mulvey and Pelletier, 2000)\footnote{Hence by our remarks in
the introduction, they are effective mathematical models for static, undirected,
space-like connections between spatial, no matter if quantum, points.}, as opposed to our
primitive $\amg$-finschemes that model a dynamical local quantum
causal topology\footnote{Hence by our
remarks in the introduction, they are sound mathematical models 
for dynamically variable, directed, time-like, local connections between
quantum spacetime events. This gauged or dynamical character of the
qauset topologies in our finschemes will become even more transparent
when we organize the latter into a topos-like structure shortly.}.

In closing this $C^{*}$-quantales {\it versus} $\amg$-finschemes comparison
we mention that it would be very interesting to compare Table 1 at the
end with a similar table presented in (Breslav {\it et al.}, 1999)
where the Rota-algebraic approach to finitary spacetime
topologies\footnote{Albeit, with not a directly causal interpretation
for these topological structures.} is
juxtaposed against an analogous $C^{*}$-algebraic non-commutative lattice approach due to
Balachandran {\it et al.} (1996)\footnote{See also (Landi, 1997, Landi
and Lizzi 1997).}.

Now we wish to shed more light on the curved geometrical and quantum
logical aspects of our primitive finschemes $\amg^{s}$ by comparing
them against the topos-theoretic models that were recently employed to
describe the `warped' and `non-objective' nature of quantum logic
which are the two main consequences of the
Kochen-Specker theorem\footnote{Or commonly known as `paradox' from the point of view of
classical logic (Redhead, 1990).} of quantum logic (Butterfield and Isham, 1998, 1999,
Butterfield {\it et al.}, 2000). First we will give a brief account of
the results from (Butterfield and Isham, 1998, 1999) based on a concise exposition of these two papers by Rawling and Selesnick (2000), then we
will cast our $\amg$-schemes in such topos-theoretic terms, so that
finally we will highlight this comparison by commenting on the entries
of table 2 attached at the end.

So, following closely (Rawling and Selesnick, 2000) we let ${\cal{L}}({\cal{H}})$ be  an
orthomodular ortholattice, a so-called `quantum logic'\footnote{Usually this is
supposed to be the lattice ${\cal{L}}$ of closed subspaces of a Hilbert state
space ${\cal{H}}$ associated with a quantum system, hence the symbol
${\cal{L}}({\cal{H}})$.}. At the `purely logical' or propositional level quantum logics are
distinguished from the lattices modeling classical
logics in that they
are fundamentally non-distributive (Birkhoff and Von Neumann,
1936). However, there is another equally fundamental and more `geometrical' difference 
between quantum and classical logics: while the second are `flat' in the
sense that any Boolean algebra $A$\footnote{Here we will  
not distinguish between classical logics and their corresponding
Boolean algebras $A$.} is isomorphic to the algebra of
global sections of a sheaf of ${\bf 2}$s\footnote{${\bf 2}$ is the 
algebra of the 
classical Boolean binary alternative $\{0,1\}\equiv\{\bot={\rm F}
,\top={\rm T}\}$. It
is the 
`trivial' Boolean subalgebra of every Boolean algebra $A$.} over the
algebra's Stone space ${\rm Spec}\, A$ (Selesnick, 1998), the first
are `warped' or
`twisted', or in a geometrical sense `curved',  
relative to their Boolean substructures\footnote{That is, relative to
their Boolean subalgebras in $W[{\cal{L}}({\cal{H}})]:=\{ A:~A\subset{\cal{L}}({\cal{H}})\}$. It is a well established fact that the $A$s in
$W$ are generated by mutually compatible elements of
${\cal{L}}({\cal{H}})$ which, in more familiar physical terms, are
eigenspaces of compatible ({\it ie}, commuting) `simultaneously
measurable 
observables' of the
quantum system.}. It is this difference that lies at the heart of the
Kochen-Specker theorem of quantum logic as Butterfield and Isham point
out.

More analytically, each Boolean subalgebra $A$ in $W$ represents a
`classical window' through which states of the quantum system may be
`observed'. $W$ is a poset category with the partial order `subset of' 
relation $A\subseteq B$
between two of ${\cal{L}}({\cal{H}})$'s Boolean subalgebras $A$ and
$B$ being interpreted physically as `increasing the power of
resolution in observing the
quantum system's states'\footnote{That is, $B$ enables one to
`observe' more finely or clearly the quantum system's states than
$A$. $A$ provides a coarser view of the system's states than $B$;
equivalently, $B$ is finer than $A$.}. The
states that can be observed through a given $A$ are the latter's `spectral
points'-elements of $A$'s Stone space ${\rm Spec}\, A$. In turn, the
latter may be identified with the set of Boolean valuations of $A$
into the `trivial' Boolean algebra ${\bf 2}$\footnote{These valuations
are homomorphic surjections of $A$ onto its Boolean subalgebra
${\bf 2}$ of classical Boolean truth values.}. Thus each state of the quantum
system that can be observed through $A$ assigns to the elements of $A$
a classical truth value in the Boolean binary alternative ${\bf
2}$. If one abides to the realist existential and non-operational ideal that the states of a quantum system
have some kind of 
`objective reality', in the sense that they `exist' independently of
the acts or operations of observation of external `macroscopic' or
`classical' 
observers, then they should be observable through each classical
window $A$ of ${\cal{L}}({\cal{H}})$ and, of course, such a view of
them should be independent of the power of resolution that one employs to
do so\footnote{That is, even if $A\subseteq B$, one should be able to
view the states of the quantum system equally clearly through either
$A$ or $B$.}. Such `objective reality states' could then be modeled
after `characteristic maps' on each classical window $A$,  
$\chi_{A}:~A\in W\rightarrow {\rm Spec}\, A$ satisfying the
restriction property

\begin{equation}
\chi_{A}=\chi_{B}|A,~(A\subseteq B);~A,B\in W
\end{equation}    

Butterfield and Isham successfully observe that the main implication 
of the Kochen-Specker theorem is that no `global' assignment
$A\mapsto\chi_{A}$ exists if ${\rm dim}({\cal{H}})>2$. The epithet
`global' pertains to the fact that the correspondence

\begin{equation}
{\bf Spec}:~W\rightarrow {\rm Set}
\end{equation}

\noindent between the poset categories $W$ and ${\rm
Set}$\footnote{${\rm Set}$ consists of classical sets partially 
ordered by inclusion. Shortly we will see how the sets in ${\rm Set}$
can be interpreted as variable classical sets, thus they should be
distinguished from the objects in the category ${\rm SET}$ which are usually
interpreted as `constant' 
classical sets
(Bell, 1988, Selesnick, 1991, Mac Lane and Moerdijk, 1992).} is a
contravariant functor or presheaf\footnote{See (Mac Lane and
Moerdijk, 1992, Raptis, 2000{\it b}).} given `point-wise' in the
respective categories by

\begin{equation}
\begin{array}{ll}
&{\rm object-wise}:~{\bf Spec}(A)={\rm Spec}(A)\in{\rm Set}\cr
&{\rm arrow-wise}:~ {\bf Spec}(A\subseteq B)={\rm
Spec}(A)\supseteq{\rm Spec}(B)\in{\rm Set}\cr
\end{array}
\end{equation}

\noindent and as a presheaf it admits no global section of the type
$A\mapsto\chi_{A}$ (again, if ${\rm dim}({\cal{H}})>2$).

Now the category ${\rm Set}^{W^{opp}}$ of these presheaves is an
example of a topos (Bell, 1988, Mac
Lane and Moerdijk, 1992, Rawling and Selesnick, 2000), and it is well known that every topos has a terminal object $1$
with respect to which global sections of its objects are
defined. For ${\rm Set}^{W^{opp}}$ in particular, arrows of the form
$1\rightarrow{\bf Spec}$ are in $1$-$1$ correspondence with global
sections of ${\bf Spec}$ which are easily seen to be assignments of the form
$A\mapsto\chi_{A}$ that do not exist by virtue of the Kochen-Specker theorem.

It must be said however that although global sections of ${\bf Spec}$
do not exist, local ones do and correspond to consistent choices of
valuations on certain Boolean subalgebras of
${\cal{L}}({\cal{H}})$. Now the category ${\rm Set}^{W^{opp}}$
qualifies as a topos, because among various structures in it there
is an object $\omg$, the so-called `subobject classifier',
relative to which subobjects of any given object ${\bf Spec}$ of the
topos are classified by injective arrows from the object into $\omg$\footnote{See (Mac Lane and
Moerdijk, 1992) for a detailed definition of $\omg$ with
examples.}. Local choices from 
or sections of ${\bf Spec}$\footnote{That is, `local Boolean
valuations'.} give rise to sub-presheaves of ${\bf Spec}$, thus lift to
global 
presheaf morphisms of the following kind

\begin{equation}
{\bf Spec}\rightarrow\omg
\end{equation}

\noindent In the context of the Kochen-Specker theorem, Butterfield
and Isham's (1998, 1999) result can be stated as follows: if the target category ${\rm
Set}$ of the contravariant functor ${\bf Spec}$ is identified with the classical Boolean topos
${\rm SET}$ of constant classical sets whose subobject classifier $\omg$ is
${\bf 2}$\footnote{The set of Boolean truth values.}, there is no global sheaf
morphism of the sort depicted in $(13)$\footnote{Object-wise, $(13)$
defines `$A$-Boolean propositions' of the form `${\rm Spec}\,
A\rightarrow{\bf 2}$'.}. However, if one interprets the objects in
${\rm Set}$ as `variable' classical sets\footnote{Varying with respect to the Boolean
subalgebras $A$ of
${\cal{L}}({\cal{H}})$ in the base poset category $W$.} like
Butterfield and Isham do, one can re-express the negative result of the
Kochen-Specker theorem in a positive way as follows

\begin{equation}
{\rm Spec}\, A\rightarrow\omg(A)
\end{equation}

\noindent Expression $(14)$ may be interpreted as some sort of
`localization or gauging of truth' in $W^{{\rm Set}^{opp}}$ whereby the trivial
Boolean algebra ${\bf 2}$ is replaced by the $A$-dependent Heyting
algebra\footnote{That is, a complete distributive lattice, as we saw
earlier in the context of quantales.} $\omg(A)$ of
`generalized truth values' that are localized or soldered on each
object 
$A$ in $W$. Thus, although global Boolean propositions or ${\bf
2}$-valued valuations do not exist on the ${\bf Spec}$ objects of $W^{{\rm Set}^{opp}}$, local
ones,  
`varying per Boolean subalgebra $A$ of ${\cal{L}}({\cal{H}})$', abound
at the expense of
requiring these `truth-assignments' to take values in $A$-dependent
`generalized truth spaces' $\omg(A)$\footnote{Butterfield and Isham call these
`context-dependent valuations'.}. 

It is a standard result in topos theory that every topos has an
`internal logic or language' associated with it that is
characteristically  
intuitionistic\footnote{A so-called `Brouwerian type theory' or
`constructivistic logic'.} (Lambek and Scott, 1986, Mac Lane and Moerdijk, 1992). This is
reflected in the subobject classifier $\omg$ of the topos that is seen
to be a complete Heyting algebra, or topologically speaking, a locale. Expression $(14)$ is a
paradigm of a deep
affinity between `classical logical universes', namely topoi, and `classical' generalized topological spaces
({\it ie}, locales) which may be stated as follows: `topoi are
(locally) localic'\footnote{Shortly we will dwell longer on this
affinity between classical generalized logical universes, namely
topoi, and classical generalized topological spaces, namely locales,
in that we are going to present a topos-like organization of our
primitive $\amg$-finschemes of qausets as being a paradigm of
structures called `quantum topoi' that are the analogues of quantales much in the same way
that classical topoi are the analogues of locales, namely, we will see
that our quantum topos of qausets is `locally quantalic'. This was
also mentioned in the context of $C^{*}$-quantales before.}.  

In the case of the topos ${\rm Set}^{W^{opp}}$ associated with the
quantum logic ${\cal{L}}({\cal{H}})$ (Butterfield and Isham, 1998,
1999) the aforementioned affinity between classical topoi and locales can be
summed up into the
following motto: `although quantum logic does not have a global
notion of Boolean two-valued truth associated with its propositions like
classical Boolean logic does, it has a local intuitionistic
many-valued truth\footnote{The points of the Heyting algebra $\omg(A)$
in $(14)$ corresponding to `multiple truth values'. As we said
earlier, $\omg(A)$ is a generalized truth space.} that still is of a 
`classical' sort in the sense that the Heyting algebra that the
$A$-local propositions of the quantum logic take their truth values still is
a distributive lattice'. This feature of quantum logic may be called
`local multi-valued realism (or classicism)', or as Butterfield and Isham succesfully
coined it, `neo-realism'\footnote{Here we will call it
`neo-classicism' following mainly (Finkelstein, 1996) who identifies a
version of 
`Platonic Realism', called `Ontism', with the basic philosophy
underlying classical physics. See also (Redhead, 1990) for a
discussion of the radical revision of the realist ideal of classical physics
that the quantum revolution brought about.}. 

In closing this
presentetation of the results from (Butterfield and Isham, 1998, 1999)
via (Rawling and Selesnick, 2000), we mention that the topos ${\rm
Set}^{W^{opp}}$ of variable sets over the base category $W$ of Boolean subalgebras of a
quantum lattice ${\cal{L}}({\cal{H}})$ is completely analogous to the
topos ${\bf Sh}(X)$ of sheaves of sets over a region $X$ of Minkowski
space $\cem$\footnote{$\cem$ regarded as a topological manifold ({\it
ie}, a $C^{0}$-manifold).} in
which all the flat classical and quantum field theories are modeled
(Selesnick, 1991). Indeed, in the latter paper it is argued how ${\bf
Sh}(X)$ may be regarded as a universe of continuously variable
classical sets\footnote{Like the elements of the target category ${\rm Set}$ in the topos
${\rm Set}^{W^{opp}}$.} varying
over $X$ which serves as a classical continuous background parameter
space indexing this variation. Selesnick emphasized that the classical sets may be
regarded as being variable in the topos ${\bf Sh}(X)$ exactly because
its internal logic is a strongly typed non-Boolean intuitionistic
Heyting algebra localized over or soldered on the points of $X$\footnote{In
${\bf Sh}(X)$ the analogue of the base poset category $W$ of ${\rm
Set}^{W^{opp}}$ is the poset category of open subsets of $X$ partially
ordered by set-theoretic inclusion `$\subseteq$'.}. Selesnick, at the
end of the (1991) paper, and mainly based on a reticular and quantal
model for spacetime structure and its dynamics, namely Finkelstein's
`quantum causal net', suggested that one should look for a quantum version
of the classical spacetime topos ${\bf Sh}(X)$ which could then be regarded as the
fundamental 
structure in which to model a conceptually sound and pragmatically finite unification of GR and quantum theory ({\it
ie}, `a finite quantum gravity')\footnote{We may call Selesnick's proposal `the quantum
spacetime topos for finite quantum gravity project' or `the quantum topos
project' for short.}.

Below we present our primitive finschemes of qausets in
topos-theoretic terms similar to the ones in which quantum logic was
presented above, so that we can compare
the respective structures and establish structural relationships
between our curved qauset theory and the `warped' quantum
logic of Butterfield and Isham. Our ultimate aim will be to take the
first steps in Selesnick's quantum topos project\footnote{The
possibility of 
approaching Selesnick's quantum topos project via curved finsheaves of qausets
was first noted at the end of (Mallios and Raptis, 2000).}. 

First we give the analogue in our theory of the presheaf expressions
$(11)$ and $(12)$ above. As it was mentioned in the previous section,
the Alexandrov-Sorkin poset category $\as=\{ P_{i},\preceq\}$ of finitary topological posets is
`anti-equivalent' or contravariant to the Rota-Zapatrin poset category 
$\rz=\{\omg_{i},\succeq\}$ of
topological incidence Rota algebras associated with them in the sense
that that there is a contravariant functor 

\begin{equation}
{\bf Spec}^{\omega}:~\as\rightarrow\rz
\end{equation}

\noindent from $\as$ consisting of finitary
substitutes $P_{i}$ and injective poset morphisms or refining relations (arrows) `$\preceq$' between them
(Sorkin, 1991, Raptis, 2000{\it b})\footnote{$\as=\{ P_{i},\preceq\}$ was
called `an inverse system or net of finitary substitutes' in these
papers and the refinement arrows were continuous injections (poset
morphisms) from coarser to finer posets (see previous section). Note
that these refinement relations `$\preceq$' between the topological
posets in $\as$ are completely analogous to the partial order
relations between the Boolean subalgebras of ${\cal{L}}({\cal{H}})$ in
the base poset category $W$ of ${\rm Set}^{W^{opp}}$ that they too
were physically interpreted as `coarsening the grain of observation'
(Rawling and Selesnick, 2000) 
or `increasing the power of resolution' (Mallios and Raptis, 2000).} to
$\rz$ consisting of incidence Rota algebras $\omg_{i}$
associated with the $P_{i}$s of $\as$ and surjective algebra
homomorphisms or coarsening relations (arrows) `$\succeq$' between them (Raptis and Zapatrin, 2000)\footnote{Write
these Rota algebra surjections as
$\omg_{i}\succeq\omg_{j}$. Parenthetically note that
the $(\omega ,p)$-duality between finitary substitutes and their
incidence algebras mentioned in the previous section entails that
whereas the 
inverse system $\as$ has an inverse limit space homeomorphic to a
continuous manifold $X$ ({\it ie}, the points of $X$ are maximally
localized at this limit) (Sorkin, 1991, Raptis, 2000{\it b}), its
contravariant 
`direct system' $\rz :=\{\omg_{i}(P_{i}),\succeq\}$ has a direct limit space that yields
the stalks of $S(X)\equiv {\cal{C}}^{0}(X)$-the sheaf of continuous functions on $X$ ({\it
ie}, the points of $S(X)$ are maximally localized at this
limit; see section 2). These two 
categorically dual (`opposite') processes of localization of finitary posets
(inverse limit) and their dual incidence algebras (direct limit) was
first noticed at the end of (Raptis, 2000{\it b}); see also (Mallios
and Raptis, 2000).}. Point-wise in
the respective categories the
${\bf Spec}^{\omega}$ correspondence reads

\begin{equation}
\begin{array}{ll}
&{\rm object-wise}:~{\bf Spec}^{\omega}(P_{i})\equiv\omega(P_{i})=\omg_{i}(P_{i})\in\rz\cr
&{\rm arrow-wise}:~ {\bf Spec}^{\omega}(P_{i}\preceq P_{j})=\omg_{i}\succeq\omg_{j}\in\rz\cr
\end{array}
\end{equation}

\noindent where in the first row we used the $\omega$-construction of
the incidence algebra from its corresponding finitary poset that was
presented in detail in the previous section. Clearly, ${\bf
Spec}^{\omega}$ is a presheaf in the same way that the
correspondence ${\bf Spec}$ in $(11)$ was seen to be
such. Furthermore, the object-wise $\omega$-correspondence between
$P_{i}$ and $\omg_{i}$ was seen to be a finitary spacetime sheaf in
the sense of (Raptis, 2000{\it b}). Thus, ${\bf Spec}^{\omega}$ is a
presheaf of finitary spacetime sheaves of topological incidence
algebras. It follows that if we evoke from (Raptis, 2000{\it a}) the
`semantic processes' of causalization
and quantization for the posets of $\as$ and their
corresponding incidence algebras in $\rz$, 
we arrive at the presheaf ${\stackrel{\longrightarrow}{\bf
Spec}}^{\vec{\omega}}$\footnote{From now on we will denote this as
`$\stackrel{\longrightarrow}{\bf Spec}$' for short. Again, the arrow over
${\bf Spec}$ reminds one of the causal meaning of the arrows of the
objects of the $\stackrel{\longrightarrow}{\bf Spec}$-related $\as$ and
$\rz$ categories (which should also be written now as $\vec{\as}$ and $\vec{\rz}$).} of finsheaves of qausets.  

Due to the close structural similarities between the ${\bf Spec}$ map of $(11)$
and the $\stackrel{\longrightarrow}{\bf Spec}$ of $(15)$, we infer 
that there is a topos-like
organization of the $\stackrel{\longrightarrow}{\bf Spec}$s similar to how the ${\bf Spec}$s were
organized into the `warped' ${\rm Set}^{W^{opp}}$. We symbolize the resulting
topos structure by
${\rm Qauset}^{\vec{\as}^{opp}}$\footnote{Note that instead of
`$\vec{\rz}$', we wrote `${\rm Qauset}$' for the target poset category
of the $\stackrel{\longrightarrow}{\bf Spec}$ objects of this topos  
just to pronounce its close similarity with ${\rm Set}^{W^{opp}}$ to
be explored shortly.}. 

We propose that ${\rm Qauset}^{\vec{\as}^{opp}}$ is a candidate for the
quantum topos that Selesnick (1991) anticipated. We support our
proposal on the following close parallels between it and ${\rm
Set}^{W^{opp}}$:

\noindent (a) The `points' in ${\rm Set}^{W^{opp}}$ are the Boolean
subalgebras $A$ of ${\cal{L}}({\cal{H}})$ (or the points in the
corresponding 
Stone spaces ${\rm Spec}\, A$ mapped by the functor 
${\bf Spec}$). The sets in ${\rm Set}$ are regarded as varying over the base poset
category $W$ consisting of points and the latter may be viewed as the
localization sites of the former. $W$ may be thought of as the
background base parameter
space `indexing' the variation of sets in ${\rm Set}$. Similarly, the
`quantum events' in ${\rm Qauset}^{\vec{\as}^{opp}}$
are the vertices of the underlying $\vec{P}_{i}$s in $\vec{\as}$ (more
precisely, the quantum point-events in the
primitive spectra ${\bf Spec}\,\amg_{i}(\vec{P}_{i})$ by
$\stackrel{\longrightarrow}{\bf Spec}$). The qausets in
${\rm Qauset}\equiv\vec{\rz}$ are regarded as varying over the base poset
category $\vec{\as}$ consisting of quantum events. $\vec{\as}$ may be thought of
as the background parameter space `indexing' the variation of qausets
in ${\rm Qauset}$.

\noindent (b) The interpretation of the sets in ${\rm Set}$ as variable
entities in the topos ${\rm Set}^{W^{opp}}$ comes from the
Kochen-Specker theorem which may be stated as follows: `the presheaf
objects 
${\bf Spec}$ in it admit no global section', or stated in a positive
way, that `the valuations or states $\chi_{A}$ on the points $A$ in the base poset category $W$ are gauged
or localized or have become context (point) $A$-dependent as ${\rm
Spec}\, A\rightarrow\omg(A)$'\footnote{The symbol for the subobject classifier should
not be confused with that used for incidence Rota algebras. Context
will make it clear what is meant by $\omg$. In any case, qausets are
symbolized by 
$\amg$s not $\omg$s.}. Similarly, we saw in the previous section that
the qausets may be regarded as being dynamically variable entities due
to their localization or gauging as primitive schemes over their
primitive spectra ${\bf Spec}\,\amg_{i}(\vec{P}_{i})$\footnote{Thus, now
objects in ${\rm Qauset}^{\vec{\as}^{opp}}$ are primitive finschemes
$\amg^{s}_{i}$ not finsheaves of qausets.}. Stated in a positive
way the latter corresponds to the definition of non-flat connection operators
${\cal{D}}_{i}=d_{i}+{\cal{A}}_{i}$ for each $\amg^{s}_{i}$ object of
the source category $\vec\as$ of the $\stackrel{\longrightarrow}{\bf
Spec}$ objects of ${\rm
Qauset}^{\vec{\as}^{opp}}$. These connections are  $\amg^{s}_{i}$-scheme
morphisms whose non-trivial part ${\cal{A}}_{i}$ are sections of the
$\amg^{1}_{i}$ subschemes. As we noted in the previous section, that the ${\cal{D}}_{i}$s are non-flat means in
turn that their non-trivial parts ${\cal{A}}_{i}$ are local ({\it ie},
not global) 
sections of the $\amg^{1}_{i}$s.      
 
\noindent (c) Let us dwell a bit on the last remarks in (b) above. We
saw in connection with $(13)$ how local sections of the presheaf objects ${\bf
Spec}$ in ${\rm Set}^{W^{opp}}$ give rise to subpresheaves of ${\bf
Spec}$ thus to global presheaf morphisms between the latter and
another presheaf object structure  
$\omg$ in ${\rm Set}^{W^{opp}}$-the topos' subobject classifier-of the following sort

$${\bf Spec}\rightarrow\omg$$

\noindent or `locally'\footnote{That is, point-wise in the base
category $W$.} as follows

$${\rm Spec}\, A\rightarrow\omg(A)$$

\noindent In complete analogy with ${\rm Set}^{W^{opp}}$, and since we
saw in the previous section how ${\cal{A}}_{i}$ is
${\cal{G}}^{s}_{i}$-valued\footnote{With ${\cal{G}}^{s}_{i}$ the
principal primitive finscheme of reticular and quantal orthochronous
Lorentz `structure group symmetries' of $\amg^{s}_{i}$. Locally, ${\cal{A}}_{i}$ is
supposed to take values in the reticular and quantal version ${\bf g}_{i}$ of the
Lie algebra of the orthochronous Lorentz group ${\bf G}=SO(1,3)^{\uparrow}$ ({\it
ie}, $sl(2,\com)_{i}$) (Mallios and Raptis, 2000).}, local
${\cal{A}}_{i}$ sections of the $\amg^{1}_{i}$s in ${\rm
Qauset}^{\vec{\as}^{opp}}$ lift to global primitive $G$-finscheme
morphisms of the following kind

\begin{equation}
\amg^{s}_{i}\rightarrow {\cal{G}}^{s}_{i}\cr
\end{equation}

\noindent or `locally' in $\amg^{s}_{i}$\footnote{Recall from the previous section that `locally in $\amg^{s}_{i}$'
means `for immediate quantum causal connections which are sections of the subfinscheme
$\amg^{1s}_{i}$ of $\amg^{s}_{i}$'.}

\begin{equation}   
 \amg^{1s}_{i}\rightarrow {\bf g}^{s}_{i}=sl(2,\com)_{i}
\end{equation}

\noindent By comparing the expressions $(17)$ and $(18)$ in ${\rm
Qauset}^{\vec{\as}^{opp}}$ with $(13)$ and $(14)$ in ${\rm
Set}^{W^{opp}}$, we infer that the analogue in the former topos of the
localized or gauged Heyting algebra subobject
classifier $\omg$ in the latter topos is a localized reticular and quantal
version of the Lorentz-spin algebra $so(1,3)^{\uparrow}\simeq sl(2,\com)$ of the orthochronous
Lorentz group $SO(1,3)^{\uparrow}$ of local symmetries of Lorentzian
gravity. Indeed, Selesnick (1994) working on Finkelstein's quantum set
theory and the quantum causal net dynamics based on it (Finkelstein,
1988, 1989, 1996)-a theory very similar to our reticular QCD as
mentioned in (Raptis, 2000{\it a}, Mallios and Raptis, 2000)-found
that a quantum version of the subobject classifier ${\bf 2}$ of the
topos ${\rm SET}$ of constant classical sets is $sl(2,\com)$, and he interpreted the latter as the local relativity group of
Finkelstein's curvaceous quantum net\footnote{This essentially shows
how appropriate is Finkelstein's choice to found the marriage of
quantum with relativity theory ({\it ie}, Quantum Relativity, 1996) on
a `Quantum Set Theory', in that the quantization of the Boolean binary
alternative `symmetry' of classical sets yields directly the local
relativity group of Lorentzian gravity.}. Here too, it is understood
that quantum causal sets coherently superpose in the stalks of the
structure sheaves of the primitive $\amg^{s}$-finscheme objects of
${\rm Qauset}^{\vec{\as}^{opp}}$ which also means that their local
structure or gauge symmetries encoded in their adjoint primitive finschemes
${\cal{G}}_{i}^{s}$ must be quantal (Mallios and Raptis,
2000)\footnote{This is our version of Finkelstein's insight in Quantum
Relativity (1996) that if a system is quantum, so must be its
symmetries. See similar comments in the previous section in connection
with footnote 118 about the FLSP and the local structure (gauge) symmetries of our
finschemes of qausets.}. Then we adjoin to every local object $\amg^{s}_{i}$
`locally' in the topos  ${\rm Qauset}^{\vec{\as}^{opp}}$ a copy of the
primitive finscheme of its reticular, causal and quantal structure
symmetries ${\cal{G}}^{s}_{i}$ as in $(17)$\footnote{This is the
analogue of $(13)$ which holds in  ${\rm
Set}^{W^{opp}}$.}, and regard the gauged or localized thus curved
character of the $\amg^{s}_{i}$s as being represented by `global'
principal primitive finscheme morphisms, between the primitive
$\amg$-finschemes of qausets and their adjoint or associated principal
primitive 
${\bf G}$-finschemes of their local structure symmetries, of the following kind {\it \`a
la} $(18)$

\begin{equation}
\begin{array}{ll}
&{\cal{D}}_{i}:~\amg^{s}_{i}\rightarrow {\cal{G}}^{s}_{i};~{\cal{D}}_{i}=d_{i}+{\cal{A}}_{i}\cr
&{\cal{A}}_{i}:~\amg^{1s}_{i}\subset\amg^{s}_{i}\rightarrow{\bf g}^{s}_{i}\subset{\cal{G}}^{s}_{i}\cr
\end{array}
\end{equation}

\noindent If one gives the epithets `quantum logical' to the topos
${\rm Set}^{W^{opp}}$ and `quantum causal' to ${\rm
Qauset}^{\vec{\as}^{opp}}$, one realizes that `local truth value of an
$A$-proposition in the complete Heyting algebra $\omg(A)$' in the
quantum logical topos ${\rm Set}^{W^{opp}}$' translates to `local
orthochronous Lorentz symmetry for the qausets in the stalk of the structure sheaf of $\amg^{s}_{i}$ over the quantum spacetime event $p_{i}$ of the object $\vec{P}_{i}$ in the base category $\vec{\as}$ which event, in turn, corresponds to a primitive ideal of $\amg_{i}(\vec{P}_{i})$ in ${\bf Spec}\,\amg_{i}(\vec{P}_{i})$'.

\noindent (d) With the remarks in (a)-(c) in mind, one realizes that
the variability of the classical sets in the target
category\footnote{As opposed to the base or source category $W$.}
${\rm Set}$ of the presheaf objects ${\bf Spec}$ in the warped
quantum logical topos ${\rm Set}^{W^{opp}}$ which is due to the
localization or
gauging or $A$-contextualization of classical Boolean ({\it ie}, ${\bf
2}$-valued) valuations on the $A$-propositions in the base category
$W$ of ${\bf Spec}$, translates to the dynamical variability of the
qausets in the target category ${\rm Qauset}$ of the objects
${\stackrel{\longrightarrow}{\bf Spec}}$ in the curved quantum
causal topos ${\rm Qauset}^{\vec{\as}^{opp}}$ which is due to the
localization or gauging of the qausets' reticular and quantal
orthochronous Lorentz symmetries over the quantum spacetime events in
the base category $\vec{\as}$ of ${\stackrel{\longrightarrow}{\bf
Spec}}$.

It is important to note however that this `localization and concomitant warping of truth' in ${\rm
Set}^{W^{opp}}$ results in a local logic or topology in the topos that is still
classical; albeit, intuitionistic: the local generalized truth or topological 
space corresponding to the Heyting algebra or locale $\omg$ assigned locally to every
Boolean subalgebra $A$ of the quantum logic ${\cal{L}}({\cal{H}})$, as
$(14)$ shows. In this sense the quantum logical topos ${\rm
Set}^{W^{opp}}$ is still
classical, or better, `neo-classical'\footnote{See footnote 193.}. 
On the other hand, we have fundamentally assumed that the qausets dwelling
in the stalks of the structure sheaves of the primitive finscheme
objects of the topos ${\rm Qauset}^{\vec{\as}^{opp}}$ coherently
superpose with each other, which implies that `the
localization and concomitant curving of causality and its
local orthochronous Lorentz
symmetries' in this topos are also in a strong sense quantum (Mallios and Raptis,
2000)\footnote{For instance, the local logic in this topos of qausets
must be `quantum' in the general non-distributive lattice sense of
Birkhoff and Von Neumann (1936), since the non-distributivity property
of quantum logic is due to the coherent superpositions (quantum
interference) of quanta-in our case, of qausets (Raptis, 2000{\it a}).}. Precisely for this we regard the quantum causal topos ${\rm
Qauset}^{\vec{\as}^{opp}}$ a sound paradigm of a quantum topos. We
further 
`justify' this observation in (e) next.

\noindent (e) We must emphasize at this point that since the curved primitive
$\amg$-finscheme objects of the target poset category ${\rm Qauset}$ of the presheaf
objects $\stackrel{\rightarrow}{\bf Spec}$ of ${\rm Qauset}^{\vec{\as}^{opp}}$ are
`quantalic' {\it \`a la} Mulvey and Pelletier (2000) as we saw
earlier, we may take the topos ${\rm Qauset}^{\vec{\as}^{opp}}$
to be an example of the `missing factor' structure in the following analogy that
has puzzled mathematicians for quite some time now

\begin{equation}
\frac{\rm topoi}{\rm locales}=\frac{\rm ?}{\rm quantales}
\end{equation}

\noindent it being understood that the elusive `quantum topos' is the
structure to 
replace the questionmark above\footnote{Jim Lambek and Steve Selesnick in
private communication. In short, we may take the analogy $(20)$ as
saying that as a classical topos is locally localic (topologically
speaking) or locally intuitionistic/neo-classical  (logically speaking) like the
${\rm Set}^{W^{opp}}$ of Butterfield and Isham (1998, 1999) or the
${\bf Sh}(X)$ of Selesnick (1991), so a quantum topos is (expected to be) locally
quantalic; that is to say, locally it is a non-commutative topological
lattice 
(topologically speaking) \underline{and} a non-distributive logical lattice
(logically speaking). Our ${\rm Qauset}^{\vec{\as}^{opp}}$ is locally
quantalic in this sense.}. 

To give more evidence for how our ${\rm
Qauset}^{\vec{\as}^{opp}}$ is a good example of such a quantum topos structure,
we refer to a topos  ${\rm Set}^{{\cal{V}}^{opp}}$ analogous to the ${\rm Set}^{W^{opp}}$ of
(Butterfield and Isham, 1998, 1999, Rawling and Selesnick, 2000)
presented in (Butterfield {\it et al.}, 2000) with the only structural
difference
between the two topoi being  
that the base
category $W$ in the second was replaced by a poset category ${\cal{V}}$ of commutative $C^{*}$
or Von Neumann algebras $V$. It is understood that if the
non-commutative $C^{*}$-algebra $B({\cal{H}})$ of bounded
linear operators on the Hilbert space ${\cal{H}}$ of a quantum system is considered, the
objects in ${\cal{V}}$ are the abelian Von Neumann subalgebras of
$B({\cal{H}})$\footnote{Interestingly enough, the maximal abelian
subalgebras of non-commutative involutive algebras of quantum
spacetime operations or actions are coined `classical frames' in (Finkelstein,
1996, Selesnick, 1998), and the theory for the quantum structure and dynamics of
spacetime called `Quantum Relativity' may be equivalently named `Frame
Relativity'. It is certainly interesting to investigate the extent to
which a topos, like the ${\rm Set}^{{\cal{V}}^{opp}}$ above for example, is a sound
mathematical model for implementing the basic ideas of Finkelstein's
Quantum Relativity 
theory on unifying the world's quantum logic with its relativistic
causal structure. Below we will give some more strong hints that our quantum topos ${\rm
Qauset}^{\vec{\as}^{opp}}$ may be regarded as such a model.}. Without giving more technical details from
(Butterfield {\it et al.}, 2000), one can use the
spectral theorem on $B({\cal{H}})$ to establish a $1$-$1$
correspondence between the poset categories $W$ and ${\cal{V}}$. Thus
the presheaf 
objects in the topos  ${\rm Set}^{{\cal{V}}^{opp}}$ were called
`spectral presheaves' and were symbolized by ${\bf\Sigma}$. In  ${\rm
Set}^{{\cal{V}}^{opp}}$ the consequence of the Kochen-Specker theorem
is that its ${\bf\Sigma}$ objects admit no global sections which, in turn,
amounts to the definition of generalized ({\it ie}, Heyting-valued)
and $V$-dependent ({\it ie}, $V$-contextualized) valuations on ${\cal{V}}$ in complete
analogy to the ${\rm Set}^{W^{opp}}$ case. Then, ${\rm
Set}^{{\cal{V}}^{opp}}$, like  ${\rm Set}^{W^{opp}}$, is seen to be (locally)
localic or neo-classical\footnote{Just for this we will argue shortly
that ${\rm
Set}^{{\cal{V}}^{opp}}$, in contradistinction to our ${\rm
Qauset}^{\vec{\as}^{opp}}$, can not be an example of the quantum topos
that Selesnick (1991) anticipated to underlie Finkelstein's curved
quantum causal net.}. 

On the other hand, as Butterfield {\it
et al.} (2000) remarked in constructing the spectra for the
commutative $*$-algebras $V$ in ${\cal{V}}$, if non-commutative
$C^{*}$-algebras $N$ were used as objects in the base category ${\cal{N}}$, their spectra would be
quantales in the sense of (Mulvey and Pelletier, 2000)\footnote{That
is, in the non-commutative $C^{*}$-algebra case the analogue of a
spectrum is the set of all closed two-sided ideals in the non-abelian
$N$s. See table 1.}. The natural conjecture
is that the resulting topos-like structure, ${\rm
Set}^{{\cal{N}}^{opp}}$ can be regarded as being `locally
quantalic' in contradistinction to ${\rm
Set}^{W^{opp}}$ or ${\rm Set}^{{\cal{V}}^{opp}}$ which is logically  
`locally Heyting' or topologically `locally localic'. Then perhaps it
too could qualify as a possible candidate for a sound model of the
aforementioned quantum topos similar to our ${\rm Qauset}^{\vec{\as}^{opp}}$\footnote{Refer to
table 1 to see in what sense our primitive $\amg$-finscheme objects of
${\rm Qauset}^{\vec{\as}^{opp}}$ represent
a localized or gauged version of the $C^{*}$-quantales of Mulvey and
Pelletier (2000).}. Since now both ${\rm
Set}^{{\cal{N}}^{opp}}$ and our topos-like ${\rm
Qauset}^{\vec{\as}^{opp}}$ allow for coherent quantum
superpositions locally, they may be viewed as being `locally
quantalic'\footnote{That is to say, as being locally non-commutative
topological lattices $\topl(\vee ,\&)$ and non-distributive quantum
logic lattices ${\cal{L}}(\vee ,\wedge)$.}, hence both are good candidates for the quantum topos
structure anticipated in $(20)$. With these remarks we come to appreciate more the
significance of adopting from (Finkelstein, 1988) the FLSP for the
dynamically variable 
qausets in the stalks of the curved finsheaves in (Mallios and Raptis,
2000).

Of course, it is understood that the topos ${\rm
Set}^{{\cal{N}}^{opp}}$ may be thought of as some kind of `Hilbert
space 
bundle or sheaf' on whose stalks ({\it ie}, Hilbert spaces) the
non-abelian local $C^{*}$-observable algebras $N$
in ${\cal{N}}$ are represented. Evidently, the base category
${\cal{N}}$ is no longer a poset category like the $W$ in 
${\rm Set}^{{\cal{W}}^{opp}}$. It is significant to note however that such a
sheaf-like structure of Hilbert spaces arises quite naturally in
another topos-theoretic approach to quantum theory and quantum
gravity, namely, Isham's version of consistent histories (Isham,
1997)\footnote{Chris Isham in private communication.}. Moreover, finite dimensional Hilbert spaces associated with quantum lattices of
consistent histories have recently been localized over the event-vertices of
causets (Markopoulou, 2000)\footnote{Thus the base spaces of these
finite dimensional 
Hilbert sheaves are again poset categories, albeit, with a directly
causal interpretation of the partial orders involved and not as
`refinement relations'.} in a manner very similar to how our
qausets in ${\rm Qauset}^{\vec{\as}^{opp}}$ are localized over
(curved) causal spaces $\vec{P}_{i}$. The resulting structures were coined `quantum
causal histories'. It is certainly worthwhile to try to relate
our quantum topos of dynamically variable qausets ${\rm Qauset}^{\vec{\as}^{opp}}$ with 
the consistent quantum causal histories of Isham (1997) and Markopoulou (2000). The first step in
this direction would be to view the finite dimensional irreps of our
qausets as acting on Hilbert spaces in much the same way that finite
dimensional Hilbert space irreps of topological incidence Rota
algebras were studied in (Zapatrin, 1998, Breslav {\it et al.},
1999). Such a study however has to be left for another paper.  

Due to (a)-(d) above, we suggest that ${\rm Qauset}^{\vec{\as}^{opp}}$
is a good candidate for the quantum topos that Selesnick (1991)
anticipated to underlie Finkelstein's quantum set and net theory
(Finkelstein, 1988, 1989, 1996), thus
it can serve as a reasonable mathematical model for the kinematical structure of a
reticular, causal and quantal version of Lorentzian gravity, as
Mallios and Raptis (2000) also anticipated at the end of that paper.

Now the reader can refer to table 2 at the back for a comparison
between the neo-classical quantum logical topos ${\rm Set}^{W^{opp}}$ of Butterfield
and Isham (1998, 1999) and our quantum causal quantum topos ${\rm
Qauset}^{\vec{\as}^{opp}}$. Conceptually, perhaps the most significant
aspect of this comparison is that in ${\rm Qauset}^{\vec{\as}^{opp}}$
the analogue of `local truth value' in ${\rm Set}^{W^{opp}}$ is
`local quantum causal symmetry', so that it follows that the formal
analogue in the quantum logical topos ${\rm Set}^{W^{opp}}$ of local
causality $\stackrel{*}{\rightarrow}$ in the base poset category $\as$
of classical (albeit curved) causets is the classical logical implication structure
`$\Rightarrow$'. In view of the fact that it is quite problematic to
define a sound implication connective in quantum logic (Rawling and
Selesnick, 2000), and because $\stackrel{*}{\rightarrow}$ or its
$\vec\omega$-equivalent $\vec{\rho}$ has become a local dynamical
quantum variable in ${\rm Qauset}^{\vec{\as}^{opp}}$ via a non-flat
${\cal{D}}$, it follows that the troubles one has in defining a
`global' quantum implication may be not due to the non-distributive
nature of quantum logic {\it per se}, but rather due to its warped or `twisted'
character as we saw above. If that turns out to be the case indeed, in
complete analogy with the intransitive curved local causality
$\stackrel{*}{\rightarrow}$ in ${\rm Qauset}^{\vec{\as}^{opp}}$ (Mallios
and Raptis, 2000), we
conjecture that $\stackrel{*}{\Rightarrow}$\footnote{This is the local
`neo-classical' intuitionistic implication $A$-context-wise in ${\rm
Set}^{W^{opp}}$.} is also intransitive\footnote{It is only transitive
locally ({\it ie}, $A$-wise).}. The implication structure of
quantum logic is `anomalous' (Rawling and Selesnick, 2000) because the
logic itself, in contradistinction to classical flat logic, is warped
hence intransitive. This possibility however will have to be examined
more thorougly in a future paper (Raptis, 2000{\it d}).

In the final section of this paper we discuss the possibility that the
fundamental quantum time-asymmetry, the so-called `quantum arrow of
time', expected of the `true quantum gravity' (Penrose, 1987), is of a
micro-local 
kinematical or structural and not of a dynamical nature as it is
usually anticipated to be. We also argue
that a quantum topos-theoretic model like our ${\rm
Qauset}^{\vec{\as}^{opp}}$ may prove to be a solid conceptual platform
for a sound unified
description of quantum logic and quantum gravity thus vindicate some
early prophetic insights of Lawvere (1975) and Finkelstein (1969,
1979, 1996).

\section*{\normalsize\bf 5. CONCLUDING REMARKS}
 
`{\it The true quantum gravity must be a time asymmetric theory}' (Penrose,
1987)\footnote{In fact, Penrose held that the true quantum gravity
$\underline{\rm is}$ a time asymmetric theory (see footnote 2), but in
view of the fact that a conceptually cogent and finite quantum gravity
has not been proposed yet, we have replaced the factual `is' by the
conjectural imperative `must be'.}. By this remark we understand in general that the fundamental
asymmetry or directedness of time must be traced in a dynamical theory
of quantum spacetime structure\footnote{That is, when quantum
spacetime is itself viewed as a dynamical variable.}. 

In the Ashtekar
formulation of self-dual canonical QGR (Ashtekar, 1986, Baez, 1994, Baez and Muniain, 1994)
or in its equivalent covariant `path-integral over self-dual
connection space' approach (Baez, 1994, Baez and Muniain, 1994) for
instance, Penrose's imperative would
translate into first formulating a time-asymmetric and quantum version
of Einstein's 
equations, then solving them to find time-asymmetric
solutions. Indeed, Ashtekar (1986) first discovered asymmetric or
chiral 
self-dual gauge connection spin-valued variables ${\cal{A}}^{+}$ and
re-wrote the Einstein equations of GR in terms of them, and 
then proceeded to a canonical quantization of this self-dual
`spinorial' GR. Subsequently, these chiral spinorial quantum
Einstein equations were cast into a covariant `sum-over-histories or path-integral over the
space of ${\cal{A}}^{+}$ connections' form, and then they were tried
for solutions. Significant step in this direction was Rovelli and
Smolin's introduction of Wilson or ${\cal{A}}^{+}$-holonomy loops to
the search for solutions of the Chern-Simons-Witten (CSW) path-integral
version of the self-dual QGR of Ashtekar (Baez and Muniain,
1994)\footnote{The so-called `loop representation' of self-dual
quantum gravity.}. A series of subsequent discoveries of certain close affinities between
the Rovelli-Smolin loop variables for quantum gravity and abstract
mathematical objects called knots and links, resulted in the proposal that
oriented or 
chiral knots may be solutions or `states' for the self-dual quantum
gravity in the CSW picture of the theory. Indeed, Kodama\footnote{For
a relatively complete list of references to the Rovelli-Smolin loop
formulation of quantum gravity, the relation between knot theory and
quantum gravity, the CSW covariant path-integral approach to quantum gravity, and to Kodama's
result, the reader is advised to look at (Baez and Muniain, 1994), as
well as to refer to the back of various relevant papers in (Baez,
1994).} 
showed how a state for CSW quantum gravity is a quantum version of
anti-deSitter space. Shortly after it was appreciated how an oriented
knot or link invariant, the so-called `Kauffman bracket', was in fact a CSW
state in the loop representation (Baez and Muniain, 1994). All in all,
with the development of the loop approach to self-dual or chiral
quantum gravity we have got a nice paradigm of Penrose's `imperative'
above. However, one should note that the kinematical 
${\cal{A}}$-connection space over which one defines the
path-integral dynamics representing quantum gravity is that of the
self-dual $sl(2,\com)$-valued part ${\cal{A}}^{+}$ of the
gravitational gauge conection $\cal A$ and not its `anti-chiral' 
anti-self-dual part ${\cal{A}}^{-}$ that takes values in 
$\overline{sl(2,\com)}$\footnote{The so-called `conjugate
representation of the orthochronous Lorentz-spin algebra'.} (Baez and Muniain,
1994)\footnote{We will see shortly how this may be regarded as an
analogue of the kinematical or structural local quantum time-asymmetry
that we will propose in ${\rm Qauset}^{\vec{\as}^{opp}}$.}. 
      
Current research in quantum gravity aside for a moment, it was a remarkable early achievement to find
classical time-asymmetric solutions to the classical time-symmetric
Einstein equations by employing a special coordinate system to chart
the (topologically) undirected spacetime manifold\footnote{As we said
in the introduction, the locally
Euclidean $C^{0}$-manifold topology of GR is based on undirected,
reversible,  
two-way, space-like connections between its point-events (Finkelstein, 1988, 1991).} 
(Finkelstein, 1958)\footnote{The coordinate system alluded to above is
known as the Eddington-Finkelstein coordinates.}. Since the smooth
spacetime coordinates have a kinematical or structural rather
than a dynamical role in GR\footnote{In the sense that in the
classical theory of gravity the
sole dynamical variable is the spacetime metric $g$ (or its affine
Christoffel connection $\Gamma$ or the latter's gauge-theoretic
analogue-the spin-Lorentzian connection $\cal A$), while the smooth real coordinates of the
$C^{\infty}$-manifold are `gauged away' by the general coordinate
transformation group
$GL(4,\R)$ or the diffeomorphism group ${\rm Diff}(M)$ thus
implementing the Principle of General (Gauge) Covariance.}, Finkelstein's
discovery may be perceived as a kinematical or structural explication of a `hidden'
time-asymmetry of classical gravity. The fact that time-asymmetric
solutions exist for the time-symmetric Einstein equations of GR on a
topologically undirected and spatial spacetime manifold, seemed quite paradoxical
to the author at that time, since they appeared to violate the Principle of
Sufficient Reason which in that case could be taken as holding that
`time-symmetric causes must have time-symmetric effects'\footnote{The
`causes' in that case being Einstein's field equations and the `effects'
their solutions.} (Finkelstein, 1958). Finkelstein resolved 
this apparent paradox by holding that it could be regarded simply as another
consequence of the
non-linear nature of gravity. Parenthetically, and in view of the
structural or kinematical quantum time-asymmetry in the quantum topos
${\rm Qauset}^{\vec{\as}^{opp}}$ that we are going to present shortly,
we note that Finkelstein's demonstration in (1958) that the exterior
singularity of the Schwarzschild solution of the Einstein equations of
GR is in fact a `unidirectional causal membrane'\footnote{That is,
causal influences in the gravitational field of a spherical point-particle can propagate strictly only to the future or only to the
past.} entailed that the universe splits into two classes of
particles: those whose gravitational fields allow only for future propagation
of causal signals\footnote{Which Finkelstein identifies with
`particles'.}, and those that allow only for a `past-propagating
causality'\footnote{Which Finkelstein identifies with
`antiparticles'.}. In fact, at the end of the paper Finkelstein notes
that ``{\it in view of the delicate nature of the choice between the
two classes, it is possible that the gravitational equations imply
that all particles in one universe belong to the same
class}''. Shortly we will essentially base the kinematical quantum
arrow of time in ${\rm Qauset}^{\vec{\as}^{opp}}$ on a similar
`initial choice' or `condition' between ${\rm Qauset}^{\vec{\as}^{opp}}$
and its quantum time-reverse $({\rm
Qauset}^{\dagger})^{(\vec{\as}^{\rm op})^{opp}}$.  

Now we would like to reinstate the Principle of Sufficient Reason in
gravity, as well as go against the current trend in quantum gravity
research, and suggest that quantum gravity must be time-asymmetric simply
`because' the kinematical structure of quantum spacetime is so. In
other words, we will argue that a local quantum arrow of time is already
present or built-into our mathematical model for the kinematics of
quantum spacetime, which time-asymmetric kinematical structure the
dynamics ({\it ie}, `quantum gravity') must then respect so as to
`propagate and conserve the kinematical quantum arrow of time'. 

Our proposal is really simple: we saw in the previous section how the
quantum topos 
${\rm Qauset}^{\vec{\as}^{opp}}$ of the curved quantum time-forward,
or gauged future quantum
causal topologies $\amg^{s}$ is a sound model of the kinematics
of a dynamically variable future local quantum causality $\vec{\rho}$. This
then defined the `conjugate quantum topos' $({\rm
Qauset}^{\vec{\as}^{\rm op}})^{\dagger}=({\rm
Qauset}^{\dagger})^{(\vec{\as}^{\rm op})^{opp}}$ consisting of the curved quantum time-backward, or
gauged past quantum causal 
topologies $\amg^{s\dagger}$ is a sound model of the kinematics of a
dynamically variable past local quantum causality
$\vec{\rho}\,^{\dagger}$\footnote{Recall from the previous section
that $\vec{\rho}\,^{\dagger}$ may be thought
of as the past local dynamical quantum causality relation holding between quantum
spacetime events as depicted in the second line of $(8)$. It is
certainly interesting in this context to mention the remark at the end of (Mulvey and
Pelletier, 2000) that if the propositional interpretation of an element `$a$' of a
Gel'fand-Hilbert $C^{*}$-quantale is as an action on (the Hilbert space states of) a
quantum system, then its adjoint or involute `$a^{*}$' is interpreted
as  
\underline{an action in reverse time}. Thus our interpretation of the
elements of ${\rm Qauset}^{\vec{\as}^{opp}}$ and $({\rm
Qauset}^{\dagger})^{(\vec{\as}^{\rm op})^{opp}}$ as local and curved quantum causal
actions opposite in (quantum) time is well justified. Of course, as we
saw in the previous section and as it is written in table 1, the
fact that our qauset
algebras $\amg_{i}$ are not involutive ({\it ie}, they are not closed under
$\dagger$) like their $C^{*}$-quantale relatives, is precisely what 
allows for this fundamental separation of qausets (and their quantum topoi) into `two universes of local
curved 
quantum causal actions', one future (${\rm
Qauset}^{\vec{\as}^{opp}}$), the other past ($({\rm
Qauset}^{\dagger})^{(\vec{\as}^{\rm op})^{opp}}$), much like the
disjoint 
future and past unidirectional causality universes in (Finkelstein,
1958) that we saw earlier (see also below).}. 

So we have effectively obtained a binary alternative (choice) for
the kinematical local quantum time-directedness which may be presented as follows

\begin{equation}
{\rm Qauset}^{\vec{\as}^{opp}}~ \underline{\bf or}~ ({\rm
Qauset}^{\dagger})^{(\vec{\as}^{\rm op})^{opp}}
\end{equation}
    
\noindent This binary decision we may assume as having being made or `fixed'
upon the creation of the universe. In other words, we regard it as being determined by
some kind of 
`initial condition'\footnote{The words `choice', `decision' or `constraint'
could also be alternatively used as synonyms to `condition'.} for the dynamics of local
quantum causality of the following sort:

\begin{quotation} 
{\it The initial kinematical quantum causal structure is}
${\rm Qauset}^{\vec{\as}^{opp}}$.
\end{quotation}

Our proposal is in complete formal analogy to Penrose's (1987) `{\it Weyl
Curvature Hypothesis}' for the fundamental quantum time-asymmetry of
the true quantum gravity which may be stated as follows:

\begin{quotation}
\noindent{\it The Weyl curvature tensor of gravity\footnote{\rm The part $W$
of Einstein's tensor $G$ (or equivalently Riemann's $R$) that measures the entropy of the
gravitational field.} was initially zero:
$W|_{t_{0}=0}=0$}.\footnote{In this way Penrose also `explains' the
origin of the so-called `thermodynamic arrow of time'. Our
considerations however are entirely quantum, not statistical/thermodynamical (Raptis
and Zapatrin, 2000, Mallios and Raptis, 2000).}
\end{quotation}

It follows that the `categorical dynamics' of the variable local quantum causality
$\vec{\rho}$ in the quantum topos ${\rm
Qauset}^{\vec{\as}^{opp}}$\footnote{That is, a locally finite, causal
and quantal version of Lorentzian gravity (Mallios and Raptis,
2000). See also section 4.} formulated entirely in terms of primitive finscheme
morphisms will preserve the `micro-local future-directedness' of
$\vec{\rho}$ thus it will in a sense `conserve' the kinematical
micro-local initial quantum arrow of time. This may be loosely called
`the principle of conservation of the initial kinematical micro-local
quantum time-asymmetry by quantum gravity'\footnote{Again, `quantum gravity'
being perceived in our theoretical scenario as the finitary dynamics
of local quantum causality $\vec{\rho}$ (Mallios and Raptis, 2000).}.
 
At this point we must emphasize that if the $\amg^{s}$ objects of
${\rm Qauset}^{\vec{\as}^{opp}}$ have a finitary and quantal version of
$sl(2,\com)$ as local structure or relativity group\footnote{See
(Mallios and Raptis, 2000) and previous section.}, then the
$\amg^{s\dagger}$s in the `quantum time-reverse' or conjugate topos $({\rm
Qauset}^{\dagger})^{(\vec{\as}^{\rm op})^{opp}}$ are expected to have have a locally finite
and quantal version of $\overline{sl(2,\com)}$\footnote{This is the
complex conjugate of the Lie algebra $sl(2,\com)$.} as their local
relativity group structure. The qausets in the $\amg^{s}$s of ${\rm
Qauset}^{\vec{\as}^{opp}}$ may be thought of as `left-handed' spinors
transforming as vectors under the fundamental or regular irrep $S$ of $sl(2,\com)$,
while the qausets in the $\amg^{s\dagger}$s of $({\rm
Qauset}^{\dagger})^{(\vec{\as}^{\rm op})^{opp}}$ may be thought of as
`right-handed' spinors transforming as vectors under the conjugate
irrep $S^{*}$ of the spin-group corresponding to
$\overline{sl(2,\com)}$\footnote{Note that $S$ and $S^{*}$ are
inequivalent irreps. Also, since $sl(2,\com)$ is isomorphic to the
orthochronous Lorentz Lie algebra $so(1,3)^{\uparrow}$,
$\overline{sl(2,\com)}$ may be regarded as being isomorphic to the
`antichronous' ({\it ie}, time-reverse) Lorentz Lie algebra
$so(1,3)^{\downarrow}$.}. Thus, the aforementioned micro-local quantum
time-asymmetry may be alternatively stated as follows:

\begin{quotation}
{\it Only left-handed causons\footnote{\rm Recall from the previous
section that we called `causons' the dynamical quanta of local
quantum causality $\vec{\rho}$.} exist in the quantum spacetime
deep}\footnote{It follows that if ${\rm Qauset}^{\vec{\as}^{opp}}$
stands for the kinematics of the $\vec{\rho}$ causons, $({\rm
Qauset}^{\dagger})^{(\vec{\as}^{\rm op})^{opp}}$ stands for the
kinematics of the $\vec{\rho}\,^{\dagger}$ `anticausons'. The latter
are viewed as the quantum relativistic antiparticles of
causons that dynamically propagate in the quantum time-reverse ({\it ie},
antichronous) direction of the orthochronous 
causons, thus they have `local (gauge) symmetry'
$\overline{sl(2,\com)}\simeq so(1,3)^{\downarrow}$.}.
\end{quotation}

\noindent Interestingly enough, Finkelstein (1988) and Selesnick (1994, 1998) have come to the
same conclusion about the `chirality' or `handedness' or `micro-local
quantum time-asymmetry' of the dynamical spinorial spacetime
quanta\footnote{The so-called 
`chronons'.} of Finkelstein's quantum causal net: they are left-handed
or future-directed quanta\footnote{This author wishes to thank Steve
Selesnick for emphasizing all along the importance of this result in
numerous  
private communications (see also below).}. We also note here that this initial
kinematical choice
for the future quantum causal universe ${\rm Qauset}^{\vec{\as}^{opp}}$
populated exclusively by causons, and against the past quantum causal universe
$({\rm
Qauset}^{\dagger})^{(\vec{\as}^{\rm op})^{opp}}$ inhabited solely by
anti-causons, is completely analogous to the time-asymmetric or `causally
unidirectional' classical gravitational fields of point particles (and
their antiparticles) mentioned earlier in connection with (Finkelstein, 1958).  
 
Now the reader can compare this view of ours about the `origin' of the
fundamental quantum time-asymmetry expected of the true quantum
gravity with the aforementioned path-integral over self-dual
$sl(2,\com)$-valued gravitational gauge connections ${\cal{A}}^{+}$ of
Ashtekar {\it et al.} The bottom line is that according to our view the quantum gravity
represented by the latter scenario is time-asymmetric `because',
structurally or kinematically speaking, only chiral left-handed
gravitational spin
variables ${\cal{A}}^{+}$ were used in the first place to define what
is variable in ({\it
ie}, the kinematics of) the theory, and not their
$P$-mirror or $T$-reverse\footnote{$P$ is the parity  and $T$
the time reversal operators of the usual (flat) QFT.} images
${\cal{A}}^{-}$\footnote{It must be emphasized however that it is
quite doubtful whether the $CPT$ theorem of the usual (flat) QFT on a
classical Minkowskian spacetime manifold still holds in the more
primordial realm of quantum gravity where spacetime itself, apart from
the fact that it is expected to be 
`curved' in some sense, is also held to be itself a quantum
system. Thus one should be careful not to rationalize {\it a priori}
about the initial quantum arrow of time in such $CPT$ terms which may
be valid theoretical notions only for the flat matter quanta of QFT
and not for the curved spacetime quanta of quantum gravity. A similar
caution is given in (Finkelstein, 1988).}.

At this point we would like to `justify' our re-establishing the
aforementioned Principle of Sufficient Reason in a quantum gravity
perceived as the dynamics of a non-commutative local quantum causal
topology $\vec{\rho}$ in the kinematical quantum topos ${\rm
Qauset}^{\vec{\as}^{opp}}$. This `justification' is based on a quantum
causal version of the inverse limit mechanism suggested in (Sorkin,
1991, Raptis and Zapatrin, 2000, Mallios and Raptis, 2000) for
recovering the locally Euclidean $C^{0}$-manifold topology of the
classical spacetime continuum on which GR essentially rests from the
Alexandrov-Sorkin poset category or `inverse net'
$\as$ of
finitary topological substrata, or from $\as$'s contravariant Rota-Zapatrin incidence algebra poset
category or `direct net' $\rz$ of quantum topological substrata. In $\rz$ this meachanism
was interpreted as Bohr's correspondence limit (principle) (Raptis and
Zapatrin, 2000) and it was contended that the classical
$C^{0}$-spacetime manifold of macroscopic experience together with the
commutative algebras $\omg^{0}$ of its local event-determinations
arise from some sort of decoherence of the fundamentally a-local
quantum topological substrata
$\rz(\as)=\{\omg_{i}(P_{i})\}$\footnote{Note that by the association
$\rz(\as)$ above, one may simply understand the contravariant functor
(presheaf) that relates these two poset categories.}. Similarly
in the quantum causal context, we may assume that the curved directed
non-commutative quantum causal topologies in ${\rm
Qauset}^{\vec{\as}^{opp}}$, at the ideal and non-pragmatic\footnote{See
(Raptis and Zapatrin, 2000, Mallios and Raptis, 2000).} limit of
infinite localization of spacetime events\footnote{See (Raptis,
2000{\it b}, Mallios and Raptis, 2000).} in the
$\stackrel{\longrightarrow}{\bf Spec}$ presheaf objects of this quantum topos, yield the classical
time-undirected (spatial) topological spacetime manifold $M$ and the commutative
algebra of coordinates of its events. Actually, it would be desirable if
one could in fact show that the correspondence limit topos arising
from object-wise decohering the quantum topos ${\rm
Qauset}^{\vec{\as}^{opp}}$ is the classical topos ${\bf Sh}(X)$ of
sheaves of sets over the region $X$ of the curved classical spacetime
manifold $M$ of $GR$. That this is indeed so, will be shown in (Raptis,
2000{\it d}). 

At the corresponding dynamical level, the time-asymmetric quantum
gravity on ${\rm Qauset}^{\vec{\as}^{opp}}$ will yield upon decoherence
of the a-local directed and curved qausets the usual time-symmetric
Einstein equations on the undirected $M$. Thus, `justification' for our kinematical quantum time-asymmetry may be the following
compelling analogy based on the kinematical mechanism\footnote{Again, the
characterization of the inverse limit above as being `kinematical' is
due to Lee Smolin and Chris Isham. See footnote 8 in the introduction.} described above:

\begin{quotation}

As the time-symmetric Einstein equations for classical gravity ({\it
ie}, GR) fundamentally rely on the undirected locally Euclidean
$C^{0}$-manifold topological space model for spacetime and the
commutative algebra of coordinates of its events, so a time-asymmetric
quantum gravity must essentially rely on the kinematical or structural
directedness of the non-commutative local quantum causality modeled
after the quantum causal topos ${\rm Qauset}^{\vec{\as}^{opp}}$.

\end{quotation}

\noindent This analogy may be diagrammatically represented as follows

\[\begin{CD}
{\footnotesize\rm t-asymmetric\,\, q-causal\,\, kinematics}@>{\rm
sufficient}>{\rm reason}>{\rm t-asymmetric\,\, q-causal\,\, dynamics}\\
@V{\rm correspondence}V{\rm limit}V        @V{\rm correspondence}V{\rm
limit}V\\
{\rm t-symmetric\,\, c-manifold\,\, kinematics}@>{\rm sufficient}>{\rm
reason}>{\rm t-symmetric \,\, c-Einstein\,\, gravity}
\end{CD}\]

\noindent with the Principle of Sufficient Reason in the horizontal
direction reading `time-(a)symmetric kinematics implies
time-(a)symmetric dynamics' and with the aforementioned `kinematical
Bohr correspondence limit 
mechanism' in the vertical direction\footnote{In the diagram above,
the prefix  
`${\rm t}$' stands for `${\rm time}$', `${\rm q}$' for `${\rm
quantum}$' and `${\rm c}$' for `${\rm classical}$'.}. Due to the discussion above,
this mechanism may be called `kinematical quantum time-asymmetry
breaking'\footnote{To be distinguished from the usual dynamical
symmetry breaking scenarios in the usual QFT of matter, as well as
certainly worthwhile to be compared with
an analogous `coherent condensation scenario' for the time-asymmetric
superconducting quantum causal nets in (Finkelstein, 1988). In
connection with the latter, Selesnick (1994, 1998) has noted that this condensation process that `decoheres' the `purely quantum' causal
net substratum to the (Minkowski space tangent to the) continuous
time-symmetric gravitational spacetime $M$ of
macroscopic experience, involves as a first
essential step the formation of coherent states of chronon-antichronon
Cooper pairs. It is the latter states that a coarse macroscopic
observer perceives as `the states of the quantum causal net' when, in
fact, as it was noted earlier, only quantum time-directed ({\it ie},
left-handed or orthochronous)
$sl(2,\com)$-chronons `exist' in the net, not right-handed or
antichronous  
$\overline{sl(2,\com)}$-antichronons. It may well be the case that the coherent
exponentiation of the micro-local Lie algebra $sl(2,\com)$ of chronons to yield
the macro-global spin-group $SL(2,\com)$ and its mirror 
image $\overline{SL(2,\com)}$ which is indistinguishable from it at
the coarse macroscopic level of resolution, is the reason for the
`doubling of macro-spinors' ({\it ie}, elements in both the $S$ and  
$S^{*}$ spaces) thus also for our coarse perception of a macroscopic
space/time ($P/T$) symmetry (Steve Selesnick in private communication).}.

Also, from this point of view the unnaturalness of GR in assuming a
Lorentzian metric-one that distinguishes in its signature a temporal from
three spatial directions-that dynamically propagates on a topologically
`spatial' spacetime manifold ($\R^{4}$)-one that regards time as
another undirected\footnote{That is to say, two-way and essentially reversible. See introduction and
(Finkelstein, 1988, Mallios and Raptis, 2000).} spatial dimension-is also exposed and in a sense remedied.
 
We conclude the present paper by remarking on the possibility that the
aforementioned `quantum topos project', as approached here from
$A$-schematic localizations, may prove to be a solid
platform for the conceptual unification of quantum logic and quantum
gravity.

>From a purely mathematical point of view, the conceptual development
that led to topos theory {\it per se} may be roughly summarized as
follows: local topological considerations gave rise to sheaf theory
which was then employed mainly by Grothendieck, who used purely
categorical tools such as `representable functors', to study ring-localizations thus effectively define `$R$-schemes' and lay the foundations of topos theory proper. 
It was mainly Lawvere in the late 60s who recognized the significance
of certain special categories, such as those that Grothendieck
considered in algebraic geometry\footnote{Which categories he (and Tierney) then called
`topoi' or `toposes'.}, for unifying until then apparently remote from each
other and very broad fields of
mathematics such as (algebraic) geometry and logic.

It is a generally accepted fact that topoi, regarded as generalized classical logical universes of
`continuously variable classical sets'\footnote{Like the ${\bf Sh}(X)$ and ${\rm
Set}^{W^{opp}}$ ones that we saw in the previous section.}, have 
surprisingly many `purely geometrical' characteristics. In fact, Lawvere
(1975) in his celebrated talk 
in the 1973 Bristol Logic Colloquium went as far as to literally identify `algebraic geometry'
with `geometric logic' in the light of topos theory. Indeed, so
remarkable is the interplay of logic and geometry in topos theory,
especially when sheaves or their descendant schemes are used in order to provide the motivating
conceptual background, that the Mac Lane and Moerdijk (1992) book
referred to 
at the back has all four words ({\it ie}, `sheaves', `geometry',
`logic' and `topos theory') in its very title.

On the other hand, it is really an amazing coincidence (in time) that
around the same time that topos theory was feverously on the make, Finkelstein
(1969, 1979) insisted from a purely physical point of view that as
Einstein showed us `the physicality and dynamical variability of the
geometry of the world', it was high-time
for us to investigate more deeply `the physicality and dynamical
variability of the logic the world', as well as ``{\it the possibility
that most of the phenomena that we see at higher levels are logical in
origin}''\footnote{It being understood that the curvaceous geometry of
the classical spacetime of macroscopic experience and GR could be
`explained' by appealing to a dynamically variable quantum physical logic.} (Finkelstein, 1969). 

A subsequent series of works of Finkelstein and collaborators
culminated in unifying the two fundamental principles of quantum
mechanics (uncertainty) and relativity (causality) on quantum
logico-algebraic grounds\footnote{That is, the development of the
Grassmann-algebraic Quantum Set Theory and its dynamical descendant
Quantum Causal Net Theory, as well as the subsequent integration of
these two theories into a more comprehensive Quantum
Relativity Theory (Finkelstein, 1996).}. Quantum Relativity in
particular (Finkelstein, 1996), intriguingly contends that ``{\it
logics come from dynamics}''.

A quantum topos, like our quantum causal ${\rm
Qauset}^{\vec{\as}^{opp}}$, when compared to its quantum logical
analogue ${\rm
Set}^{W^{opp}}$ of Butterfield and Isham (1997, 1998), exemplifies
precisely the close conceptual interplay between quantum logic and
(the time-asymmetric kinematics of) 
quantum gravity\footnote{Again, perceived as the dynamical theory of a
local non-commutative quantum causal topology $\vec{\rho}$.}, thus it
vindicates Finkelstein's (and Lawvere's) vision in physics (and
mathematics) that physical (mathematical) logic and physical
(mathematical) geometry are not that different afterall. In effect, in
the quantum causal topos ${\rm Qauset}^{\vec{\as}^{opp}}$, especially
when its close affinities with the quantum logical topos ${\rm
Set}^{W^{opp}}$ of Butterfield and Isham are highlighted in an
approach like ours via sheaf and scheme theory, we get the
first significant hints about how the warped quantum logic of the
world may be `determined' by the dynamics of a finitary quantum
causality, that is to say, by a locally finite, causal and quantal
Lorentzian gravity. ${\rm Qauset}^{\vec{\as}^{opp}}$ is not that far
from being a sound mathematical   
model of Finkelstein's deep insight mentioned above that ``{\it logics come from
dynamics}'' (Finkelstein, 1996). However, a more thorough examination
of this deep connection between quantum logic and quantum gravity 
is left for a paper currently in preparation (Raptis, 2000{\it d}).

In closing we must stress that the quantum topos project is yet
another instance of the general tendency in current theoretical physics of looking at the
problem of the quantum structure and dynamics of spacetime from an entirely algebraic point of
view. We feel strongly that the `innately algebraic' language of
sheaf and scheme theory, as well as of their categorical outgrowth, topos theory, is well suited to implement such an `algebraization
of quantum gravity' (Crane, 1995, Isham, 1997, Raptis, 1998, Butterfield and Isham,
2000), thus once again vindicate the prophetic words of Einstein (1956)\footnote{The quotation
below can also be found in (Mallios and Raptis, 2000). The reader may
also refer to (Crane, 1995) for more quotations of Einstein about 
`an entirely algebraic description of physical reality'.}: 

\begin{quotation}

{\footnotesize One can give good reasons why
reality cannot at all be represented by a continuous field. From the
quantum phenomena it appears to follow with certainty that a finite
system of finite energy can be completely described by a finite set
of numbers (quantum numbers). This does not seem to be in accordance
with a continuum theory, and must lead to an attempt to find a purely
algebraic theory for the description of reality. But nobody knows how
to obtain the basis of such a theory.}

\end{quotation}

\section*{\normalsize\bf ACKNOWLEDGMENTS}

The author wishes to thank cordially Jim Lambek, Anastasios Mallios, Steve
Selesnick and Roman Zapatrin for numerous technical exchanges over the years on sheaves, topoi
and algebra localizations, but most importantly, for stressing their
potential significance for quantum gravity. Freddy Van Oystaeyen's {\it impromptu} but timely communication about
scheme theory and non-commutative algebraic geometry is also acknowledged. The author is also grateful to
Chris Isham for being always available to communicate and ever ready to discuss his
latest work on applications of topos theory to quantum logic and
quantum gravity. Finally, this author is indebted to David Finkelstein and Ray
Sorkin for having emphasized all along that causality is actually a more
physical concept than topology. The present work was supported by
a postdoctoral research fellowship from the Mathematics Department of
the University of Pretoria, Republic of South Africa.

\section*{\normalsize\bf REFERENCES}

\noindent Alexandrov, P.S. (1956). {\it Combinatorial Topology}, 
Greylock,
Rochester, New York.

\noindent Ashtekar, A. (1986). {\it New Variables for Classical and
Quantum Gravity}, Physical Review Letters, {\bf 57}, 2244.

\noindent Baez, J. C. (1994). {\it Knots and Quantum Gravity}, Oxford
University Press, Oxford.

\noindent Baez, J. C. and Muniain, J. P. (1994). {\it Gauge Fields,
Knots 
and Quantum Gravity}, World Scientific, Singapore.

\noindent Balachandran, A. P., Bimonte, G., Ercolessi, E., Landi, G.,
Lizzi, F., Sparano, G. and Teotonio-Sobrinho, P. (1996). {\it
Noncommutative Lattices as Finite Approximations}, Journal of Geometry
and Physics, {\bf 18}, 163; e-print: hep-th/9510217.

\noindent Bell, J. L. (1988). {\it Toposes and Local Set Theories},
Oxford University Press, Oxford.

\noindent Birkhoff, G. and Von Neumann, J. (1936). {\it The Logic of
Quantum Mechanics}, Annals of Mathematics, {\bf 37}, 823.

\noindent Bombelli, L., Lee, J., Meyer, D. and Sorkin,
R. D. (1987). {\it Space-Time as a Causal Set}, Physical Review
Letters, {\bf 59}, 521. 

\noindent Breslav, R. B., Parfionov, G. N. and Zapatrin, R. R. (1999). 
{\it Topology Measurement within the Histories Approach}, Hadronic
Journal, {\bf 22}, 225.

\noindent Breslav, R. B. and Zapatrin, R. R. (2000). {\it Differential
Structure of Greechie Logics}, International Journal of Theoretical
Physics, {\bf 39}, 1027. 

\noindent Butterfield, J. and Isham, C. J. (1998). {\it A Topos Perspective on the Kochen-Specker Theorem: I. Quantum States as Generalized Valuations}, International Journal of Theoretical Physics, {\bf 37}, 2669.

\noindent Butterfield, J. and Isham, C. J. (1999). {\it A Topos Perspective on the Kochen-Specker Theorem: II. Conceptual Aspects and Classical Analogues}, International Journal of Theoretical Physics, {\bf 38}, 827.

\noindent Butterfield, J. and Isham, C. J. (2000). {\it Some Possible Roles for Topos Theory in Quantum Theory and Quantum Gravity}, Imperial/TP/98-99/76., to appear; e-print: gr-qc/9910005.

\noindent Butterfield, J., Hamilton, J. and Isham, C. J. (2000). {\it A Topos Perspective on the Kochen-Specker Theorem: III. Von Neumann Algebras as the Base Category}, International Journal of Theoretical Physics, to appear; e-print: quant-ph/9911020. 

\noindent Connes, A. (1994). {\it Noncommutative Geometry}, Academic
Press, New York.

\noindent Crane, L. (1995). {\it Clock and Category: Is Quantum Gravity Algebraic ?}, Journal of Mathematical Physics, {\bf 36}, 6180.

\noindent Dimakis, A., and M\"uller-Hoissen, F. (1999). {\it Discrete Riemannian
Geometry}, Journal of Mathematical Physics, {\bf 40}, 1518.

\noindent Einstein, A. (1924). {\it \"Uber den \"Ather},
{\it Schweizerische Naturforschende Gesellschaft Verhanflungen},
{\bf 105}, 85 (English translation by Simon Saunders:
`{\it On the Ether}' in {\it The Philosophy of Vacuum}, Eds. Saunders, S. and 
Brown, H., Clarendon Press, Oxford, 1991).

\noindent Einstein, A. (1956). {\it The Meaning of Relativity}, Princeton University Press, Princeton.

\noindent Finkelstein, D. (1958). {\it Past-Future Asymmetry of the
Gravitational Field of a Point Particle}, Physical Review, {\bf 110}, 965.

\noindent Finkelstein, D. (1969). {\it Matter, Space and Logic}, in
{\it Boston Studies in the Philosophy of Science}, {\bf 5},
Eds. Cohen, R. S. and Wartofsky, M. W., Dordrecht, Holland; reprinted
in {\it The Logico-Algebraic Approach to Quantum
Mechanics $II$}, Ed. Hooker, C. A., Reidel, Dordrecht, Holland (1979).

\noindent Finkelstein, D. (1979). {\it The Physics of Logic}, in {\it The Logico-Algebraic Approach to Quantum
Mechanics $II$}, Ed. Hooker, C. A., Reidel, Dordrecht, Holland (1979).

\noindent Finkelstein, D. (1988). {\it `Superconducting' Causal Nets}, International Journal of Theoretical Physics, {\bf 27}, 473.

\noindent Finkelstein, D. (1989). {\it Quantum Net Dynamics},
International Journal of Theoretical Physics, {\bf 28}, 441.

\noindent Finkelstein, D. (1996). {\it Quantum Relativity: A Synthesis of the Ideas of Einstein and Heisenberg}, Springer-Verlag, New York.

\noindent Finkelstein, D. and Hallidy, W. H. (1991). {\it Q: An
Algebraic Language for Quantum-Spacetime Topology}, International
Journal of Theoretical Physics, {\bf 30}, 463.

\noindent Haag, R. (1990). {\it Fundamental Irreversibility and the
Concept of Events}, Communications in Mathematical Physics, {\bf 132}, 245.

\noindent Hartshorne, R. (1983). {\it Algebraic Geometry}, 3rd edition, Springer-Verlag, New York-Heidelberg-Berlin.

\noindent Isham, C.J. (1989). {\it Quantum Topology and the Quantisation on the Lattice of Topologies}, Classical and Quantum Gravity, {\bf 6}, 1509. 

\noindent Isham, C. J. (1997). {\it Topos Theory and Consistent Histories: The Internal Logic of the Set of All Consistent Sets}, International Journal of Theoretical Physics, {\bf 36}, 785.

\noindent Lambek, J. and Scott, P. J. (1986). {\it Introduction to
Higher Order Categorical Logic}, Cambridge University Press,
Cambridge.

\noindent Landi, G. (1997). {\it An Introduction to Noncommutative
Spaces and their Geometry}, e-print: hep-th/9701078; (a revised and
enlarged version has been published, in the Lecture Notes in Physics,
Monograph 51, Springer-Verlag, Berlin-Heidelberg; e-print: hep-th/9701078).

\noindent Landi, G. and Lizzi, F. (1997). {\it Projective Systems of
Noncommutative Lattices as a Pregeometric Substratum}, in {\it Quantum
Groups and Fundamental Physical Applications}, Eds. Kastler, D. and
Rosso, M., Nova Science Publishers, USA; e-print: math-ph/9810011.

\noindent Lawvere, F.W. (1975). {\it Continuously Variable Sets:
Algebraic Geometry$=$Geometric Logic}, in {\it Proceedings of the Logic
Colloquium in Bristol (1973)}, North-Holland, Amsterdam.

\noindent Mac Lane, S. and Moerdijk, I. (1992). {\it Sheaves in Geometry and 
Logic: A First Introduction to Topos Theory}, Springer-Verlag, New York.

\noindent Mallios, A. (1998). {\it Geometry of Vector Sheaves: An Axiomatic Approach to Differential Geometry}, vols. 1-2, Kluwer Academic Publishers, Dordrecht (volume 3, with further physical applications to Yang-Mills theories and gravity, is currently in preparation).

\noindent Mallios, A. and Raptis, I. (2000). {\it Finitary Spacetime Sheaves 
of Quantum Causal Sets: Curving Quantum Causality}, paper submitted to the 
International Journal of Theoretical Physics.

\noindent Markopoulou, F. (2000). {\it Quantum Causal Histories}, Classical and Quantum Gravity, 
{\bf 17}, 2059.

\noindent Mulvey, C. and Pelletier, J. W. (2000). {\it On the
Quantization of Points}, Journal of Pure and Applied Algebra, to
appear.

\noindent Penrose, R. (1987). {\it Newton, Quantum Theory and Reality}, 
in {\it 300 Years of Gravitation}, Eds. Hawking, S. W. and Israel, W., Cambridge 
University Press, Cambridge.

\noindent Raptis, I. (1998). {\it Axiomatic Quantum Timespace
Structure: 
A Preamble to the Quantum Topos Conception of the Vacuum},
Ph.D. Thesis, 
University of Newcastle upon Tyne, UK.

\noindent Raptis, I. (2000{\it a}). {\it Algebraic Quantization of Causal Sets}, 
International Journal of Theoretical Physics, {\bf 39}, 1233.

\noindent Raptis, I. (2000{\it b}). {\it Finitary Spacetime Sheaves}, 
International Journal of Theoretical Physics, {\bf 39}, 1699.

\noindent Raptis, I. (2000{\it
c}). {\it Locally Finite, Causal and Quantal Einstein Gravity}, in preparation.

\noindent Raptis, I. (2000{\it d}). {\it Non-Classical Linear Xenomorph as Quantum Causal Space and the Quantum Topos of Finitary Spacetime Schemes of Quantum Causal Sets}, in preparation.

\noindent Raptis, I. and Zapatrin, R. R. (2000). {\it Quantization of 
Discretized Spacetimes and the Correspondence Principle},
International  
Journal of Theoretical Physics, {\bf 39}, 1. 

\noindent Rawling, J. P. and Selesnick, S. A. (2000). {\it Orthologic
and Quantum Logic. Models and Computational Elements}, Journal of the
Association for Computing Machinery, {\bf 47}, 721.

\noindent Redhead, M. (1990). {\it Incompleteness, Non-Locality and
Realism}, Clarendon Press, Oxford.

\noindent Rideout, D. P. and Sorkin, R. D. (2000). {\it A Classical
Sequential Growth Dynamics for Causal Sets}, Physical Review D, {\bf
61}, 024002; e-print: gr-qc/9904062.

\noindent Rosenthal, K. I. (1990). {\it Quantales and their Applications}, Longman Scientific and Technical, Essex, England.

\noindent Rota, G.-C., (1968). {\it On The Foundation Of Combinatorial Theory,
I. The Theory Of M\"obius Functions}, Zeitschrift f\"ur
Wahrscheinlichkeitstheorie, {\bf 2}, 340.

\noindent Selesnick, S. A. (1991). {\it Correspondence Principle for the 
Quantum Net}, International Journal of Theoretical 
Physics, {\bf 30}, 1273.

\noindent Selesnick, S. A. (1994). {\it Dirac's Equation on the
Quantum Net}, Journal of Mathematical Physics, {\bf 35}, 3936.

\noindent Selesnick, S. A. (1998). {\it Quanta, Logic and Spacetime: 
Variations on Finkelstein's Quantum Relativity}, World Scientific, 
Singapore.

\noindent Shafarevich, I. R. (1994). {\it Basic Algebraic Geometry 2}, 2nd edition, Springer-Verlag, Berlin Heidelberg New York.

\noindent Sorkin, R. D. (1990{\it a}). {\it Does a Discrete Order Underlie Spacetime and its Metric}, in Proceedings of the Third Canadian Conference on General Relativity and Relativistic Astrophysics, Eds. Cooperstock, F. and Tupper, B., World Scientific, Singapore. 

\noindent Sorkin, R. D. (1990{\it b}). {\it Spacetime and Causal Sets}, in the Proceedings of the SILARG VII Conference, Cocoyoc, Mexico (pre-print).

\noindent Sorkin, R. D. (1991). {\it Finitary Substitute for Continuous 
Topology}, International Journal of Theoretical Physics, {\bf 30}, 923.                 

\noindent Sorkin, R. D. (1995). {\it A Specimen of Theory Construction from 
Quantum Gravity}, in {\it The Creation of Ideas in Physics},
Ed. Leplin, J., Kluwer Academic Publishers, Dordrecht.

\noindent Van Oystaeyen, F. (2000{\it a}). {\it Algebraic Geometry for Associative Algebras}, Marcel Dekker, New York.

\noindent Van Oystaeyen, F. (2000{\it b}). {\it Is the Topology of Nature 
Noncommutative ?} and {\it A Grothendieck-type of Scheme Theory for Schematic 
Algebras}, research seminars given at the Mathematics Department of the 
University of Pretoria, Republic of South Africa on 2-3/2/2000 (notes available).

\noindent Van Oystaeyen, F. and Verschoren, A. (1981). {\it
Non-Commutative Algebraic Geometry}, Springer Lecture Notes in
Mathematics, {\bf 887}, Springer-Verlag, Berlin-Heidelberg.

\noindent Wheeler, J. A. (1964). In {\it Relativity, Groups and Topology}, Eds. De Witt, C. and De Witt, B. S., Gordon and Breach, London.

\noindent Zapatrin, R. R. (1998). {\it Finitary Algebraic Superspace},
International Journal of Theoretical Physics, {\bf 37}, 799.

\noindent Zapatrin, R. R. (2000). {\it Incidence Algebras of
Simplicial Complexes}, Pure Mathematics and its Applications, to appear; e-print: math.CO/0001065.

\newpage

\thispagestyle{empty}  

\begin{table}[h!t]
\begin{tabular}{p{0.45\textwidth}%
@{\hspace{0.05\textwidth}}p{0.45\textwidth}}
\hline \\
\hskip 0.7in{\underline{\large\bf ${\bf C^{*}}$-quantales}} &\hskip 0.7in{\underline{\large\bf ${\bf\amg}$-schemes}}\\
&\\
\multicolumn{2}{c}{\bf The algebras}\\
&\\
infinite dimensional non-abelian $C^{*}$-algebras $A$ closed under $*$ & finite
dimensional non-abelian algebras $\amg$ with $\dagger$-conjugates $\amg^{\dagger}$\\
&\\
\multicolumn{2}{c}{\bf The relevant representations}\\
&\\
\hskip 0.1in equivalence classes of $*$-irreps& \hskip 0.5in equivalence classes of irreps\\
&\\
\multicolumn{2}{c}{\bf The relevant spectra}\\
&\\
\hskip 0.6in maximal: ${\rm Max}\, A$ & \hskip 0.8in primitive: ${\bf Spec}\,\amg(\vec{P})$\\
&\\
\multicolumn{2}{c}{\bf The quantum points}\\
&\\
closed linear subspaces of $A$ or closed two-sided ideals in
$A$ & kernels of irreps of $\amg$ or primitive ideals in $\amg$\\
&\\
\multicolumn{2}{c}{\bf The non-commutative topology}\\
&\\
based on the non-commutative `\&' operation on quantum points& based on the
non-commutative `$\circ$' operation on quantum points \\
&\\
\multicolumn{2}{c}{\bf The underlying logic and geometry}\\
&\\
quantum and not localized/ungauged, hence flat &
quantum and localized/gauged, hence curved\\
&\\
\multicolumn{2}{c}{\bf The physical interpretation}\\
&\\
undirected quantal spatial topologies between quantum space points& directed
quantum causal topologies between quantum spacetime events\\
&\\
\hline \\
\end{tabular}
\caption{Comparison between $C^{*}$-quantales and primitive $\amg$-finschemes}
\end{table}

\newpage

\pagestyle{empty}
\setlength{\topmargin}{0.2cm}
\begin{table}[h!t]
\begin{tabular}{p{0.45\textwidth}%
@{\hspace{0.05\textwidth}}p{0.45\textwidth}}
\hline \\
{\underline{\bf The neo-classical topos ${\bf 
Set}^{\bf W^{opp}}$}} &\hskip 0.1in{\underline{\bf The quantum topos
${\bf Qauset}^{\bf\vec{\as}^{opp}}$}}\\
&\\
\multicolumn{2}{c}{\bf The objects}\\
&\\
\hskip 0.5in `spatial' presheaves ${\bf Spec}$ &\hskip 0.3in `causal' presheaves
$\stackrel{\longrightarrow}{\bf Spec}$ \\
&\\
\multicolumn{2}{c}{\bf The subobject classifiers}\\
&\\
a complete distributive lattice (Heyting algebra or locale) $\omg$& a reticular causal and quantal
orthochronous Lorentz-spin algebra ${\bf g_{i}}=sl(2,\com)_{i}$ \\
&\\
\multicolumn{2}{c}{\bf The arrows or morphisms}\\
&\\
\hskip 0.4in ${\bf Spec}$-presheaf morphisms&\hskip 0.2in$\stackrel{\longrightarrow}{\bf
Spec}$-presheaf morphisms\\
&\\
\multicolumn{2}{c}{\bf The relevant quantum points}\\
&\\
Stone spaces ${\rm Spec}\, A$ of Boolean subalgebras $A$ of a quantum
lattice ${\cal{L}}({\cal{H}})$ in the base poset category $W$ of ${\bf
Spec}$ & primitive spectra ${\bf Spec}\,\amg_{i}(\vec{P}_{i})$ in the
base poset category $\vec{\as}$ of $\stackrel{\longrightarrow}{\bf Spec}$\\
&\\
\multicolumn{2}{c}{\bf The relevant localizations}\\
&\\
local sections of ${\bf Spec}$ yield ${\rm Spec}\,
A\rightarrow\omg(A)$ localizations of (truth in) ${\rm Set}$ over $W$& local sections of $\stackrel{\longrightarrow}{\bf
Spec}$ yield ${\cal{A}}_{i}:~\amg^{1}_{i}\rightarrow{\bf g_{i}}$
gauging of (symmetries of) ${\rm Qauset}$ over ${\vec{\as}}$\\
&\\
\multicolumn{2}{c}{\bf The internal logic and topology; twisted geometry}\\
&\\
neo-classical and locally localic, and spatial (non-relativistic); localized, hence warped,
thus intransitive local intuitionistic implication
$\stackrel{*}{\Rightarrow}$&locally quantalic ({\it ie},
non-distributive logical and non-commutative topological), and causal (relativistic); gauged, hence curved,
thus intransitive local quantum causality $\vec{\rho}$\\
&\\
\multicolumn{2}{c}{\bf The physical interpretation}\\
&\\
contextualized constructivistic valuations for variable sets or
`variable many-valued intuitionistic truth'&kinematics
for dynamically
variable finitary
quantum causal topologies or `gravitational quantum causality'\\
&\\
\hline \\
\end{tabular}
\caption{Comparison between ${\rm 
Set}^{W^{opp}}$ and ${\rm  
Qauset}^{\vec{\as}^{opp}}$}
\end{table}
  
\end{document}